\documentclass[twocolumn,showpacs,amsmath,amssymb]{revtex4}

\usepackage{graphicx,rotating}
\usepackage{dcolumn}
\usepackage{bm}
\usepackage{epsfig}
\usepackage{longtable}

\begin{document}

\title{Spin-dependent Seebeck effect and huge growth of
thermoelectric parameters at band edges in H- and F-doped graphene,
free-standing and deposited on $4H$-SiC(0001) C-face}

\author{Ma\l{}gorzata~Wierzbowska}\email{malgorzata.wierzbowska@fuw.edu.pl}
\affiliation{%
Institute of Theoretical Physics, Faculty of Physics,
University of Warsaw, ul. Ho\.za 69, 00-681 Warszawa, Poland
}%

\author{Adam Dominiak}\email{adam.dominiak@itc.pw.edu.pl}
\affiliation{%
Institute of Heat Engineering, Faculty of Power
and Aeronautical Engineering, Warsaw University of Technology,
ul. Nowowiejska 21/25, 00-665 Warszawa, Poland}

\date{\today}

\begin{abstract}
Graphene halfly doped with H or F possesses local magnetization at the undoped C sites.
Thus the Seebeck coefficient is different for each spin channel and its sign also
changes depending on the spin polarization. 
Deposition of doped graphene on the C-face $4H$-SiC(0001) with two buffer layers
substantially varies the electronic and thermoelectric properties. 
These properties are efficiently calculated from the semiclassical Boltzmann equations,
using the maximally-localized Wannier-functions interpolation 
of the band structures obtained with the density-functional theory.
Our results indicate large growth of the thermopower and the ZT efficiency at the band edges. 
We show in the model discussion that this phenomenon is more general and applies
also to other systems than graphene. 
It gives prospect for developing new spintronic devices working in the band-edge regime.
\end{abstract}


\keywords{spin-dependent Seebeck effect, doped graphene, SiC, thermopower, 
MLWF, Boltzmann equations}

\maketitle

\section{Introduction} 

The direct conversion of heat into electrical energy is an intensively 
studied field for last half century. The crucial issue in this 
research is finding new materials with specific thermal and electric properties. 
The semiconductor materials development extends applications concerning 
thermoelectric effects like Seebeck and Peltier, which are the basic phenomena 
among other non-equilibrium thermoelectric effects.
Seebeck effect is the electric voltage generation in a conductor  
under a temperature gradient, and it is used in the heat sink applications. 
With the inverse effect, namely Peltier, an active cooling can be received. 
Recent studies on spintronics and spin-caloritronics \cite{spincalo} 
revealed a spin analogue of the Seebeck effect, which can occur in magnets,
also insulating \cite{spinS1,spinS2}.
Spin-Seebeck effect can be longitudinal e.g. caused by magnon-induced spin current 
injected from a ferromagnet \cite{long1,long2}, 
or transverse e.g. from the substrate \cite{transv1, transv2}.
Spin-dependent Peltier effect has been also observed \cite{spinP}.  

The dimensionless figure of merit $ZT = \sigma S^{2}T / \kappa$
gives a measure of the system efficiency in the thermoelectric effects.
The Seebeck coefficient ($S$) describes the electric voltage 
generation under the temperature gradient stimulus through the sample with no 
electrical current. 
It is an intrinsic parameter of each material and depends on temperature ($T$). 
Higher electrical conductivity ($\sigma$), 
lower thermal conductivity ($\kappa$),
and higher Seebeck coefficient are necessary for better and more effective  
thermoelectric materials. 
Finding new materials with higher ZT factor is difficult, 
because the thermal conductivity is strongly positively related 
with the electrical conductivity, due to the fact that both 
are depended on crystal and electronic structures and carrier concentration. 
These physical properties stay in conflict to each other.
The ZT factor easily allows to compare the performance of thermoelectric devices
based on particular materials. For the thermoelectric devices which could be
competitive with the mechanical devices (power generation or cooling), 
the  ZT should be about 3. The value of ZT equal to 1 was reached in 1990s,
the value of 2 is now technically available but at high temperatures 
only \cite{ZT1,ZT2}. 

About two-thirds of all energy used worldwide is dissipated as heat and the
demand for an alternative technology to reduce fossil fuels use, 
along with their environmental impact, leads to important regiments of research fields
including that of direct thermal-electrical energy conversion
via thermoelectricity. 
Seebeck and Peltier effects applied within proper materials can solve part
of the problems faced ahead sustainable growth. 
As the semiconductor materials are promising for waste-heat recovery,
the graphene has been also studied \cite{graf}.
Gas flow sensor with the use of Seebeck effect and its correlation 
with Bernoulli law was presented for single layer graphene \cite{single}
and multilayer graphene \cite{multi}. 
These results demonstrate that graphene has great potential for flow 
sensors and energy conversion devices. 
For the graphene nanoribbons, the thermoelectric power 
including the Nernst and Seebeck effects were  
studied as well, using the nonequilibrium Green’s function approach \cite{Green}
and by molecular dynamics methods \cite{anomal,armchair}. 
The calculations of graphene behavior in the magnetic field and the related 
anomalous thermoelectric effects were reported in \cite{magn}. 
Semiclassical approach for the electronic-density dependence of 
the dc-conductivity, the optical conductivity, the thermal conductivity,
the thermopower, and the classical Hall effect in graphene has been 
proposed in \cite{semiclass}. The  first-principles band-structure calculations 
were also performed for the spin-caloritronics 
in the zigzag graphene nanoribbons \cite{zigzag}.
Nanocarbon related systems were investigated as thermoelectric-parameter 
enhanced materials \cite{COOH}.

The effect of the carbon defect in the graphene structure 
\-- hydrogen doping and/or carbon vacancy \-- 
on magnetism has been proposed in \cite{defect}.
Recent theoretical studies of halfly doped graphene with hydrogen
and also fluorine concluded that the magnetism in this system, as it is usual
for the triangular lattice, is probably frustrated and the ferromagnetic state
has a bit higher energy \cite{Rudenko}. 
Nevertheless, the local magnetic domains usually form in such systems. 
It will be interesting to investigate the magnetic states in H- and F-adsorbed graphene
at various doping patterns, e.g. rectangular or linear or random. 
Recently, the dynamics of H formation on the bilayer graphene and the mechanism
of the long-range ferromagnetism in various graphene multilayered structures
have been studied theoretically \cite{palacios1,palacios2}.  
Spin-polarization in graphene can be also obtained by random substitutional doping
with B or N, but only sites belonging to one sublattice \cite{spin-BN}.  

In this work, we focus on the ferromagnetic state of halfly doped graphene,
free-standing and deposited on the C-face $4H$-SiC(0001) substrate with two
buffer layers.
We calculated the thermoelectric properties by means of the first-priciples methods,
and found that the Seebeck effect is spin dependent in both the H- and F-doped systems.
We call the effect described here as the spin-dependent Seebeck effect. 
Our effect is purely intrinsic, 
i.e. it originates from the band structures of two spin-channels in the ferromagnetic graphene. 
Interestingly, we found large increase of the Seebeck coefficient and also of the ZT efficiency
at the band edge. Similar phenomenon has been very recently predicted in the model calculations
by Sharapov and Varlamov \cite{Sharapov} for gapped graphene, and by Hao and Lee \cite{Hao}
for bilayer graphene with the band-gap tuned by the external electric field.
Unusual fact is that, we obtained larger growth of the ZT efficiency not at high temperatures,
as common findings in the high-ZT materials, but at lower temperatures.

The paper is organized as follows: we give details of the calculations in section 2,
present the results  in section 3 \-- for the free-standing systems in subsection 3.1
and deposited systems in subsection 3.2, analyse the mathematical expressions
for the thermoelectric properties using 1D, 2D and 3D models of the band structure
in section 4, and conclude in section 5. 

\begin{figure}
\centerline{ \includegraphics[scale=0.35]{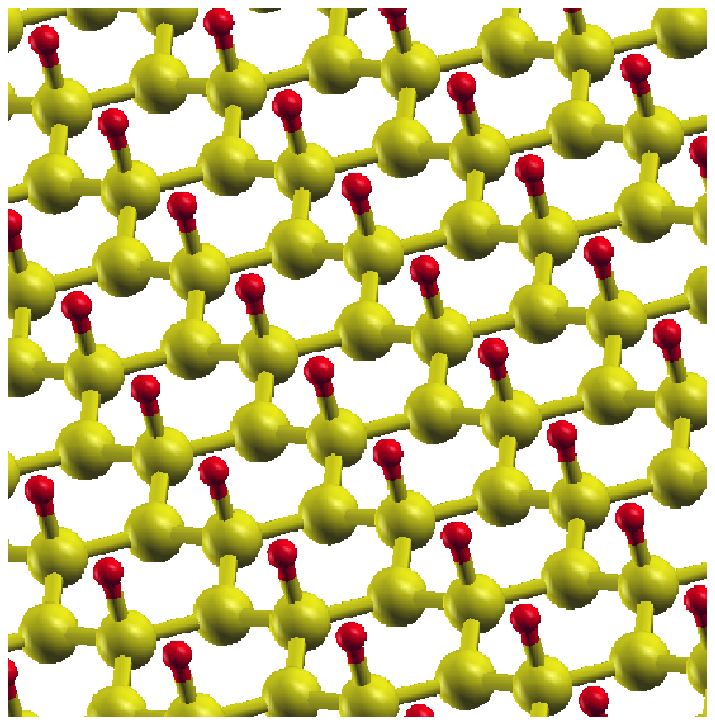}}
\centerline{ \includegraphics[scale=0.65]{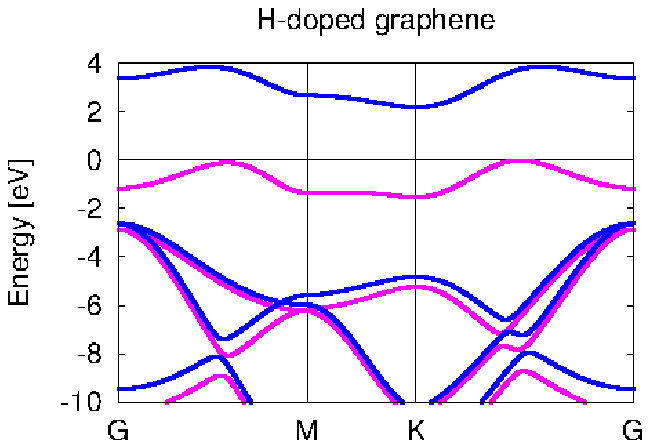} }
\centerline{ \includegraphics[scale=0.65]{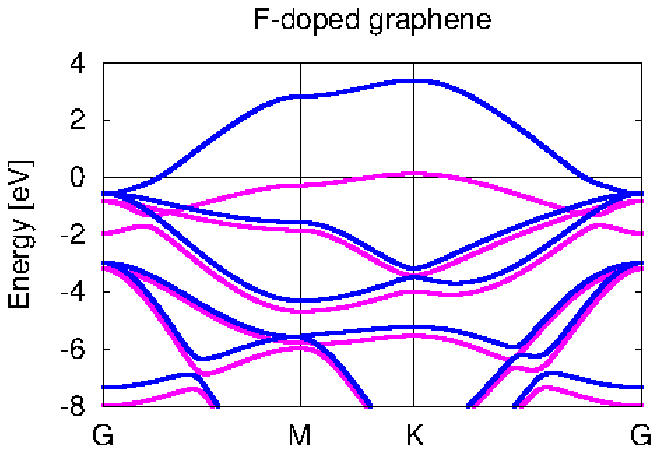} }
\caption{The atomic structure of graphene halfly doped with H or F,
and the band structures of H- and F-doped graphene.
The majority spin is printed with the magenta color and the minority spin
with the blue color.}
\label{b1}
\end{figure}

\section{Theoretical details} 

We performed the density-functional theory calculations of the band structures, and 
applied the {\sc Quantum ESPRESSO} suit of codes \cite{qe} which uses the 
plane-wave basis set and the pseudopotentials for the core electrons.
The exchange-correlation functional was chosen for the gradient corrected
Perdew-Wang(91) parametrization. The ultrasoft pseudopotentials
were used with the energy cutoffs 20 Ry and 200 Ry for the plane-waves and
the density, respectively. 
The Monkhorst-Pack uniform k-mesh in the Brillouin zone has been set to
$40\times 40\times 1$ for the free-standing and $10\times 10\times 1$ for the deposited
graphene sheets. The vacuum separation between the periodic slabs was around 50 \AA. 

In order to fit the band structure on the very fine k-mesh, 
i.e. $1600\times 1600$ points in the graphene plane,
we employed the wannier90-2.0.0 package  \cite{w90-2}\footnote{www.wannier.org}
which interpolates bands
using the maximally-localized Wannier functions \cite{RMP}.
The thermoelectric distribution function
and the thermoelectric properties were obtained from the BoltzWann 
post-processing code \cite{BW} with the constant relaxation time set to 10 fs.

The geometric parameters of buckled graphene under H- and F-doping were taken
from the work by Rudenko et al. \cite{Rudenko}. For the geometry of the doped graphene
deposited on the C-face $4H$-SiC(0001) surface, we used the 
$\sqrt{3}\times\sqrt{3}R30^{\circ}$
supercell to match the graphene sheets with the substrate. Between the doped graphene
and the surface C-atoms, we put two buffer layers in the ABA stacking including
the doped top-layer. The interlayer distances were taken from the experimental
work by Borysiuk et al. \cite{Borys}. 

\begin{figure}
\leftline{ \includegraphics[scale=0.35]{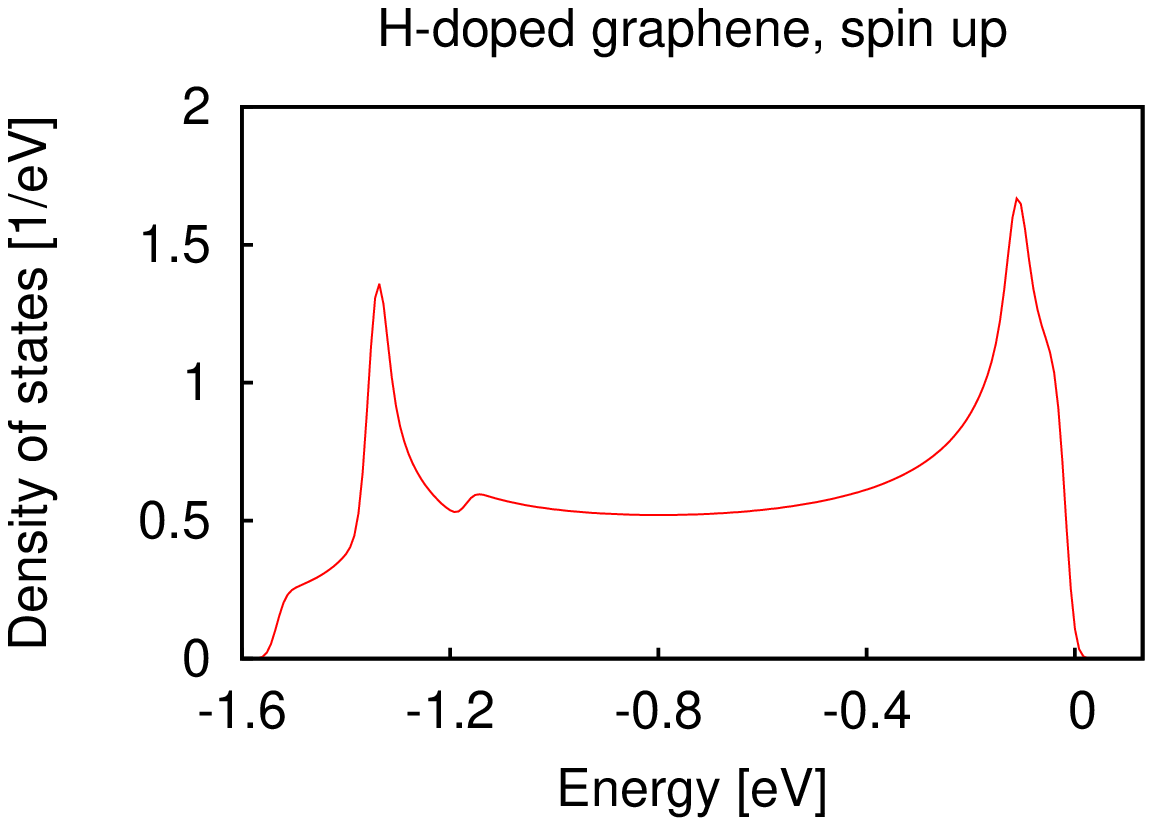}
\includegraphics[scale=0.35]{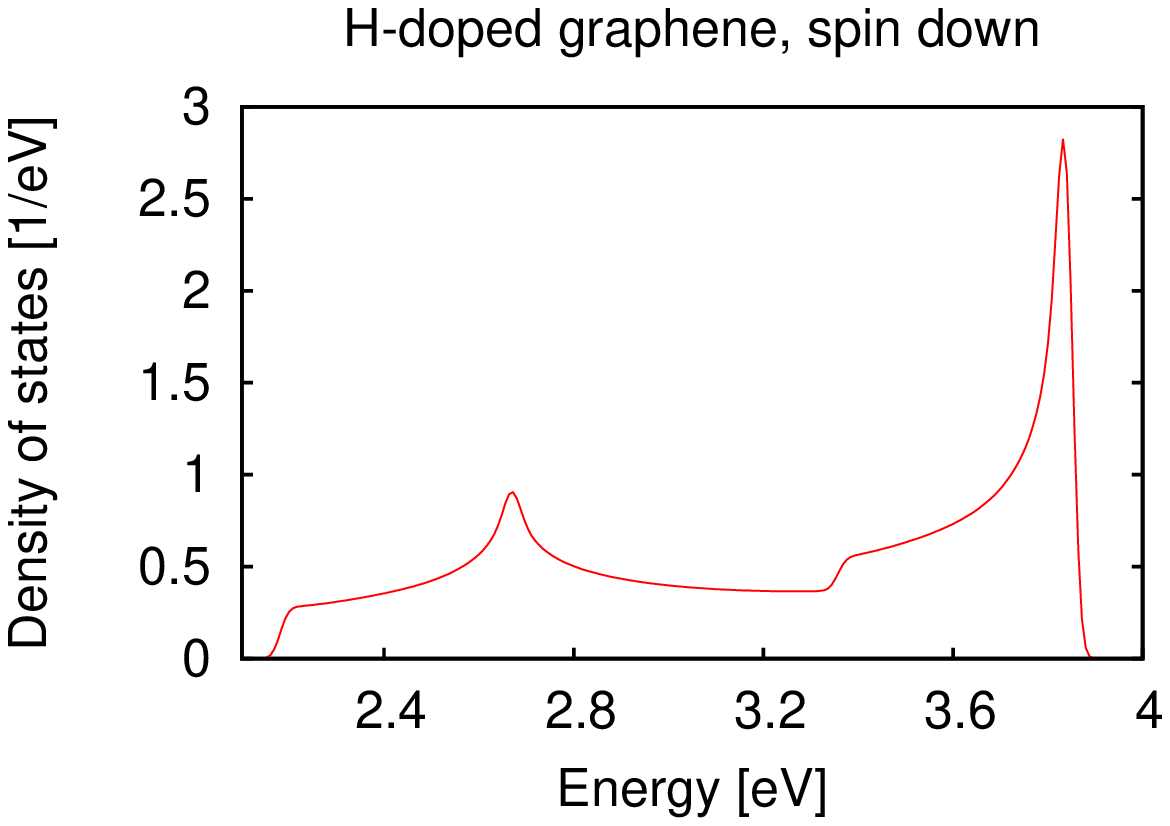}}
\vspace{2mm}
\leftline{ \includegraphics[scale=0.35]{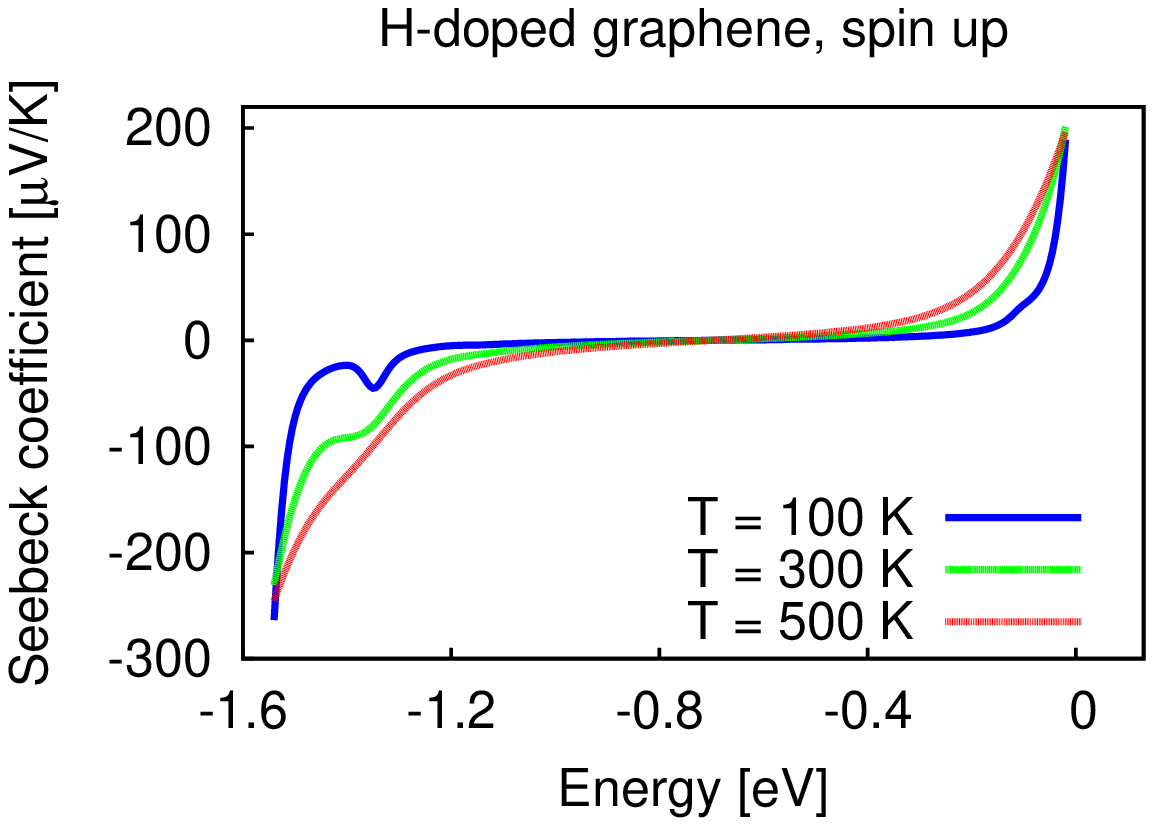}
\includegraphics[scale=0.35]{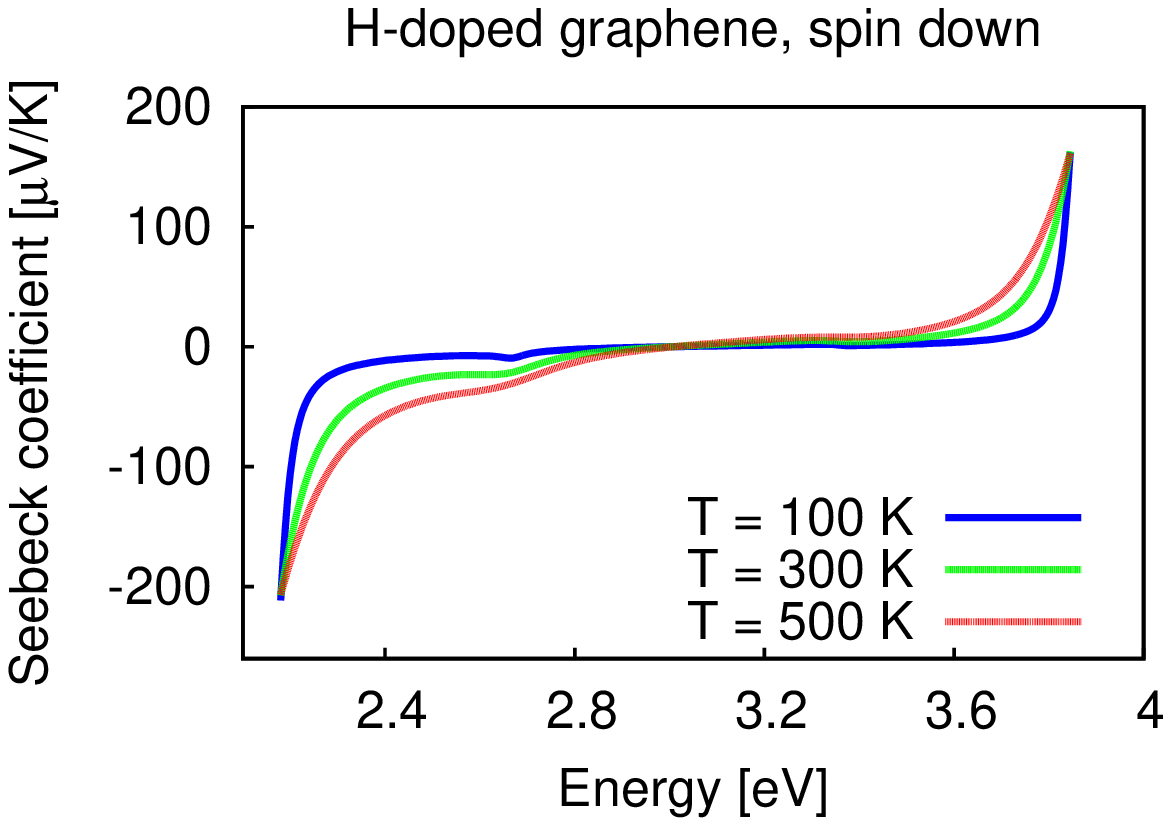}}
\vspace{2mm}
\leftline{ \includegraphics[scale=0.35]{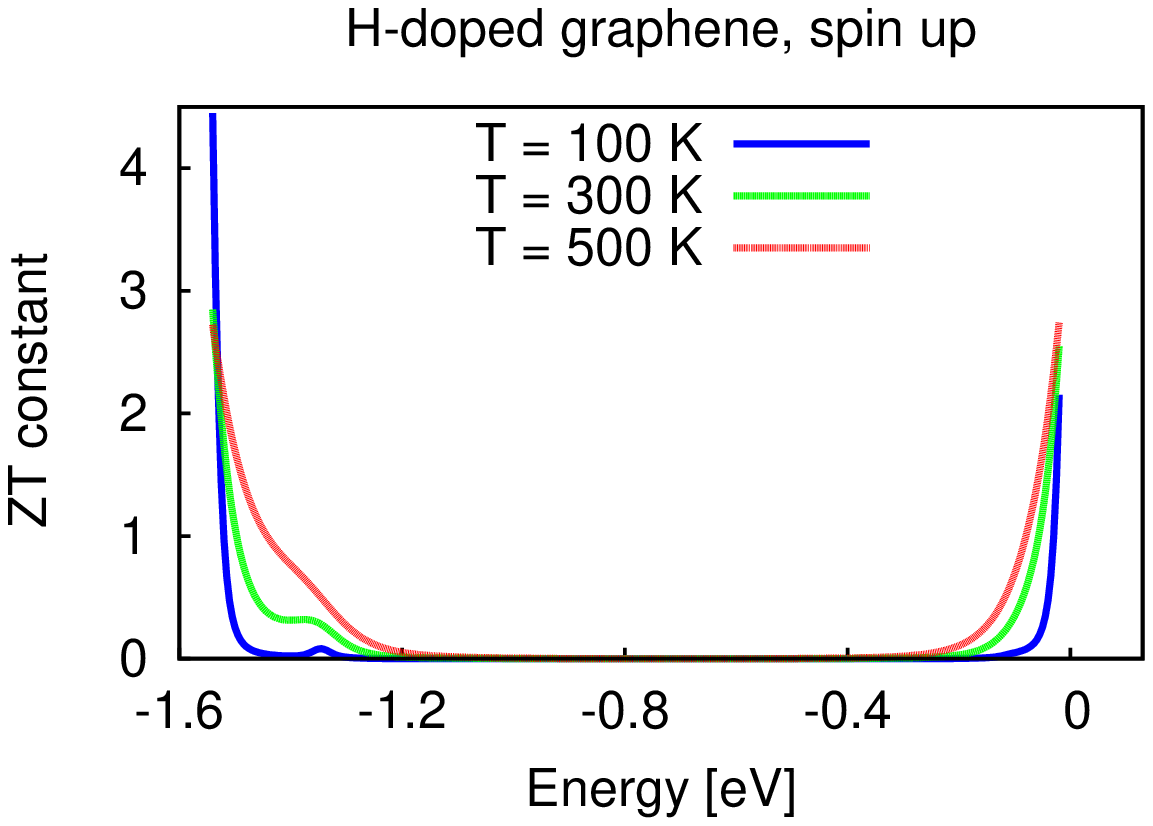}
\includegraphics[scale=0.35]{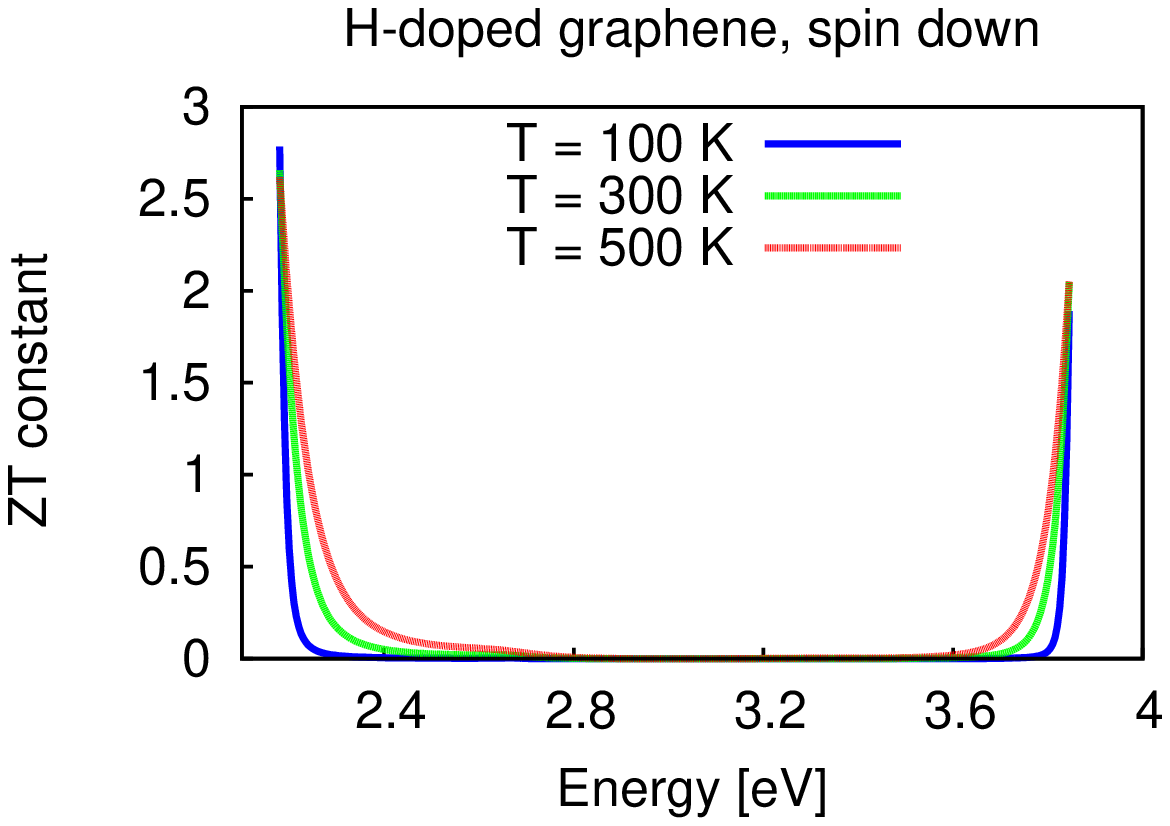}}
\caption{The density of states, the Seebeck coefficient,
and the ZT figure of merit for H-doped graphene.}
\label{b2}
\end{figure}

\begin{figure}
\leftline{ \includegraphics[scale=0.35]{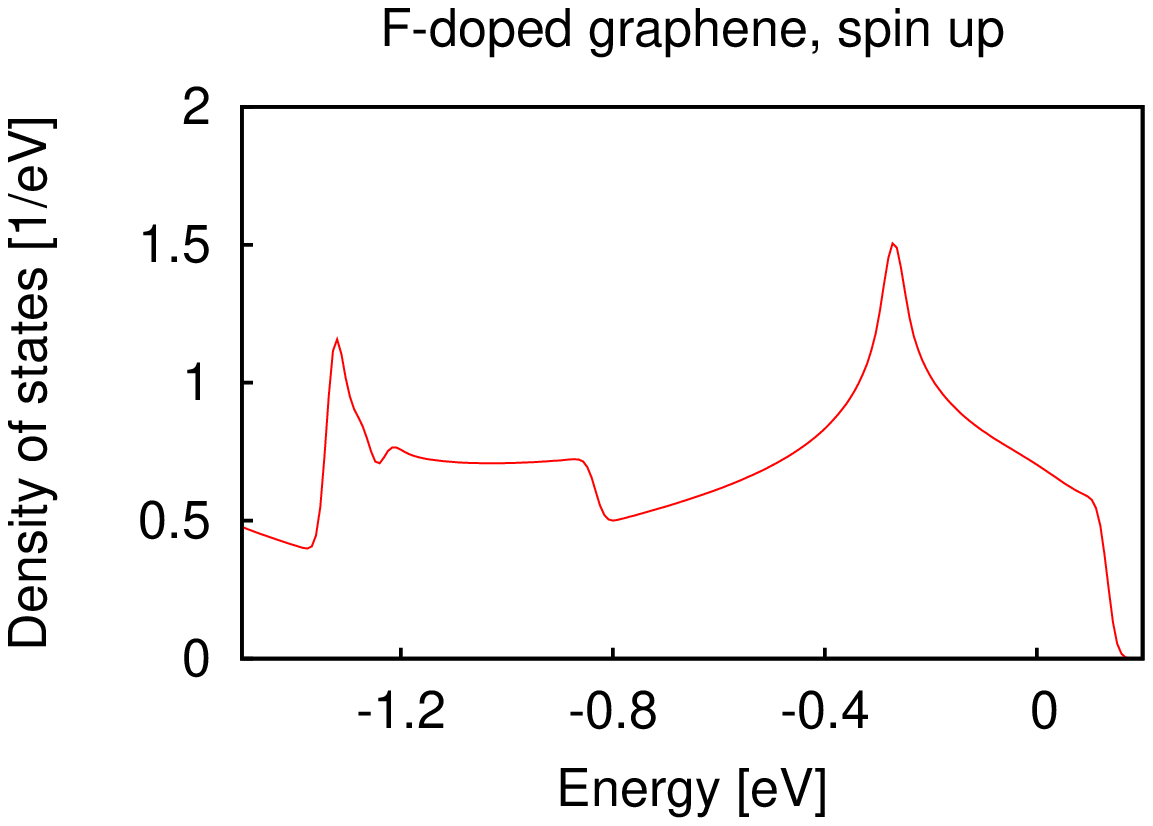}
\includegraphics[scale=0.35]{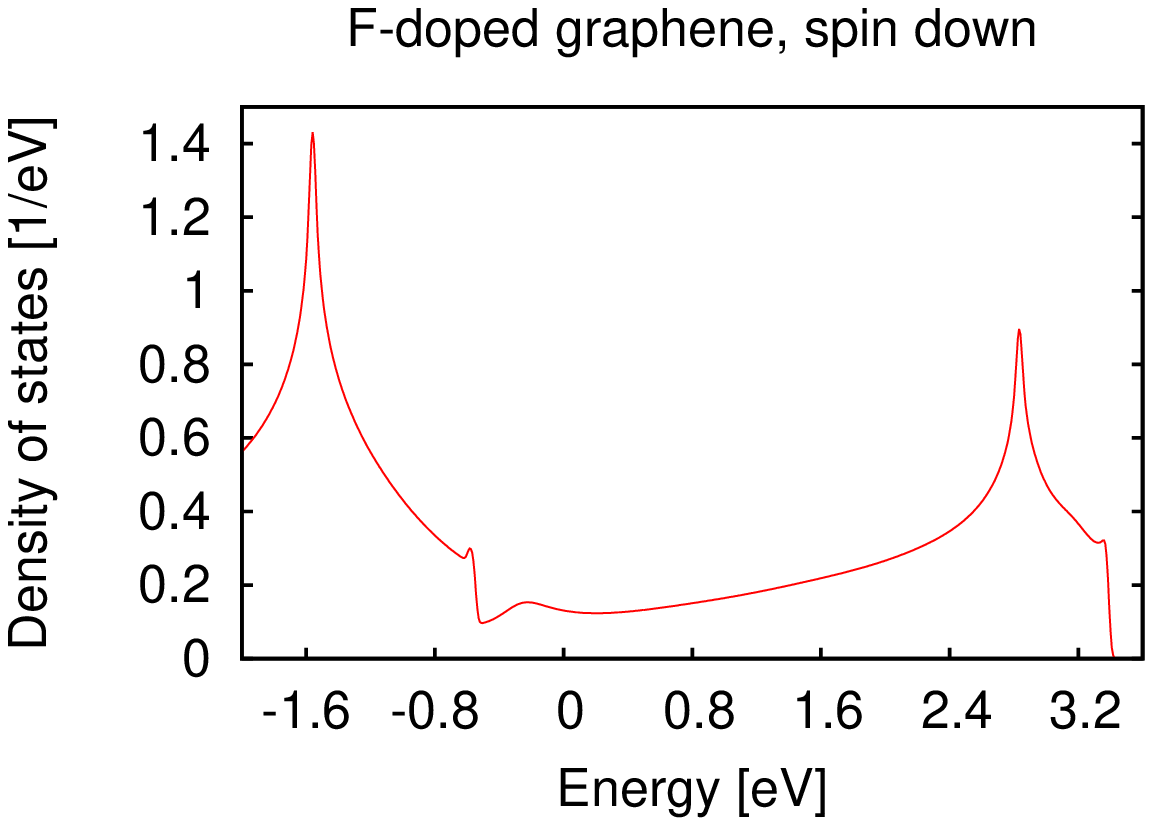}}
\vspace{2mm}
\leftline{ \includegraphics[scale=0.35]{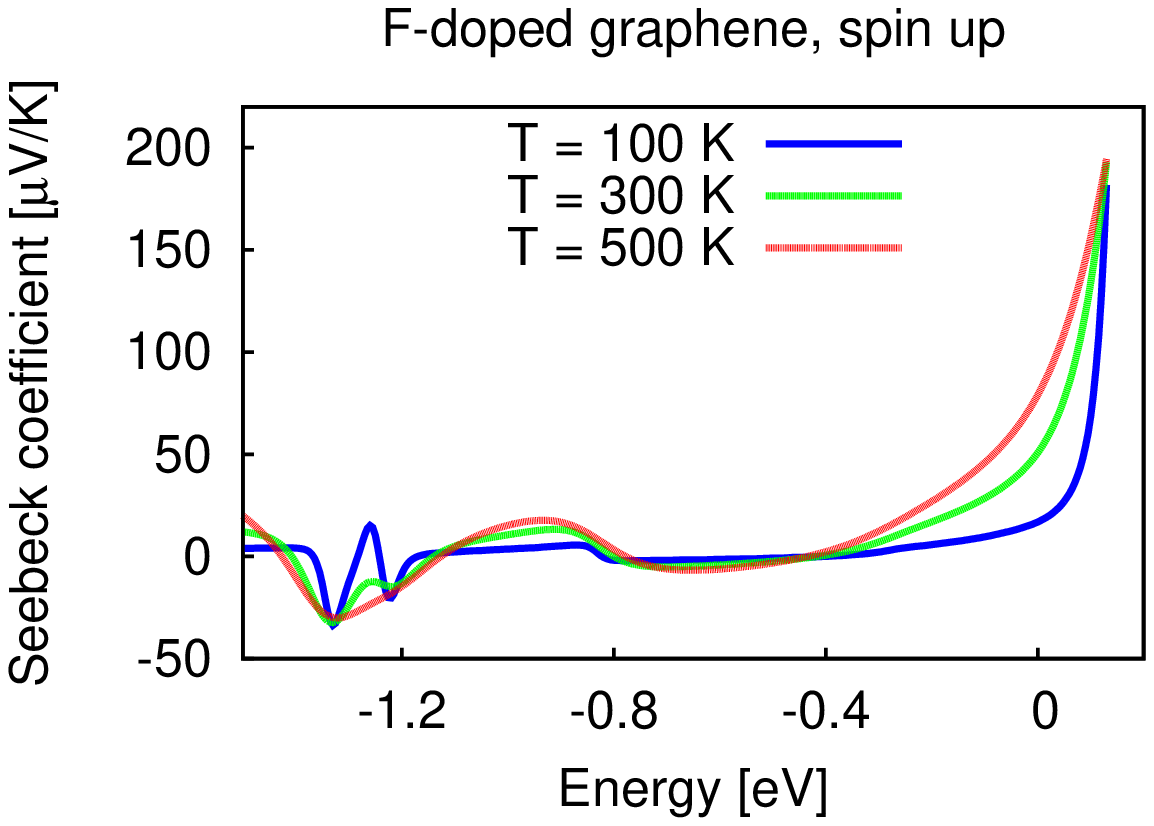}
\includegraphics[scale=0.35]{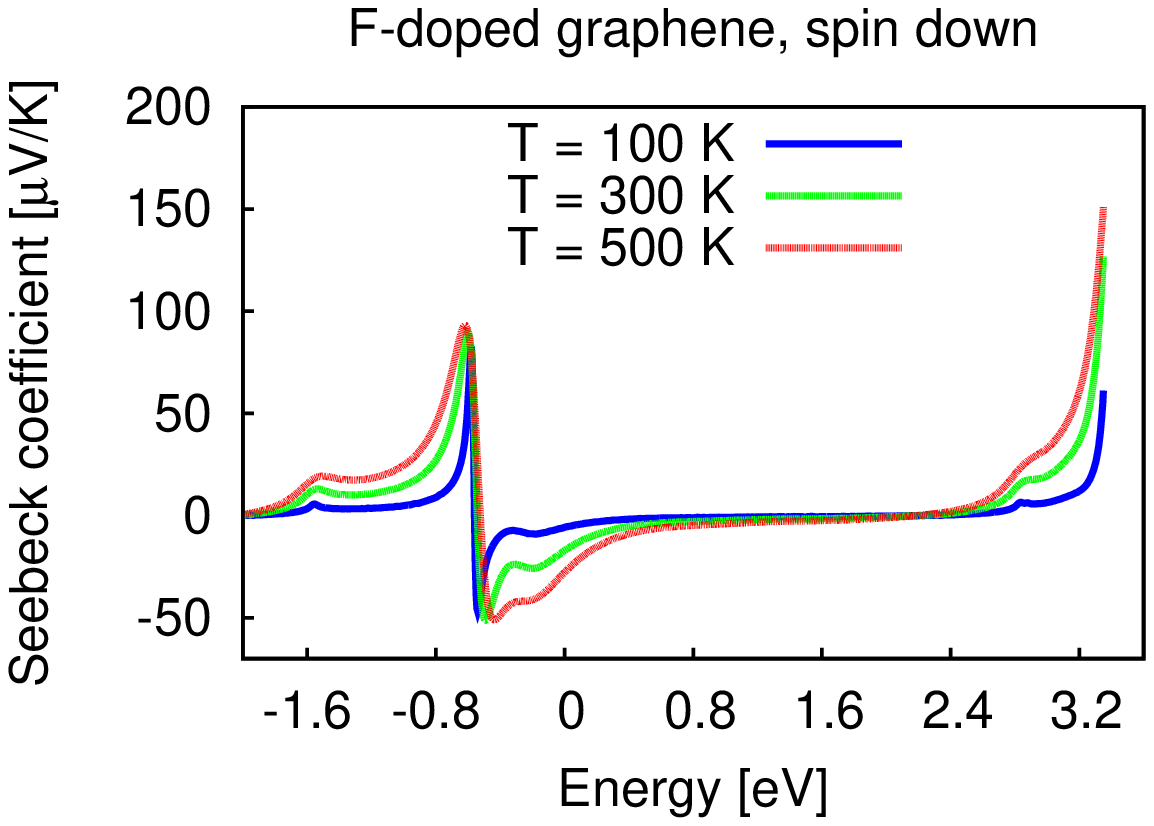}}
\vspace{2mm}
\leftline{ \includegraphics[scale=0.35]{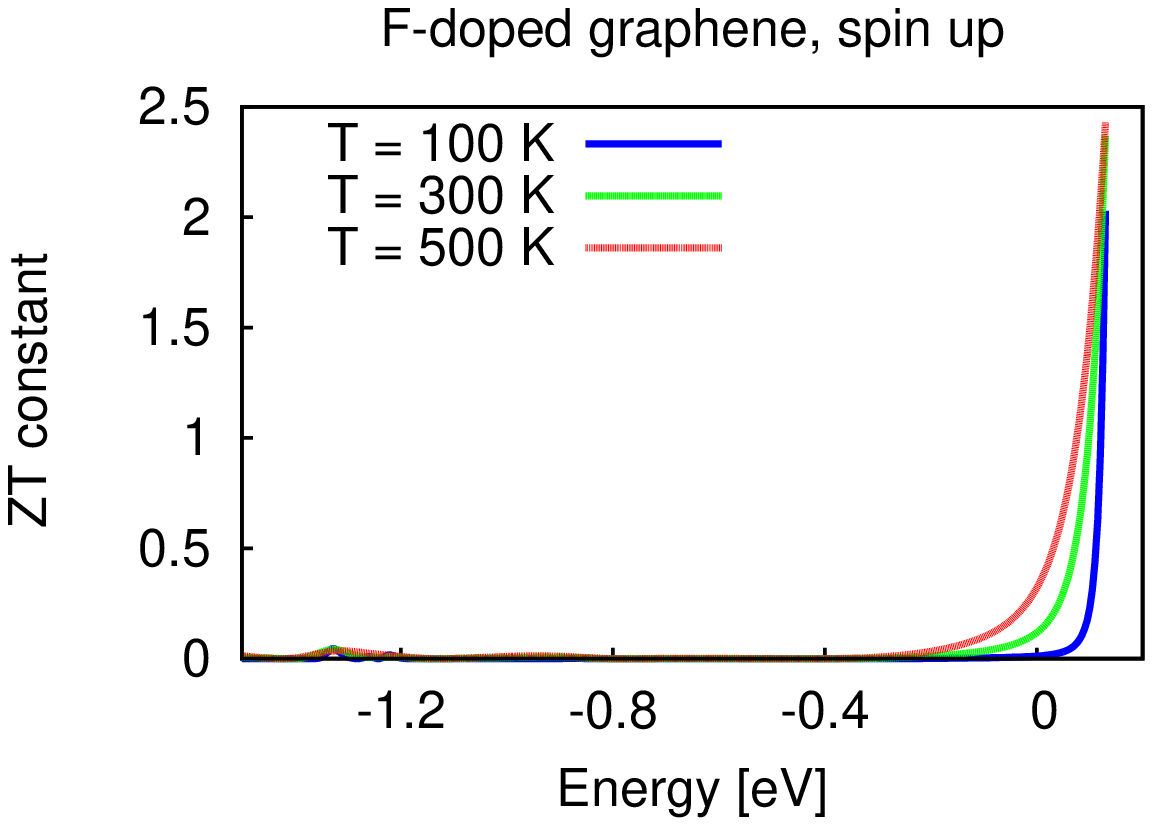}
\includegraphics[scale=0.35]{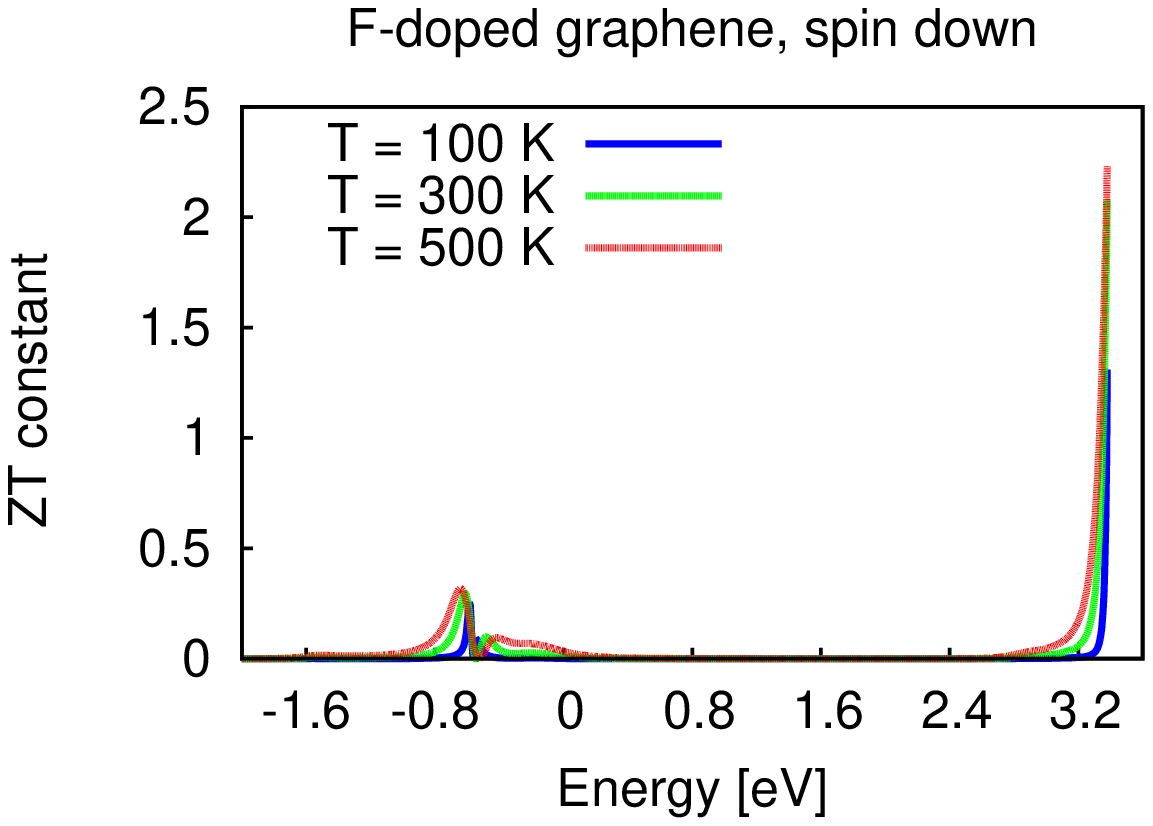}}
\caption{The density of states, the Seebeck coefficient,
and the ZT figure of merit for F-doped graphene.}
\label{b3}
\end{figure}

\section{Results} 

\subsection{Free-standing doped graphene monolayer} 

The adsorption of hydrogen atom on graphene produces the spin-polarization of neighboring
carbon atoms \cite{defect}. The same occurs for half-doping by hydrogens or 
fluorines. The magnetic moments per $p_z$-state of the free C-atom are about 
0.76 $\mu_B$ and 0.59 $\mu_B$ for H- and F-doping, respectively, obtained
with the local density approximation in Ref. \cite{Rudenko}. Our calculations
with the generalized gradient approximation give, as usually for this scheme,
higher magnetic moments per C-atom: 1.0 $\mu_B$ for H-doping and 
0.84 $\mu_B$ for F-doping. The top view of the doped graphene layer and 
the spin-polarized band structures are displayed in Fig.~1.
The impurity states in the H-doped case are completely separated on 
the occupied- and the unoccupied-states sides for the spin-up and spin-down channels, 
respectively.  The experimental confirmation 
for existence of the separated impurity band in the hydrogenated graphene is given
in angle-resolved photo-emission spectroscopy (ARPES) studies in Refs. \cite{Dan1,Dan2}.
It is interesting to mention that the local hydrogen doping causes large structural folding
of the graphene sheets, which finds various applications \cite{origami}. 
Also the thermal conductivity has been studied for the strained that way graphene
sheets \cite{H-strain}.
For the F-doped system, the impurity spin-states are separated below and above the Fermi
level around the K and M symmetry-points, but close to 
the middle of the Brillouin zone both spin-states are merged with the valence band.
We will show that the above band-structure details have consequences 
for the thermoelectric properties. 

\begin{figure}
\centerline{ \includegraphics[scale=0.45]{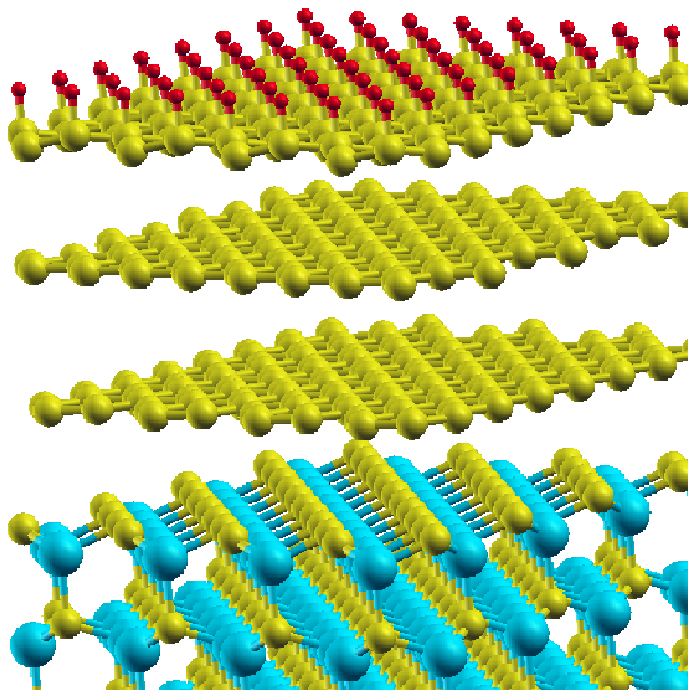}}
\centerline{ \includegraphics[scale=0.45]{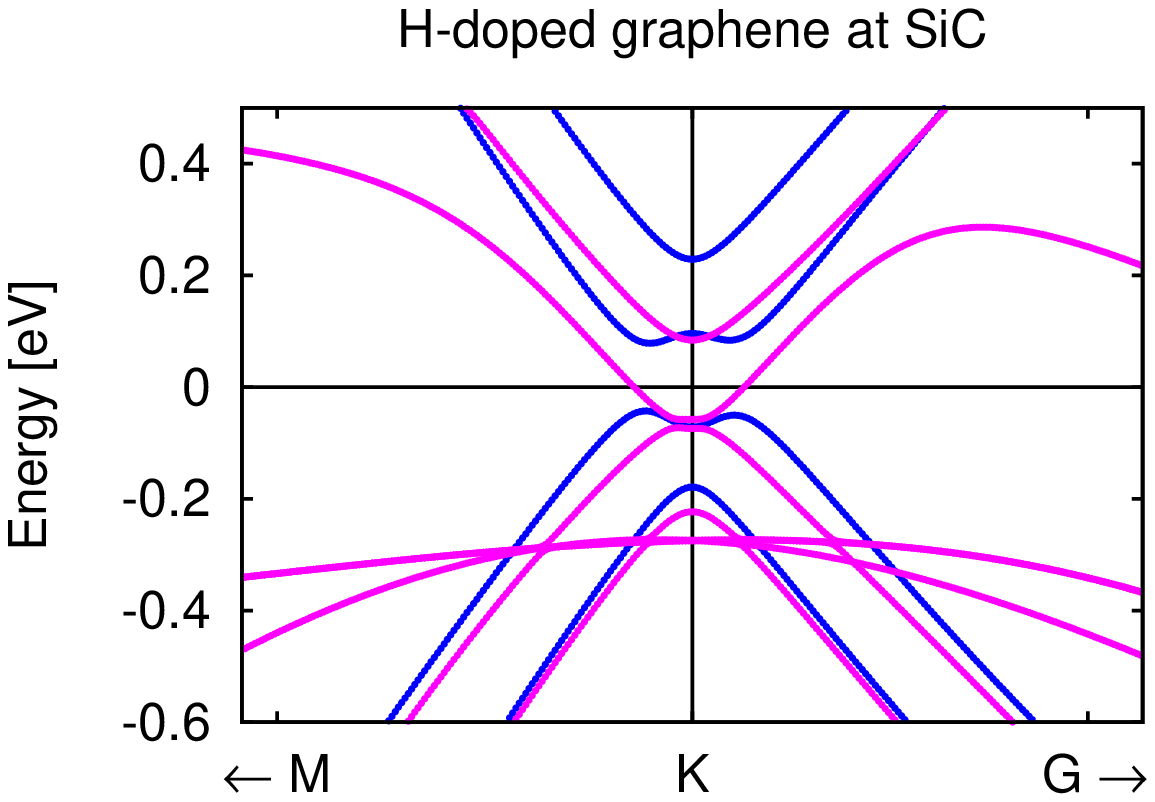} }
\centerline{ \includegraphics[scale=0.45]{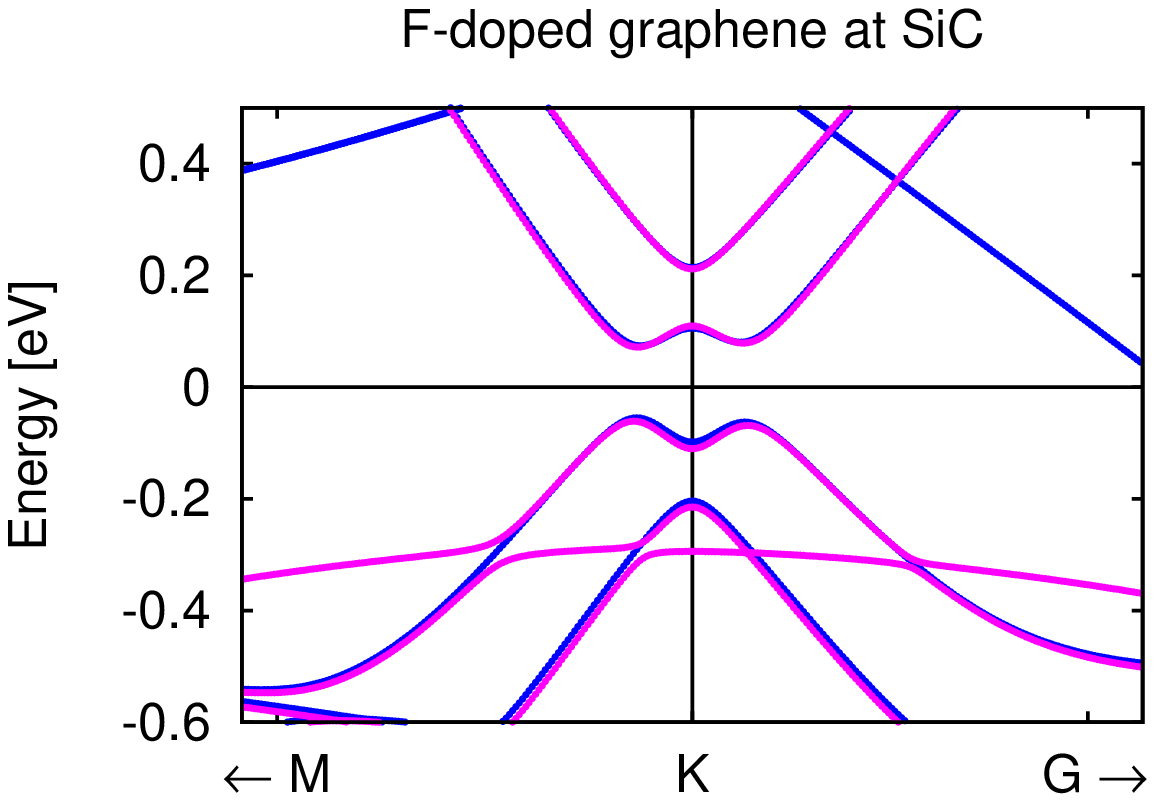} }
\caption{The atomic structure of graphene doped with H or F and deposited on
the C-face SiC(0001) surface with two buffer layers in ABA stacking,
and the band structures of H- and F-doped graphene.
The majority spin is printed with magenta color and
the minority spin with blue color.}
\label{b4}
\end{figure}

In Fig.~2, we present the density of states (DOS), the Seebeck coefficient and 
the ZT efficiency for the H-doped graphene, and in Fig.~3 the same properties 
of the F-doped graphene. 
In the middle of the impurity bandwidth for both spins 
and both H- and F-doping, the thermoelectric parameters are very low.
But at the separated band edges, the thermopower and the ZT efficiency grows  
to very high values. This is for both the high- and the low-energy ends of 
the H-impurity bands, and for the F-impurity bands only at the high-energy band edge.
The thermoelectric parameters at the low-energy parts of the F-impurity bands
are low, since in that region, the impurity bands touch the valence band manifold. 
For the spin-down channel of the F-doped graphene, 
we observe the Seebeck-coefficient and the ZT-efficiency extrema similar to these obtaned
for the pure graphene close to the band crossing at the Fermi level 
near the K-point \cite{noi-arxiv}.
Less pronounced features of the same shape in the thermopower curve 
are visible below -1.2 eV for the spin-up case, 
and they originate from the band crossings close to the G-point towards 
the M and K symmetry-points.
Since the desired thermoelectric properties occur at the band-edges, we check
the shape of the electronic density function. The DOS of all investigated free-standing 
systems is high or moderate close to the impurity band edges. 
Thus the enhancement effect of the thermoelectric parameters  
should be visible in the experiments. 

\subsection{Doped graphene monolayer deposited on the C-face $4H$-SiC(0001) 
with two buffer-layers} 

For the technological reasons, the substrate at which the electronically active layer
was deposited or grown is as important as the upper layer itself.
The C-face SiC deposited graphene characterizes by very good technological parameters,
such as the electron mobility up to 1800 cm$^2$/Vs, and  
the defect-free large wafers even 150~mm in diameter can be grown \cite{Strup}.
Moreover, the thermopower parameter of SiC is very good, about -480 $\mu$V/K.
Therefore SiC can transfer heat from the electronic device  \cite{Ssic}. 
All that made our choice for the substrate for testing the spin-dependent
thermoelectric properties of the deposited H- and F-doped graphene.

\begin{figure}
\leftline{ \includegraphics[scale=0.35]{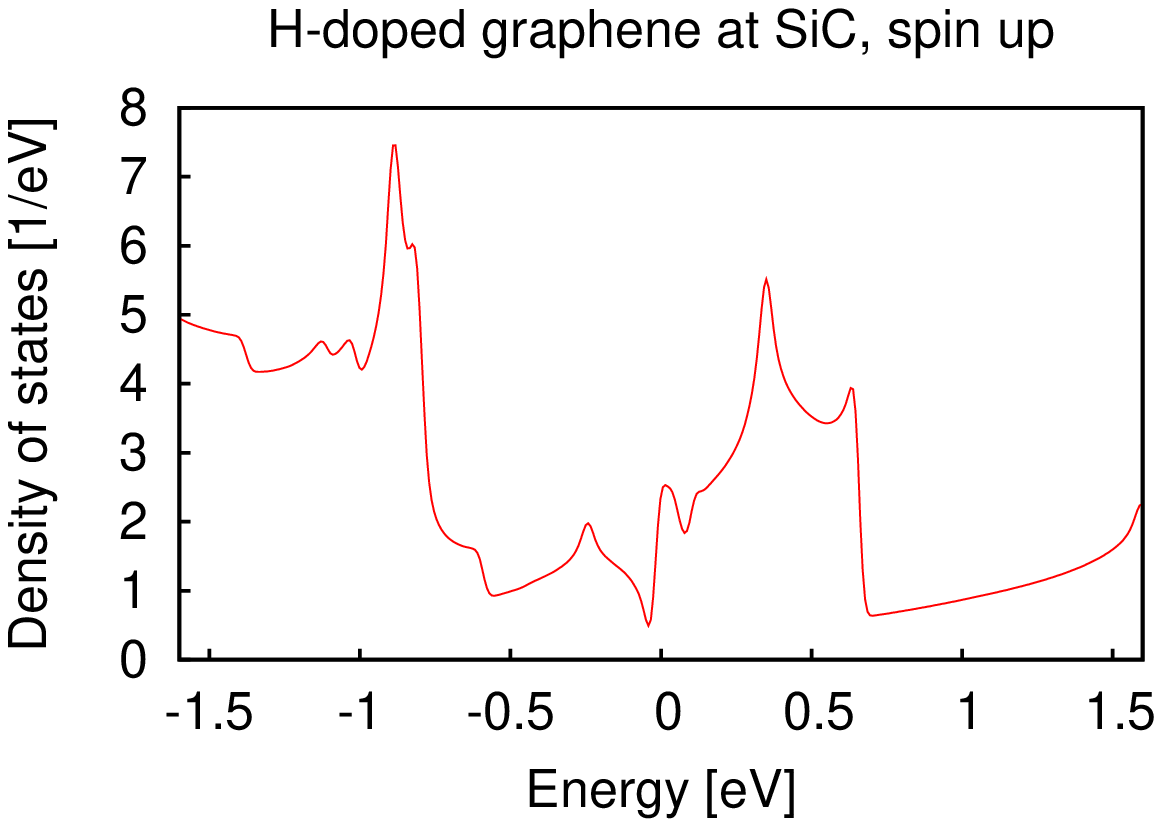}
\includegraphics[scale=0.35]{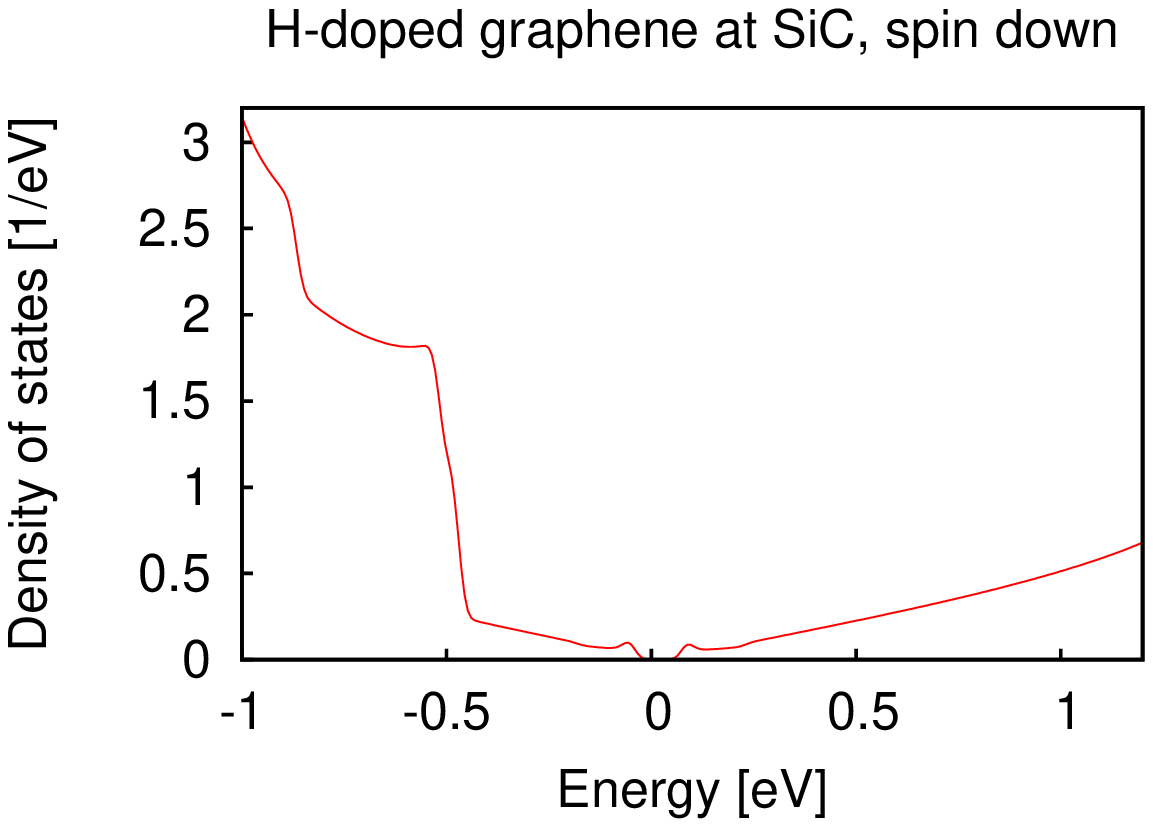}}
\vspace{2mm}
\leftline{ \includegraphics[scale=0.35]{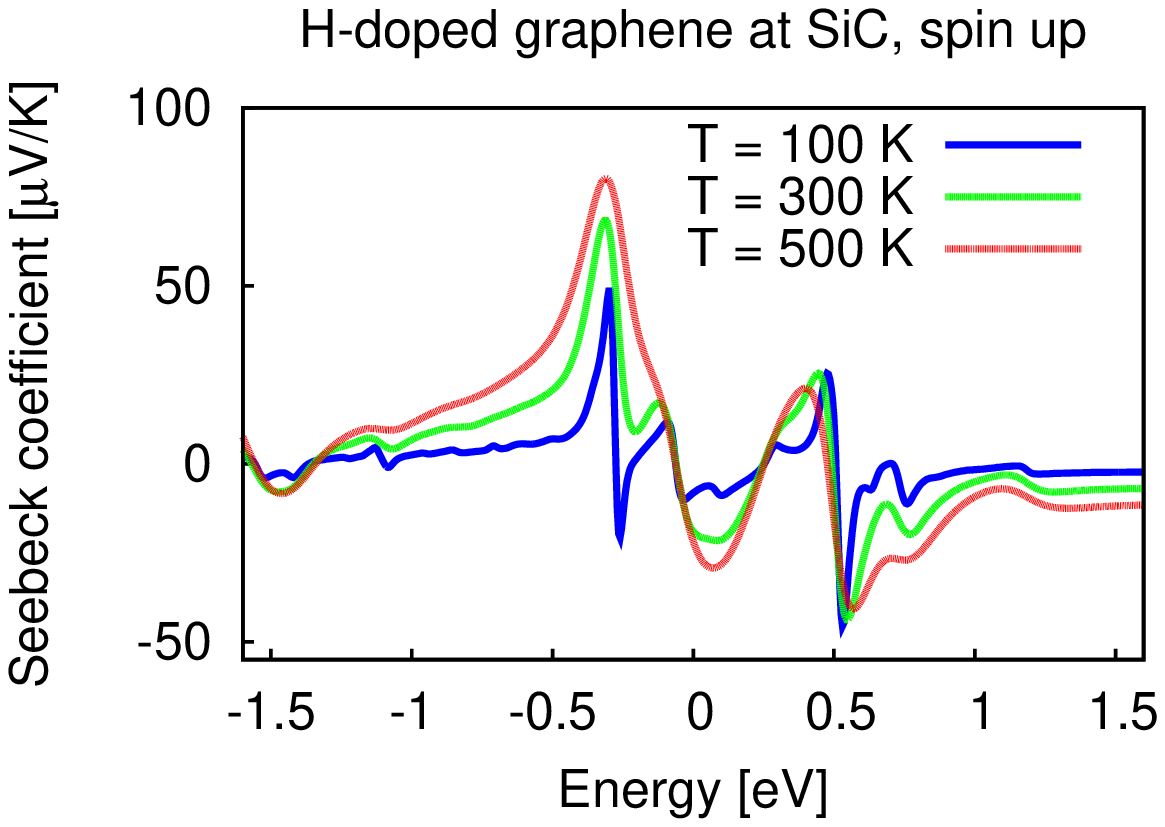}
\includegraphics[scale=0.35]{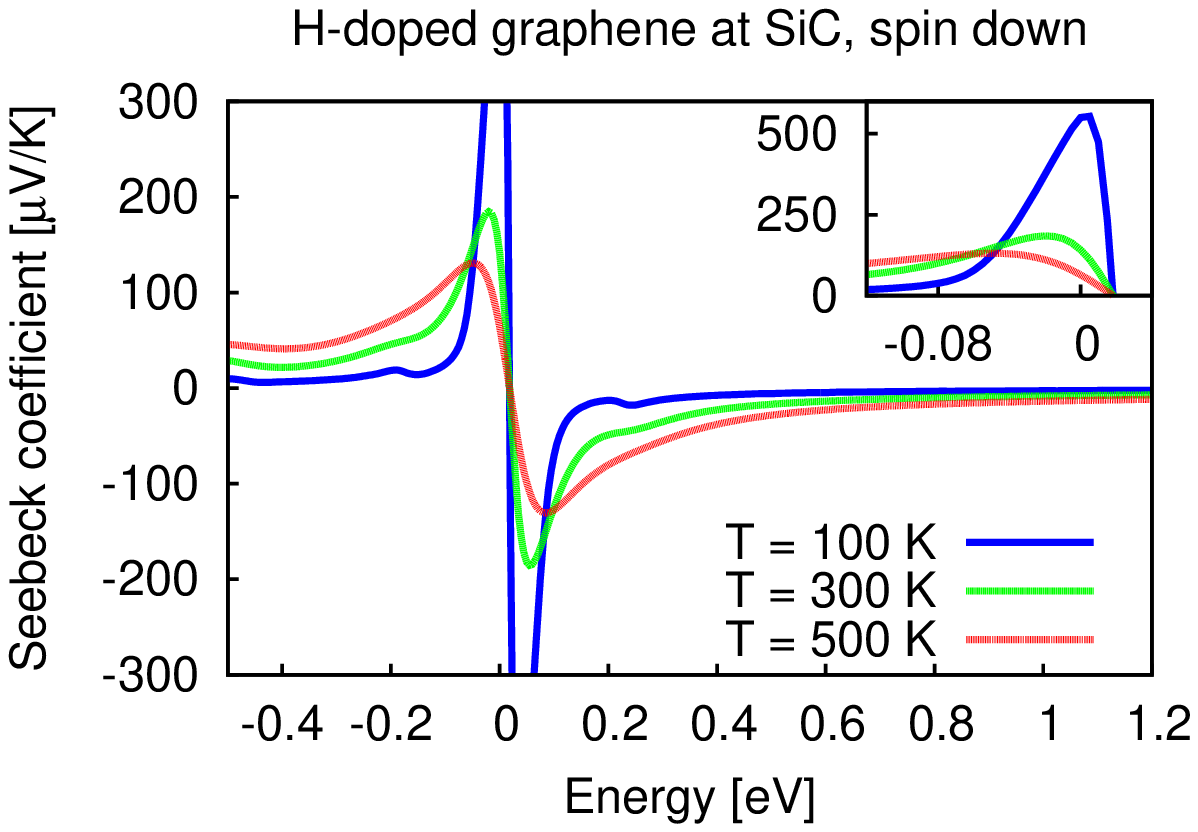}}
\vspace{2mm}
\leftline{ \includegraphics[scale=0.35]{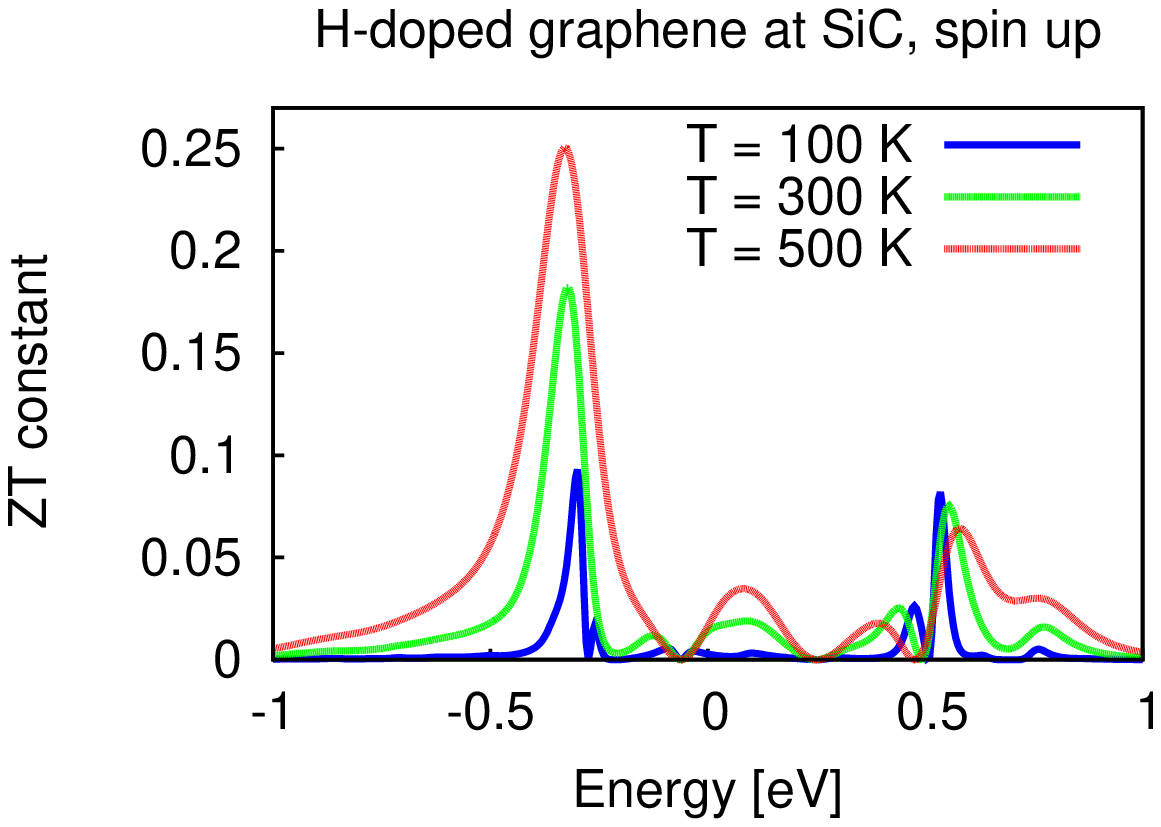}
\includegraphics[scale=0.35]{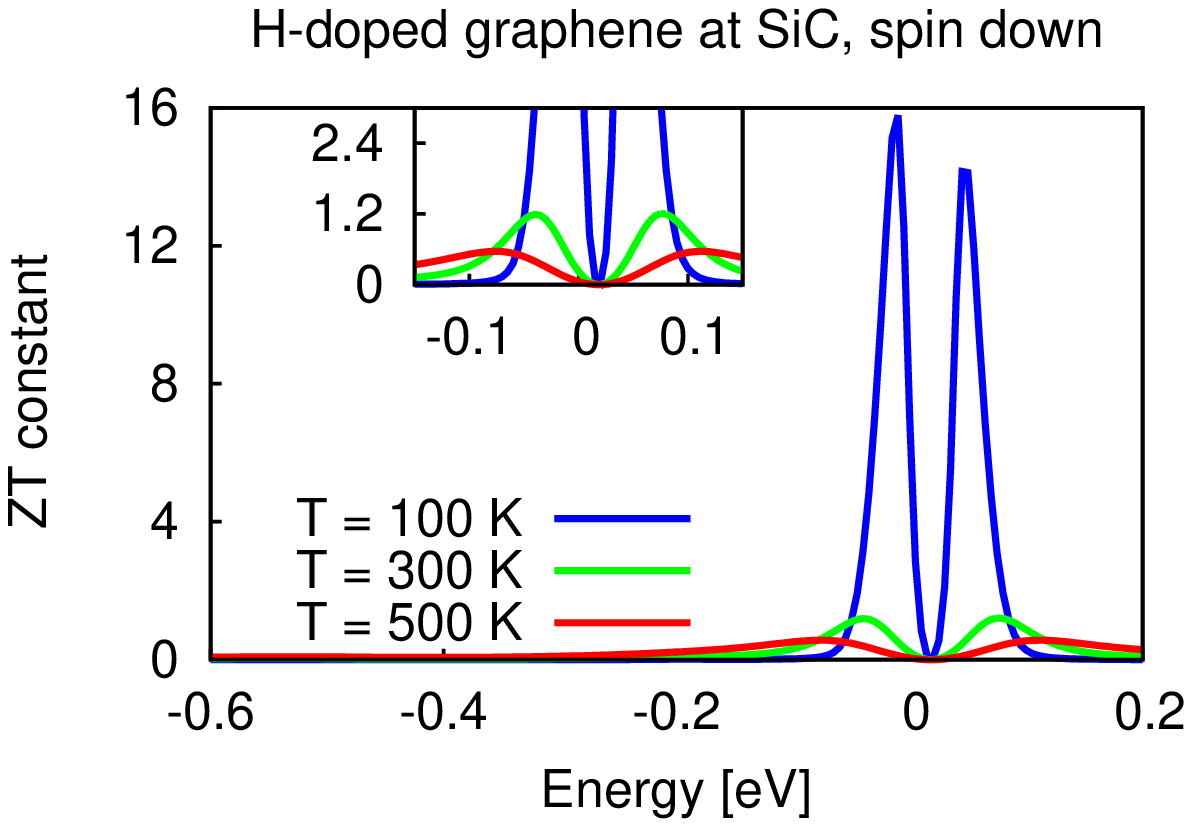}}
\caption{The density of states, the Seebeck coefficient,
and the ZT figure of merit of H-doped graphene deposited on
the C-face SiC(0001) surface with two buffer layers.}
\label{b5}
\end{figure}

\begin{figure}
\leftline{ \includegraphics[scale=0.35]{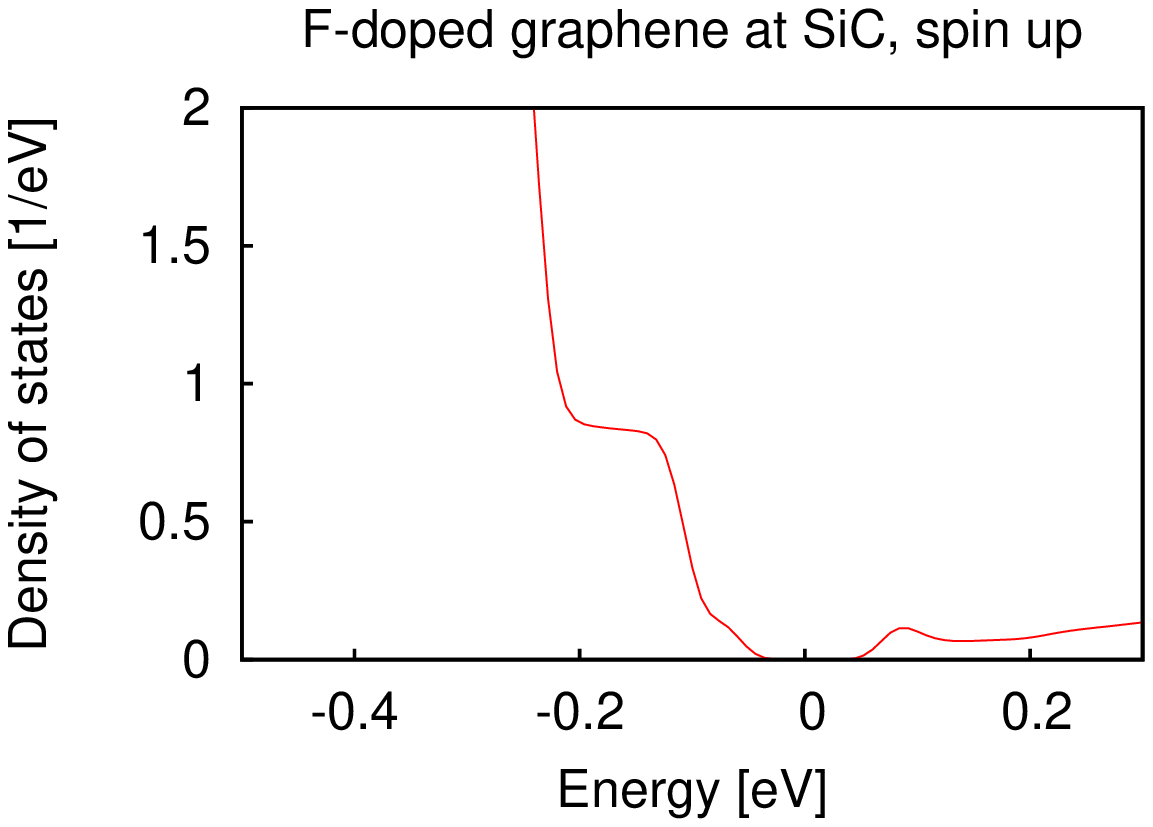}
\includegraphics[scale=0.35]{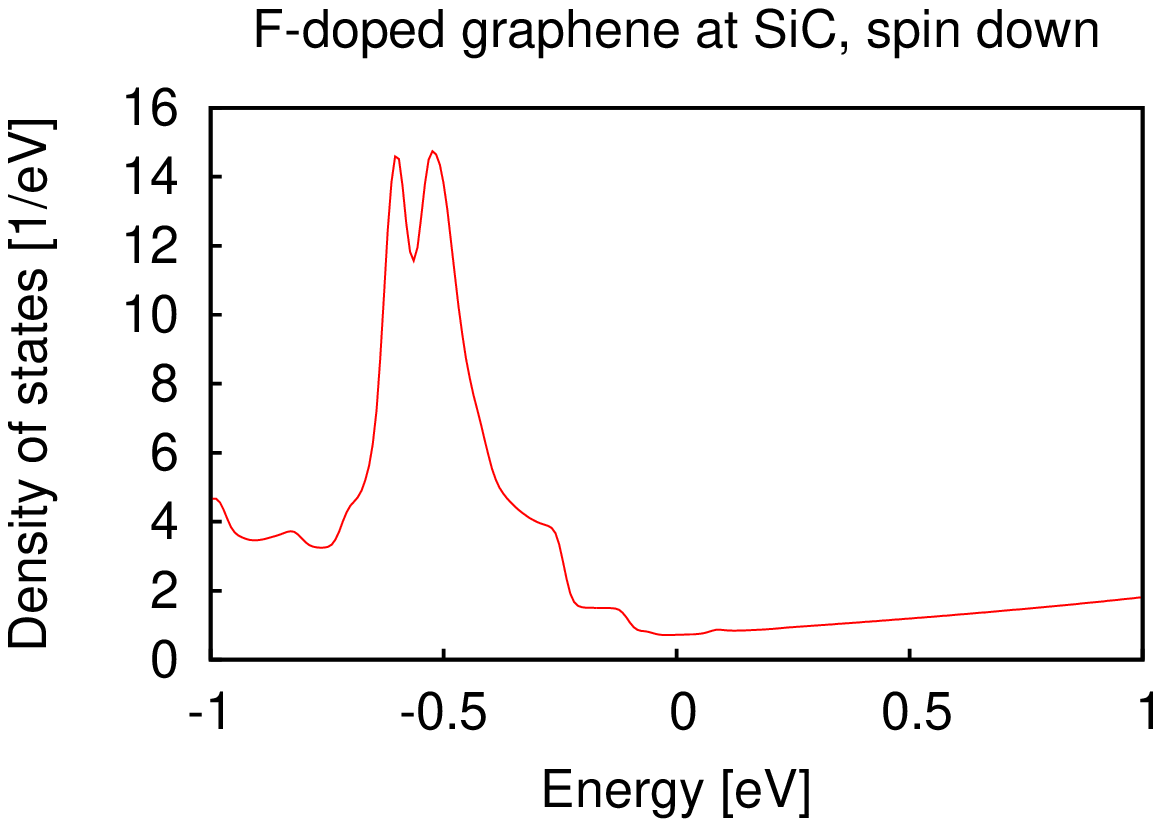}}
\vspace{2mm}
\leftline{ \includegraphics[scale=0.35]{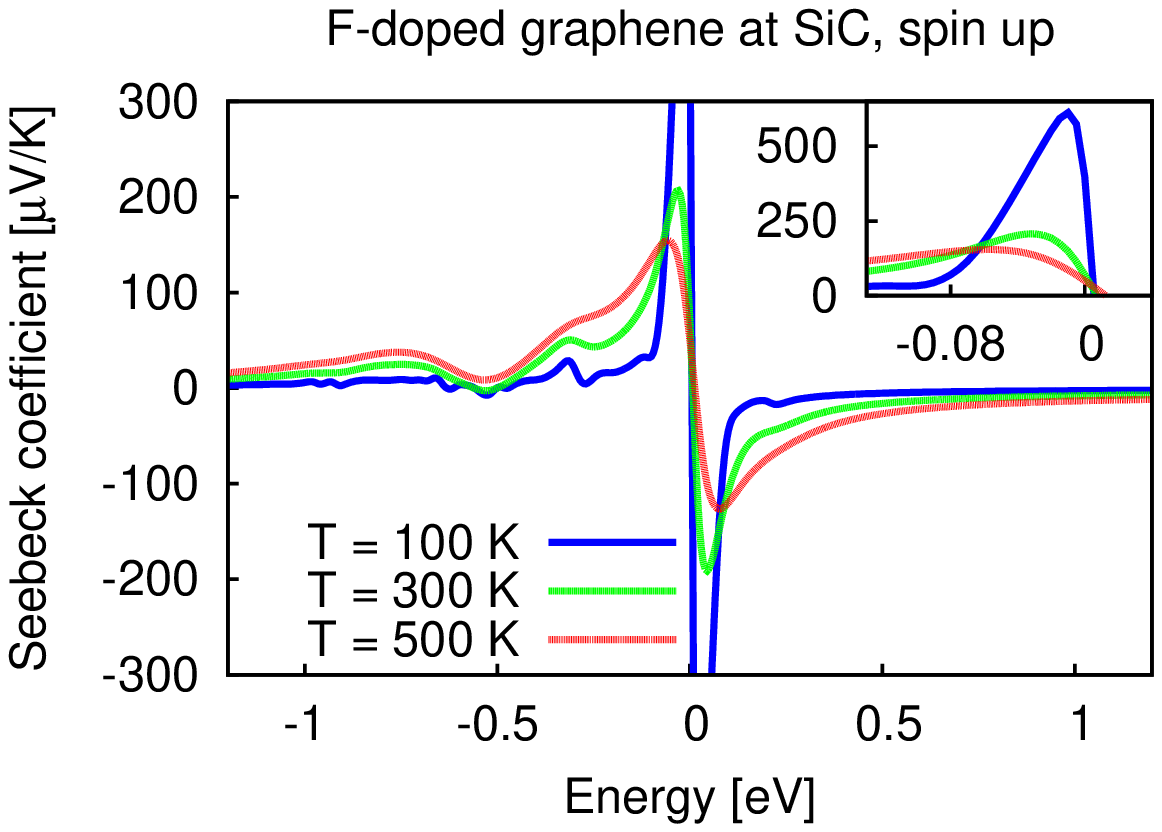}
\includegraphics[scale=0.35]{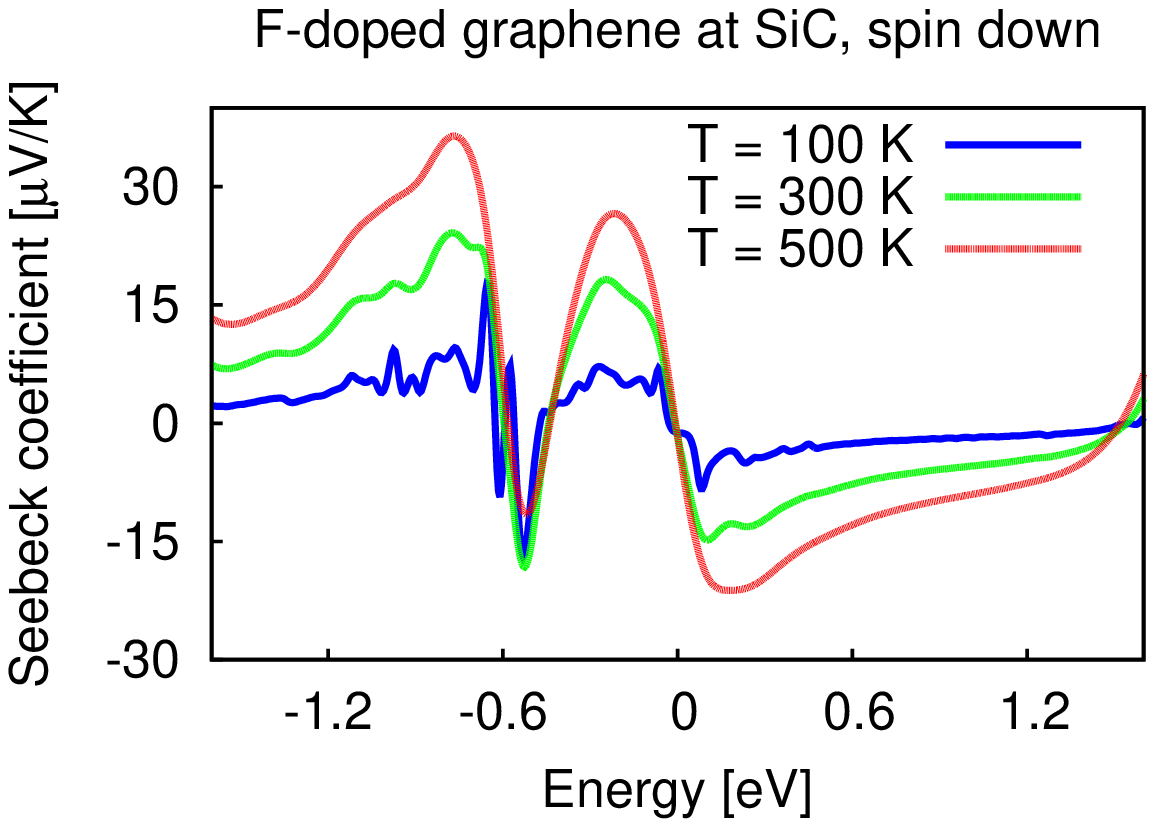}}
\vspace{2mm}
\leftline{ \includegraphics[scale=0.35]{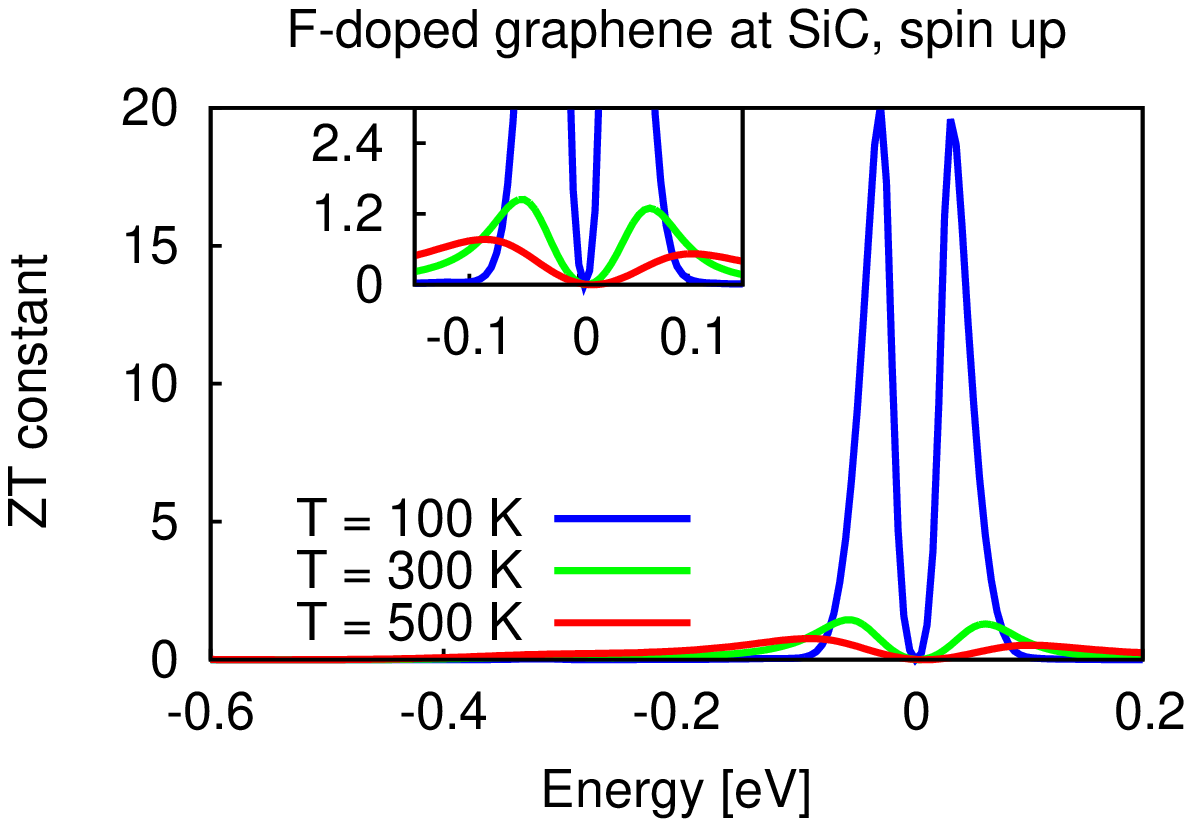}
\includegraphics[scale=0.35]{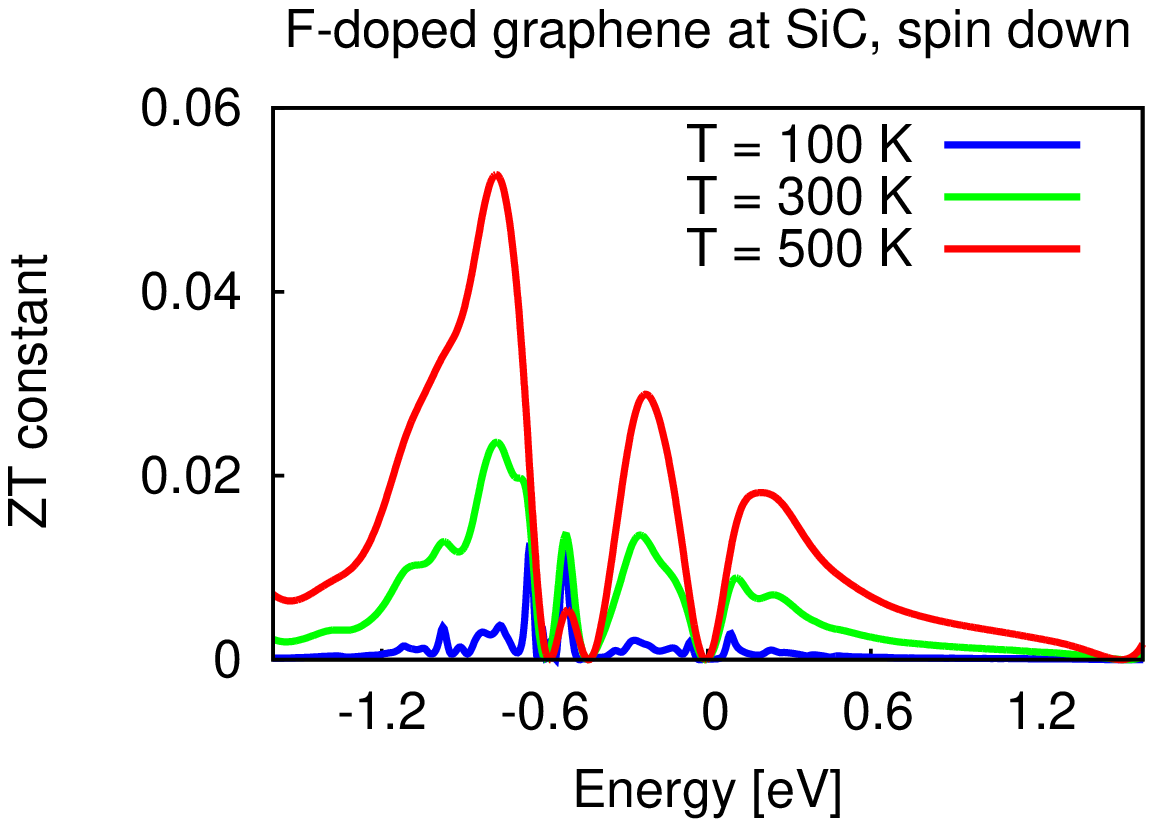}}
\caption{The density of states, the Seebeck coefficient,
and the ZT figure of merit of F-doped graphene deposited on
the C-face SiC(0001) surface with two buffer layers.}
\label{b6}
\end{figure}

The atomic structure of doped graphene sheet on top of two buffer layers
above the C-face $4H-$SiC(0001) surface is displayed in Fig.~4, together with
the band structures of H- and F-doped systems.
In the case of hydrogen doping, the majority spin-bands close to the K-point are
similar to the bands of bilayer graphene in the AB stacking 
or graphite \cite{Borys,noi-arxiv} \-- with the difference that here
the parabolic bands touch each other below the Fermi level, and
the minority spin-bands open the gap close to the K-point. In the case of fluorine doping, 
the band gap opens at the K-point in both spin channels.
For the minority spin, additional bands cross the Fermi level at G and M-point.
These details of the band structure rule the thermoelectric properties.  

The DOS, the thermopower and the ZT efficiency for the H-doped graphene are presented  
in Fig.~5, and for the F-doped graphene in Fig.~6.
When the gap opens then the thermoelectric parameters jump very high.
This situation takes place for the spin-up
channel in H-doping and the spin-down channel in F-doping cases. 
The effect might be more difficult to measure than in the free-standing cases, 
since the DOS of the deposited doped graphene is lower close to the band edges.
The Seebeck coefficient at the temperatures 100, 200, 300, 400 K reaches at maximum 
values 554, 270, 185, 148 $\mu$V/K, respectively, 
for the H-doped graphene in the spin-down channel, and
in the case of the F-doped graphene in the spin-up channel, the thermopower achieves
values of 593, 297, 207, 171 $\mu$V/K respectively. 
The ZT efficiency is 15.78, 2.88, 1.19, 0.74 for the H-doped graphene 
in the spin-down channel at 100, 200, 300, 400 K, respectively, and
correspondingly, for the F-doped graphene in the spin-up channel, 
the ZT figure of merit is 20.10, 3.45, 1.45, 0.95, respectively.
It is peculiar and of great application potential, that the thermopower and
the ZT figure of merit at the band edges achieve very high values for lower temperatures
\-- in contrast to all other cases known so far in the high-ZT physics. 

Our finding of the band-edge growth of the Seebeck coefficient 
and the ZT figure of merit is in agreement with 
the model predictions by Sharapov and Varlamov \cite{Sharapov} for pure gapped graphene.
They explained the effect of giant growth of the thermopower at the band gap
in graphene assuming the strong dependence of the relaxation
time on the energy, i.e. the band details. 
Similar work has been published by Hao and Lee \cite{Hao} for the tunable 
gapped bilayer graphene.
In this work, we show in the next section that the band-edge growth of the thermopower
is caused by vanishing band velocities, 
independently on the model used for the relaxation time.
In both, our work and the model consideretations by Sharapov and Varlamov, 
the break of the symmetry equivalence between two sublattices in graphene takes place.
However,  it will be explained below that the effect is more general and applies
to all gapped systems independently on their lattices.  

\section{Model studies of a fingerprint of the band-structure features 
in the thermopower curve} 

We analyse how the band-structure features such as local extrema, band edges 
and Van Hove singularities
are reflected in the Seebeck-coefficient curve.
We begin with the collection of the mathematical expressions within the
Boltzmann framework and the results for the model band structures. 
Later, we discuss also the Seebeck coefficent obtained from the Mott formula
for the band structures in 1D, 2D and 3D.       

\begin{figure*}
\leftline{ \includegraphics[scale=0.4]{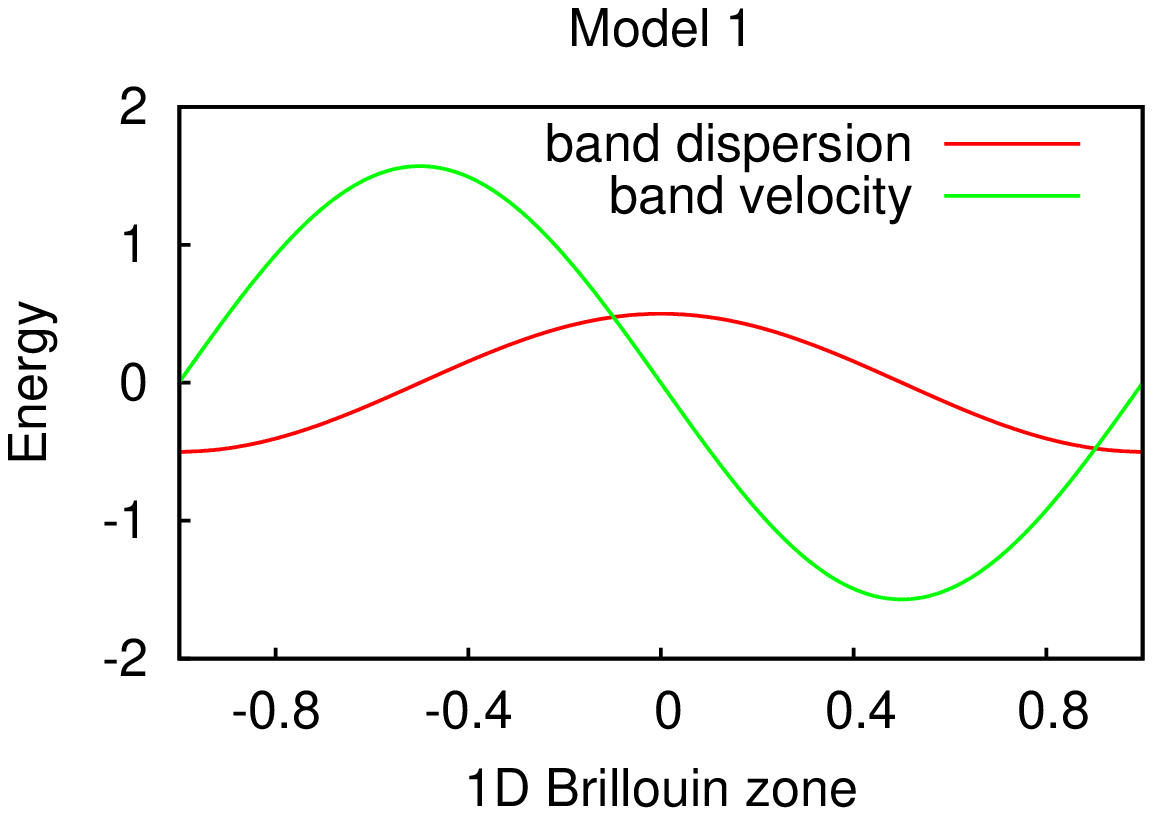}
\includegraphics[scale=0.4]{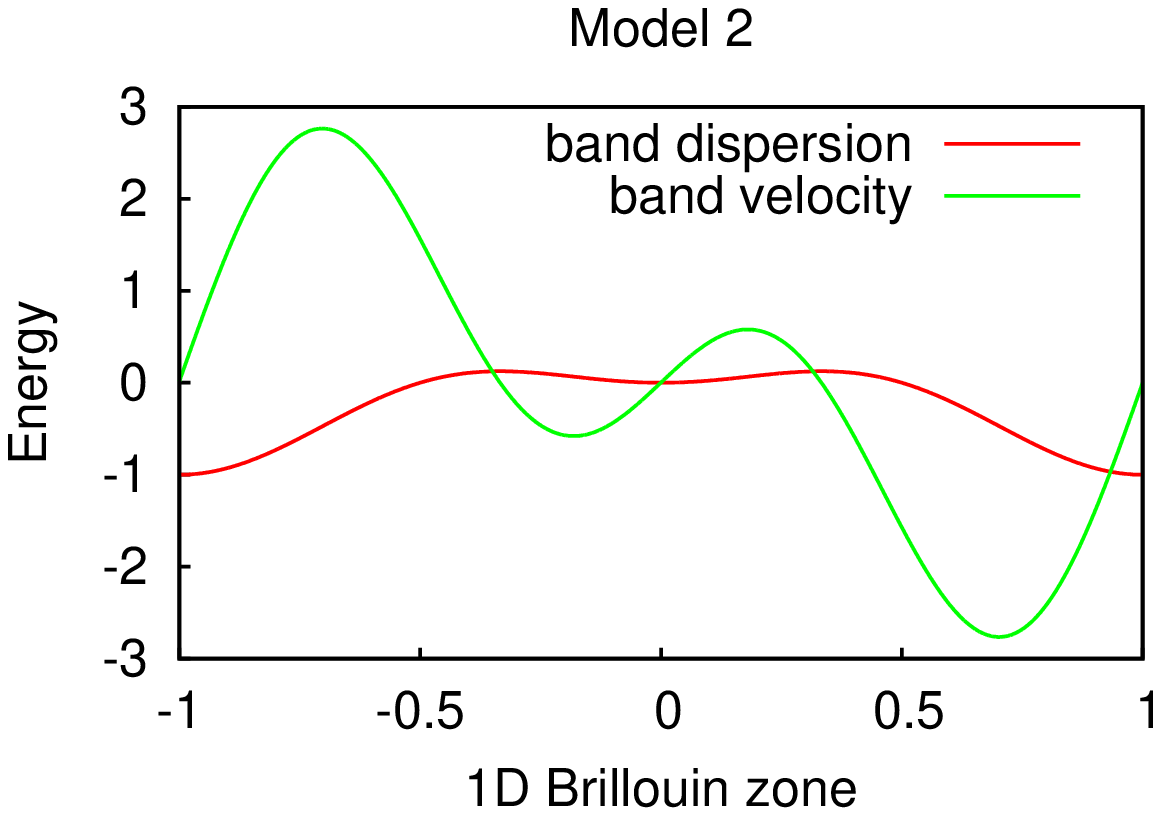}
\includegraphics[scale=0.4]{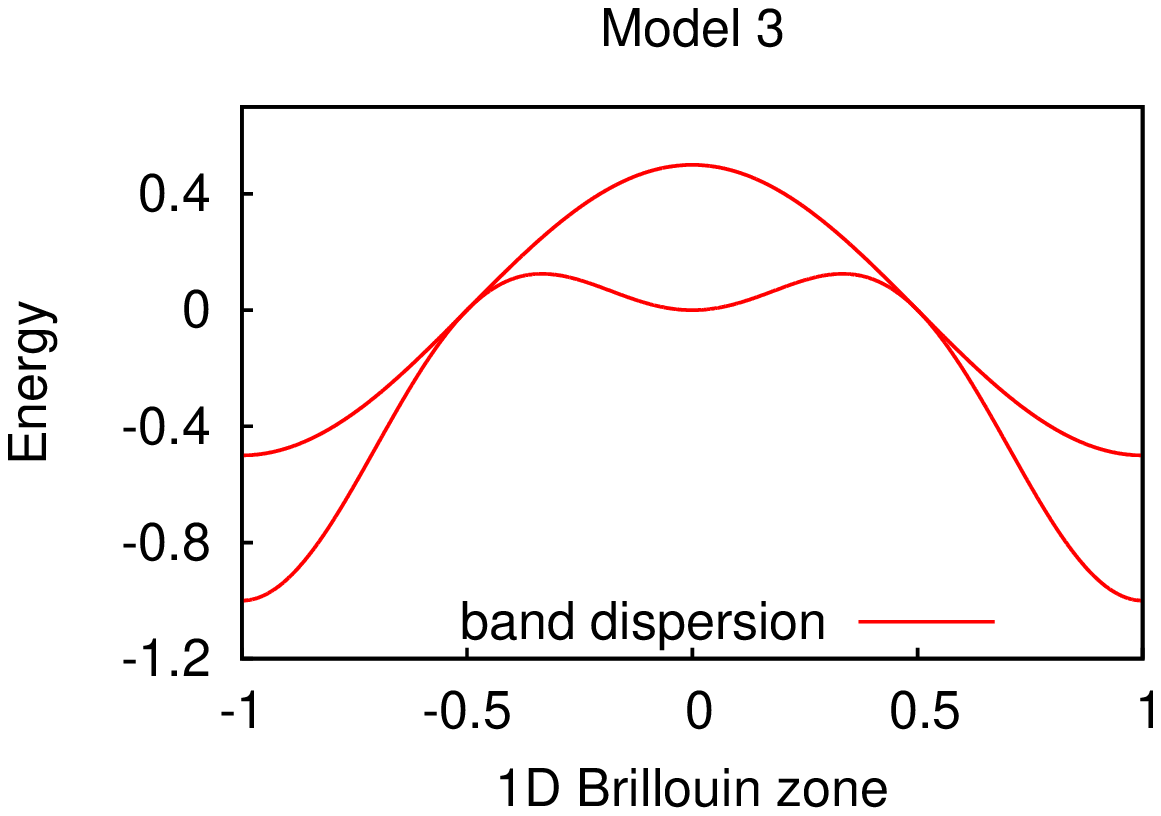}}
\vspace{2mm}
\leftline{ \includegraphics[scale=0.4]{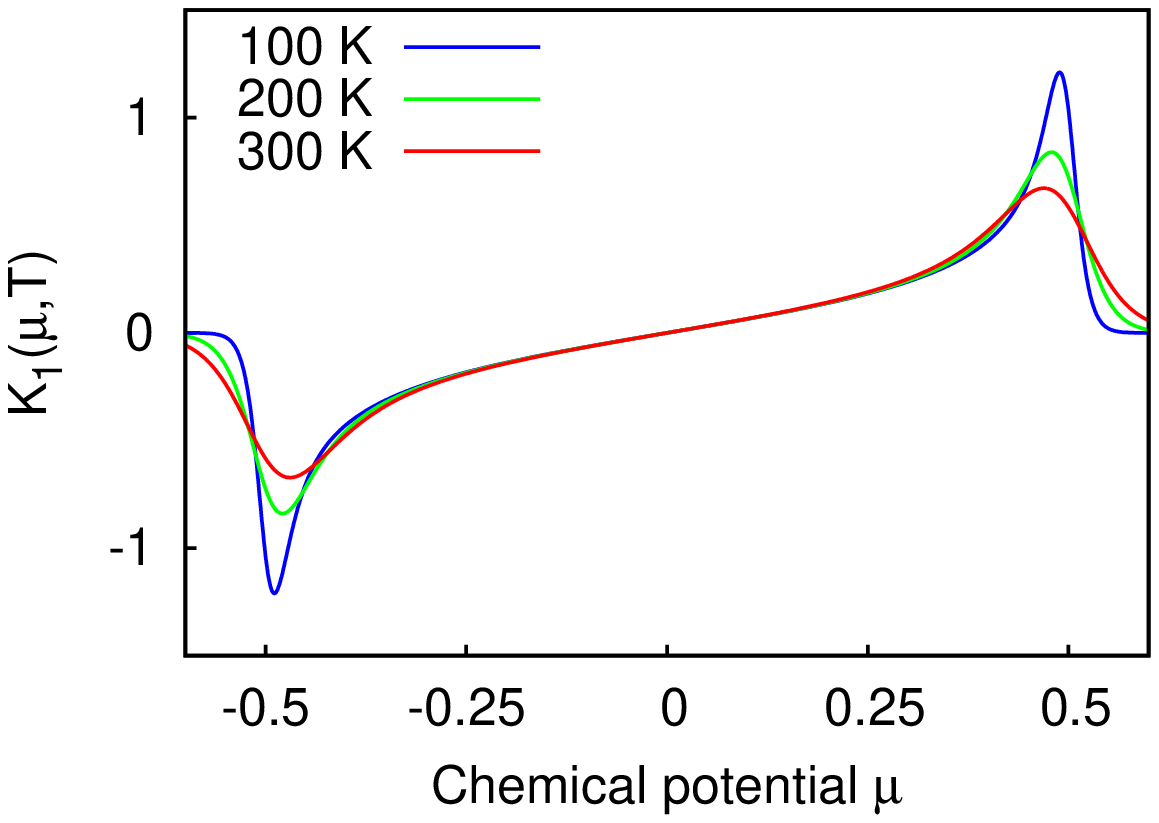}
\includegraphics[scale=0.4]{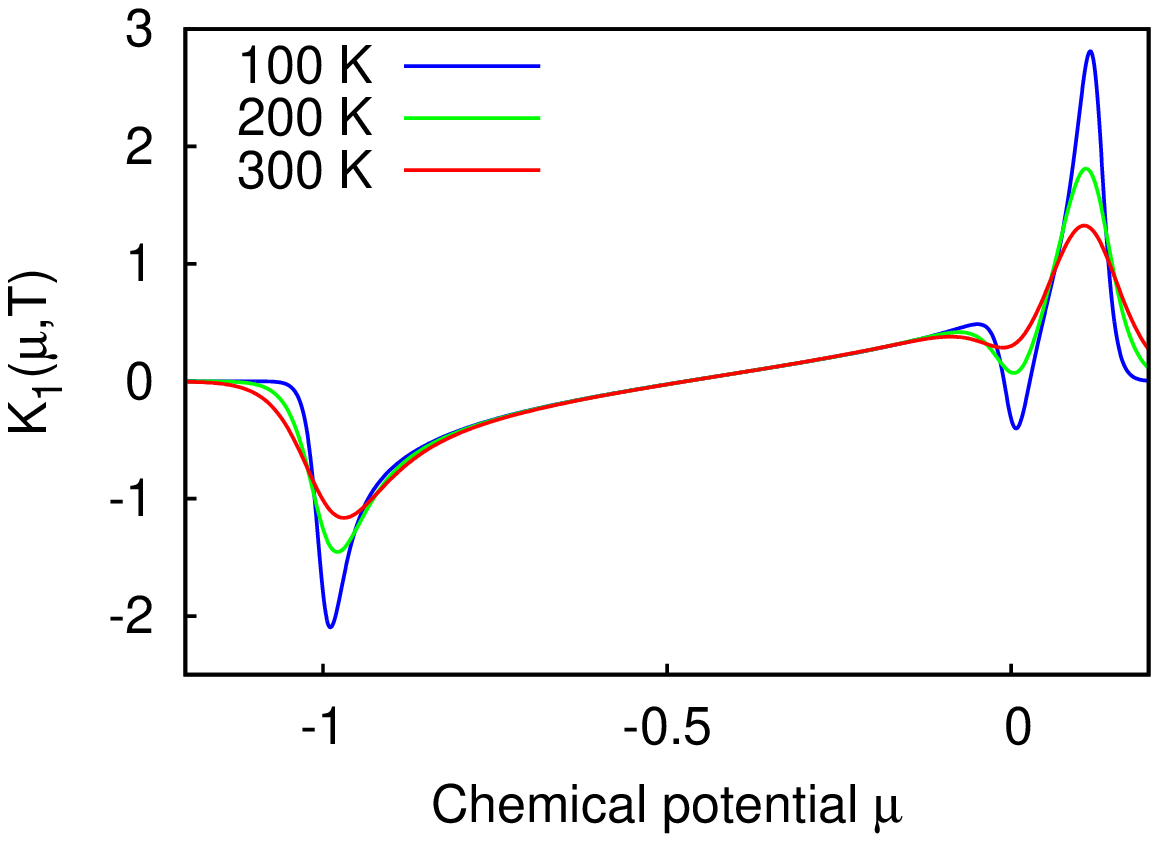}
\includegraphics[scale=0.4]{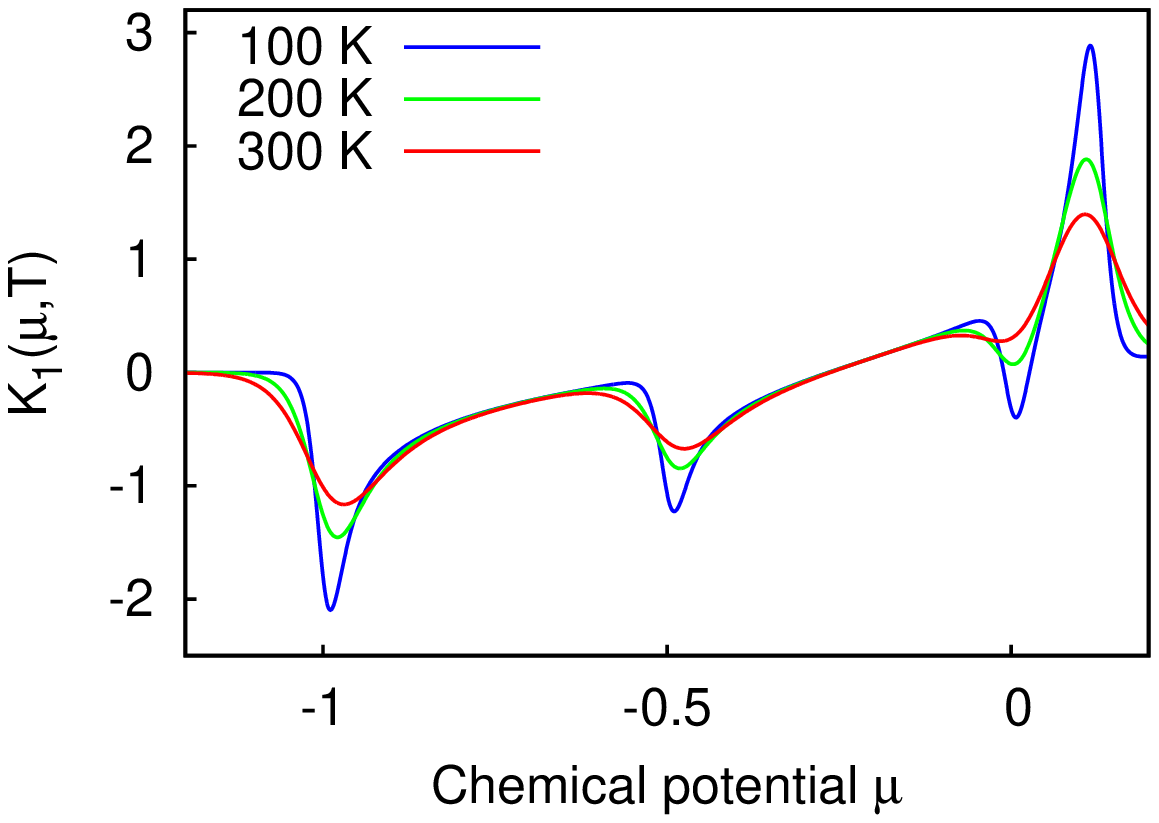}}
\vspace{2mm}
\leftline{ \includegraphics[scale=0.4]{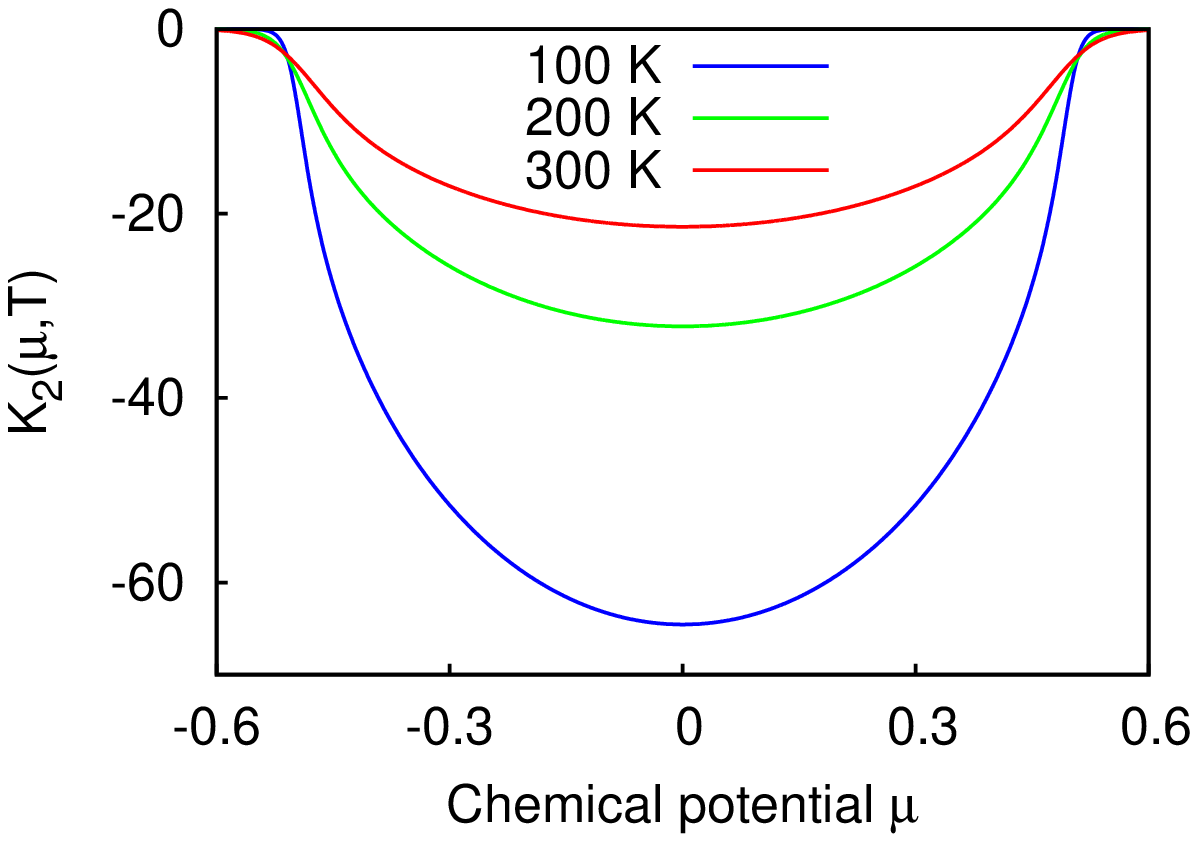}
\includegraphics[scale=0.4]{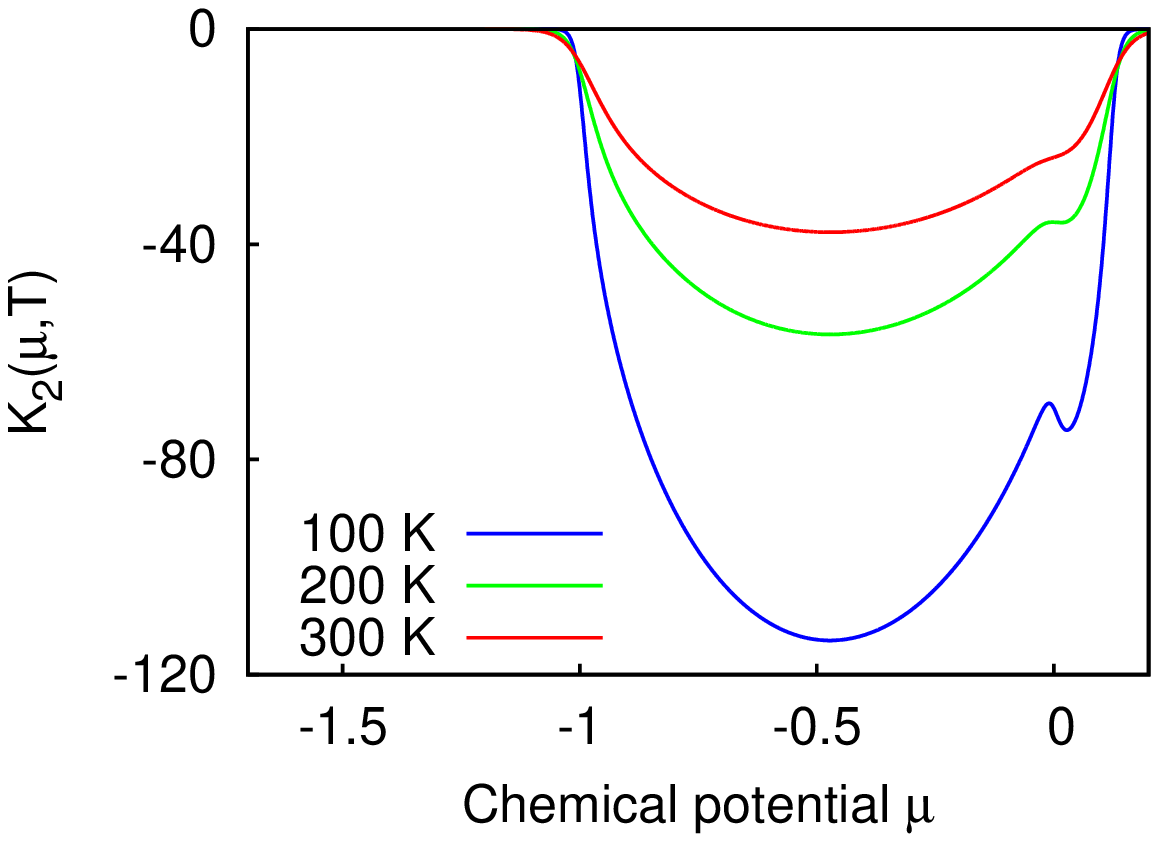}
\includegraphics[scale=0.4]{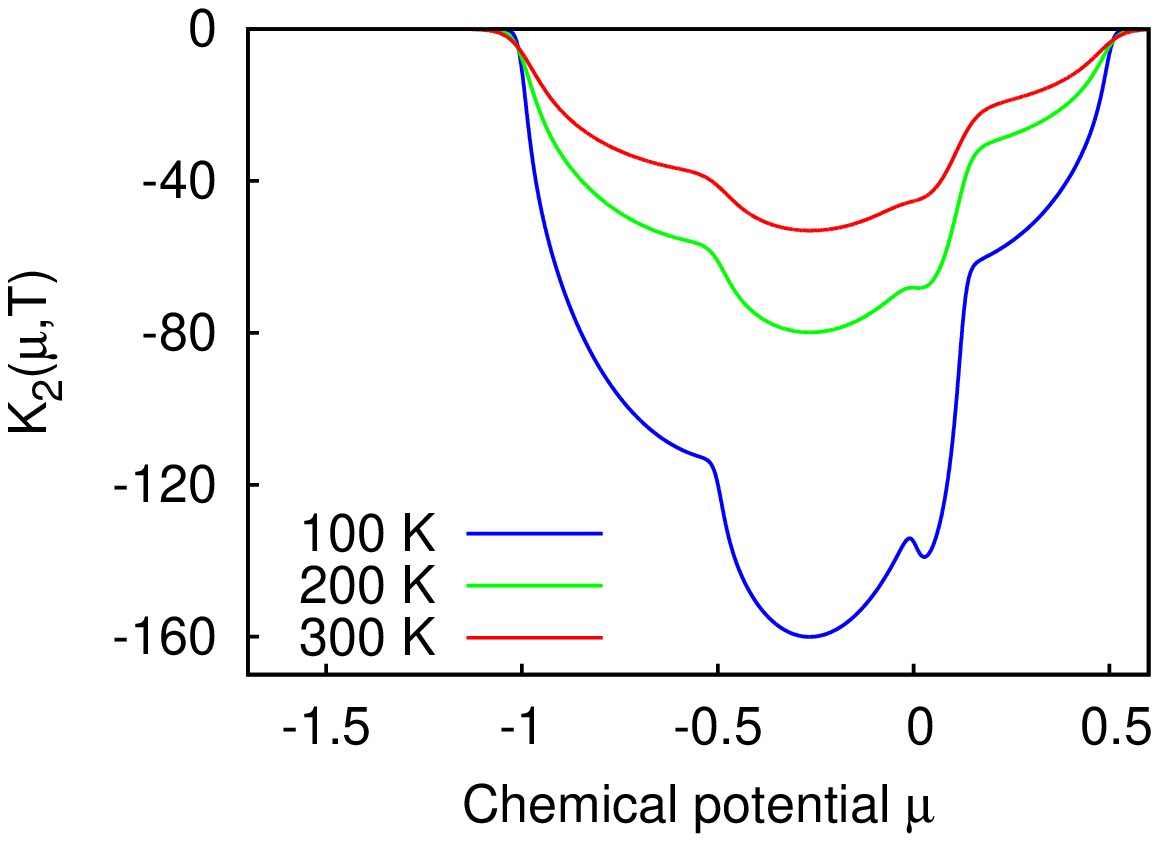}}
\vspace{2mm}
\leftline{ \includegraphics[scale=0.4]{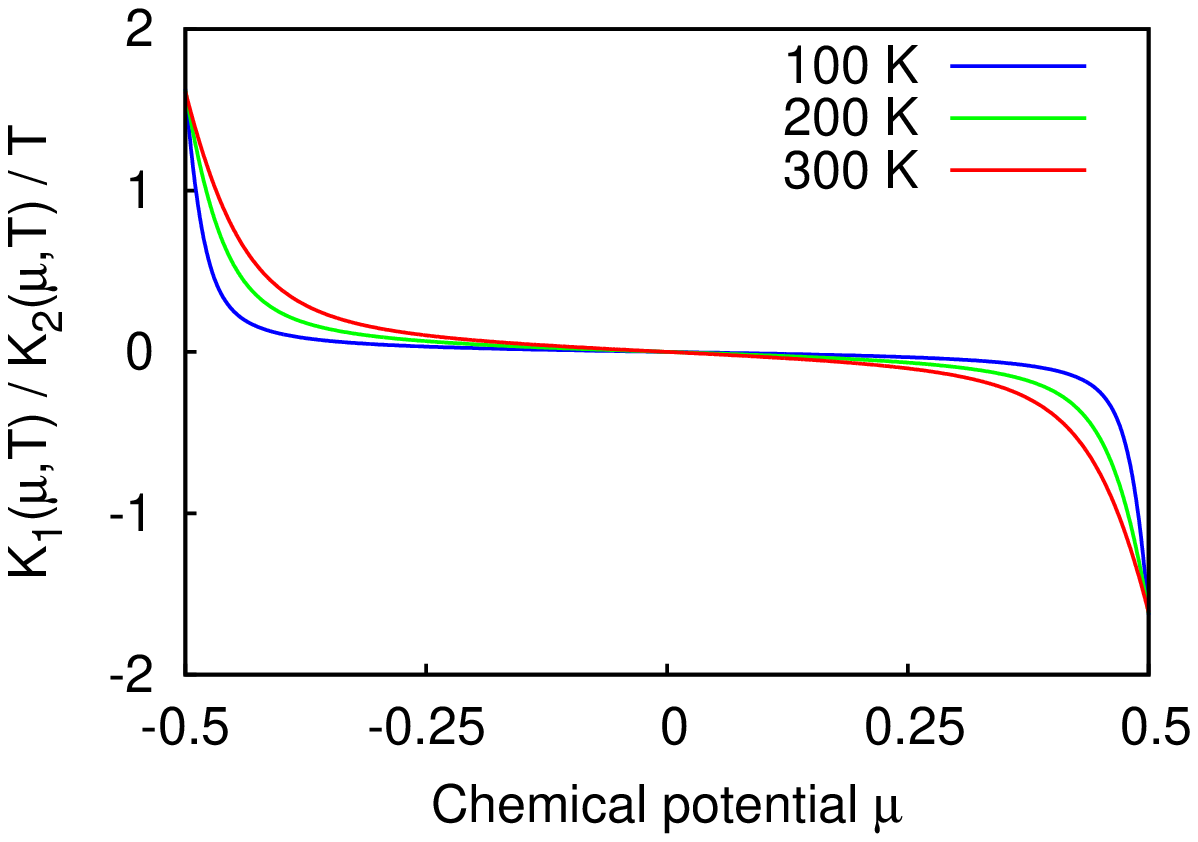}
\includegraphics[scale=0.4]{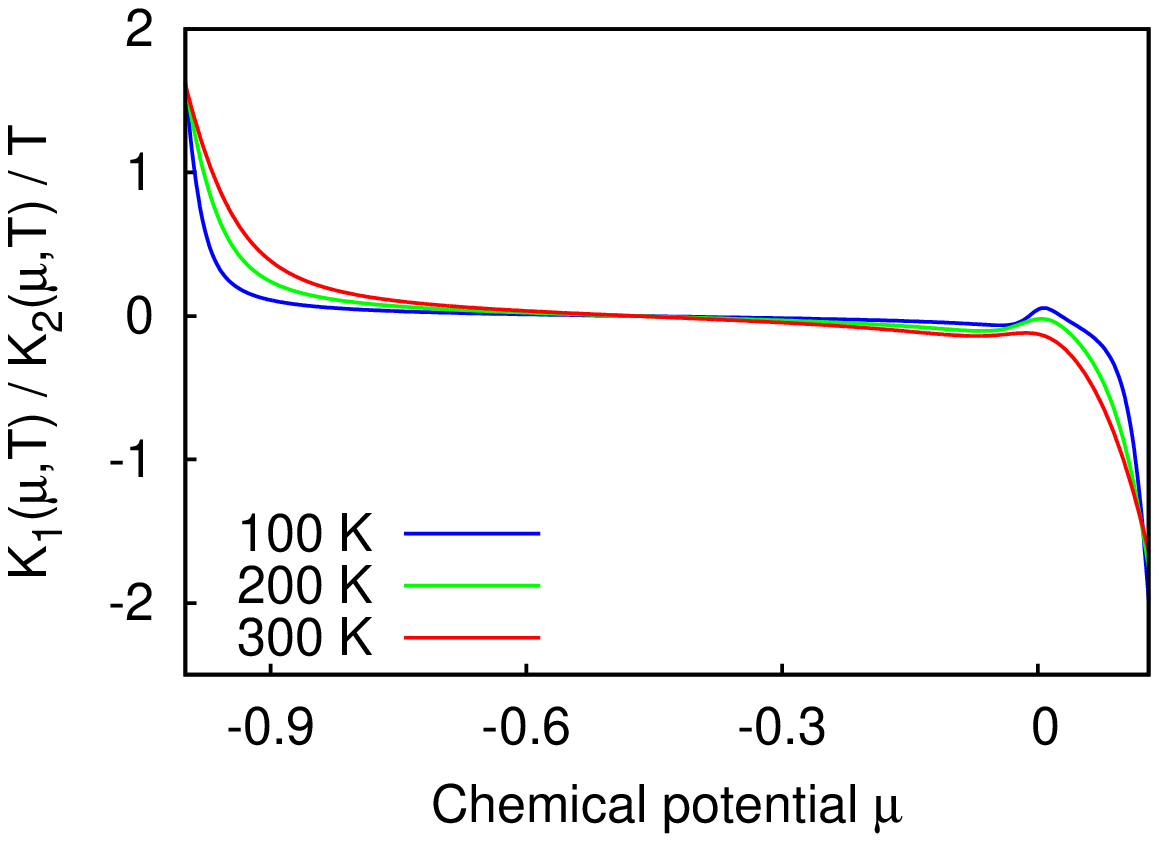}
\includegraphics[scale=0.4]{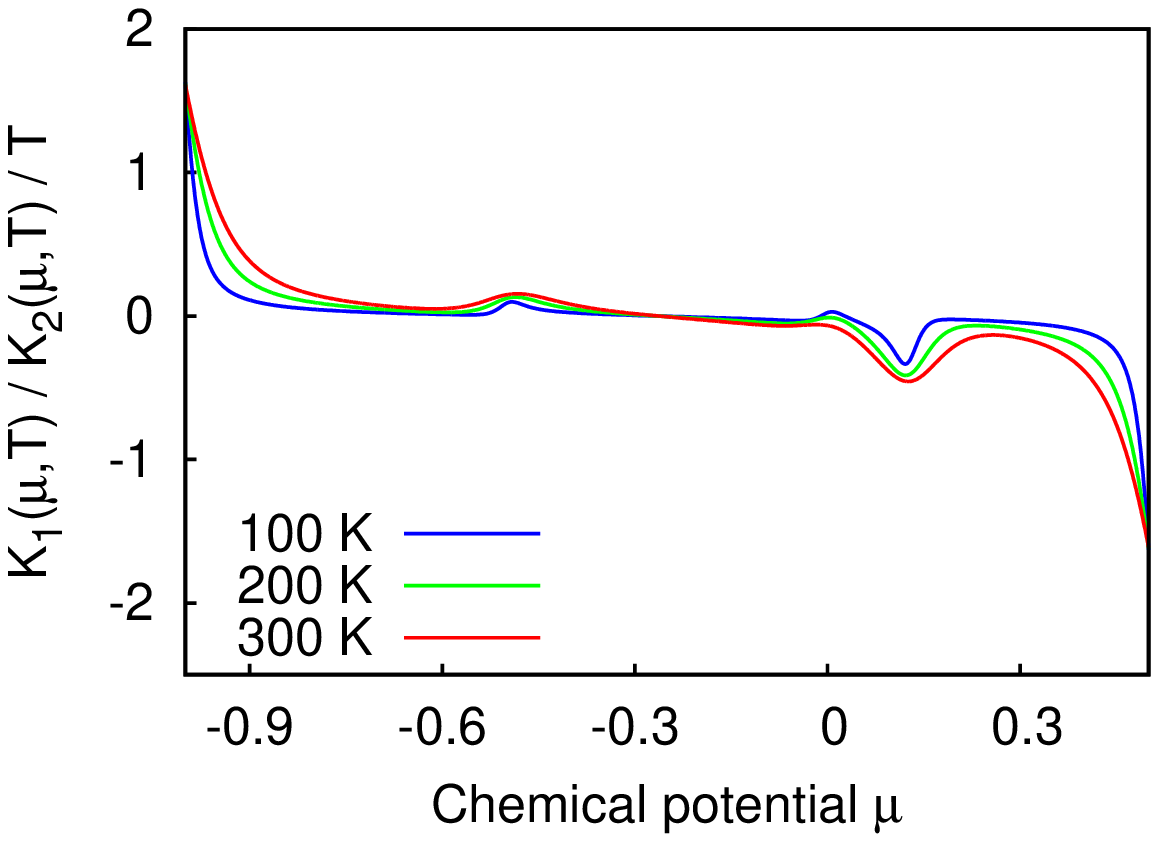}}
\caption{Model band structures in 1D Brillouin zone and kernels of the
expression for the Seebeck coefficient [in arbitrary units].
The constant relaxation time $\tau$ was assumed.}
\label{b7}
\end{figure*}

\begin{figure*}
\leftline{ \hspace{2mm} \includegraphics[scale=0.16]{f8-11.eps}
\hspace{5mm} \includegraphics[scale=0.16]{f8-12.eps}
\hspace{4mm} \includegraphics[scale=0.16]{f8-13.eps}}
\vspace{2mm}
\leftline{ \includegraphics[scale=0.4]{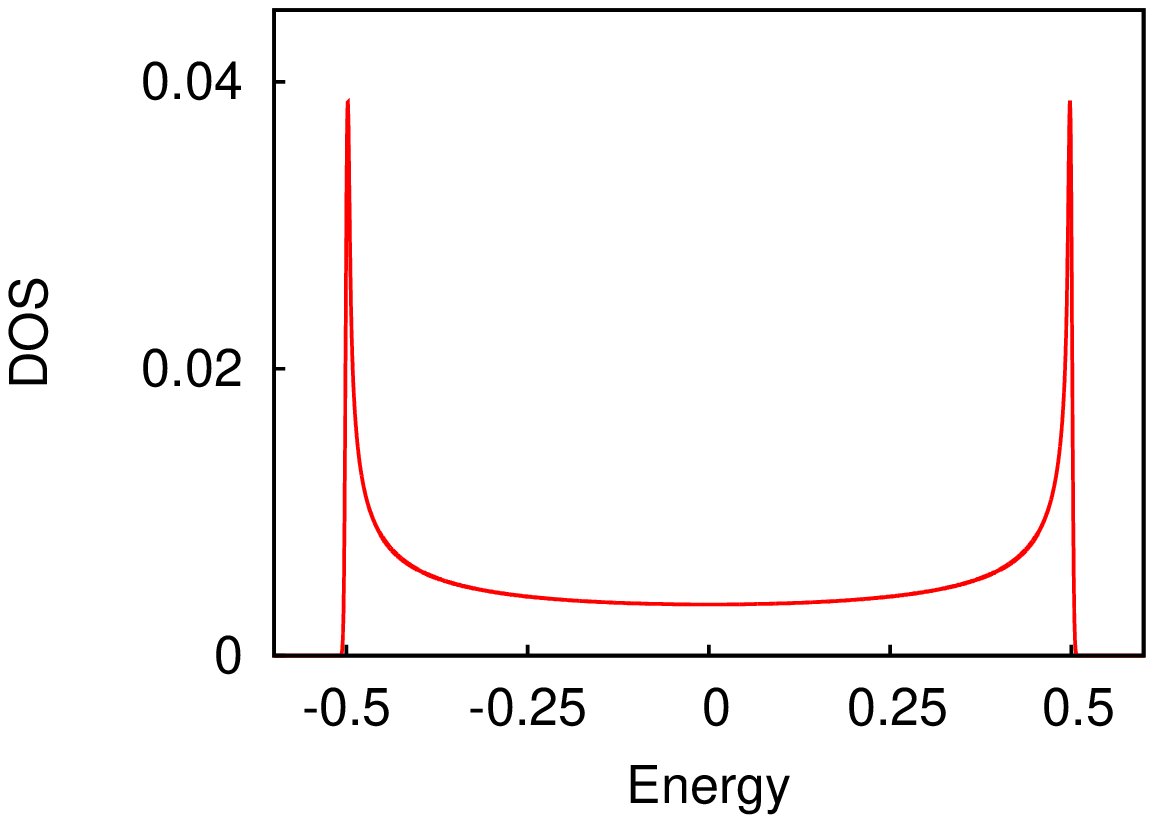}
\includegraphics[scale=0.4]{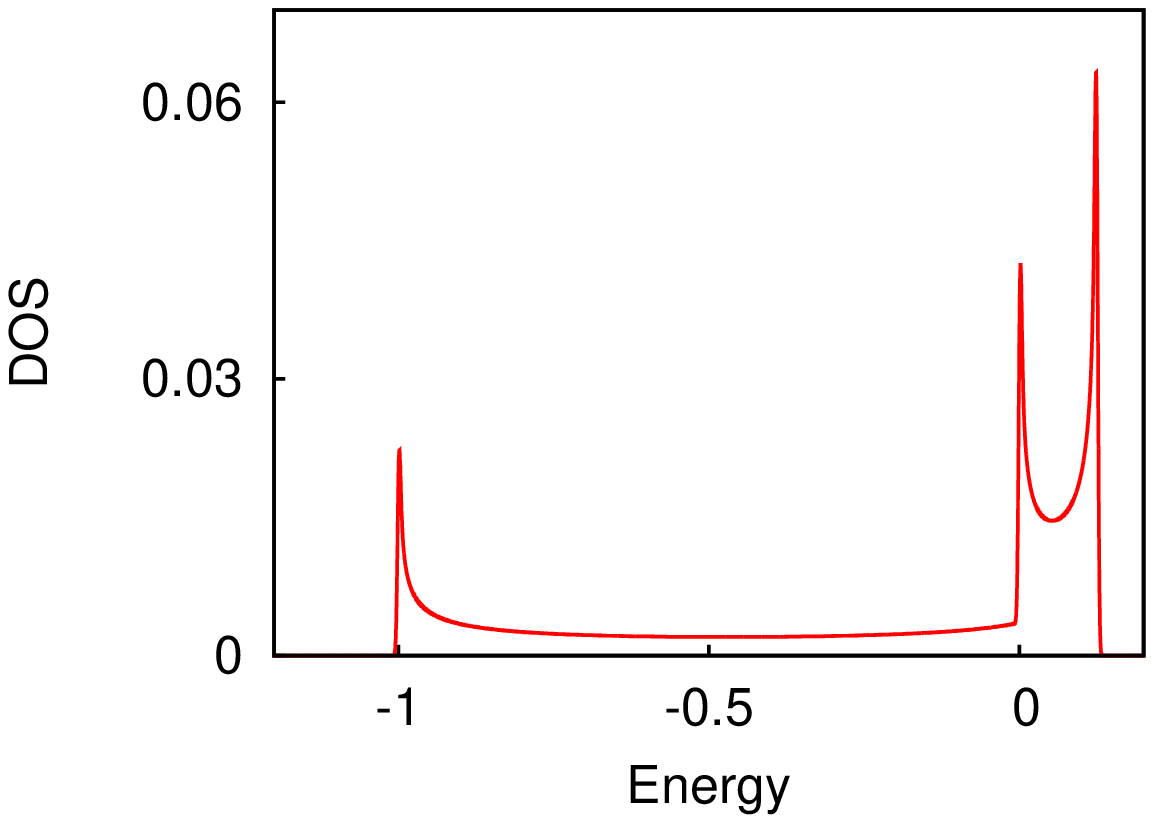}
\includegraphics[scale=0.4]{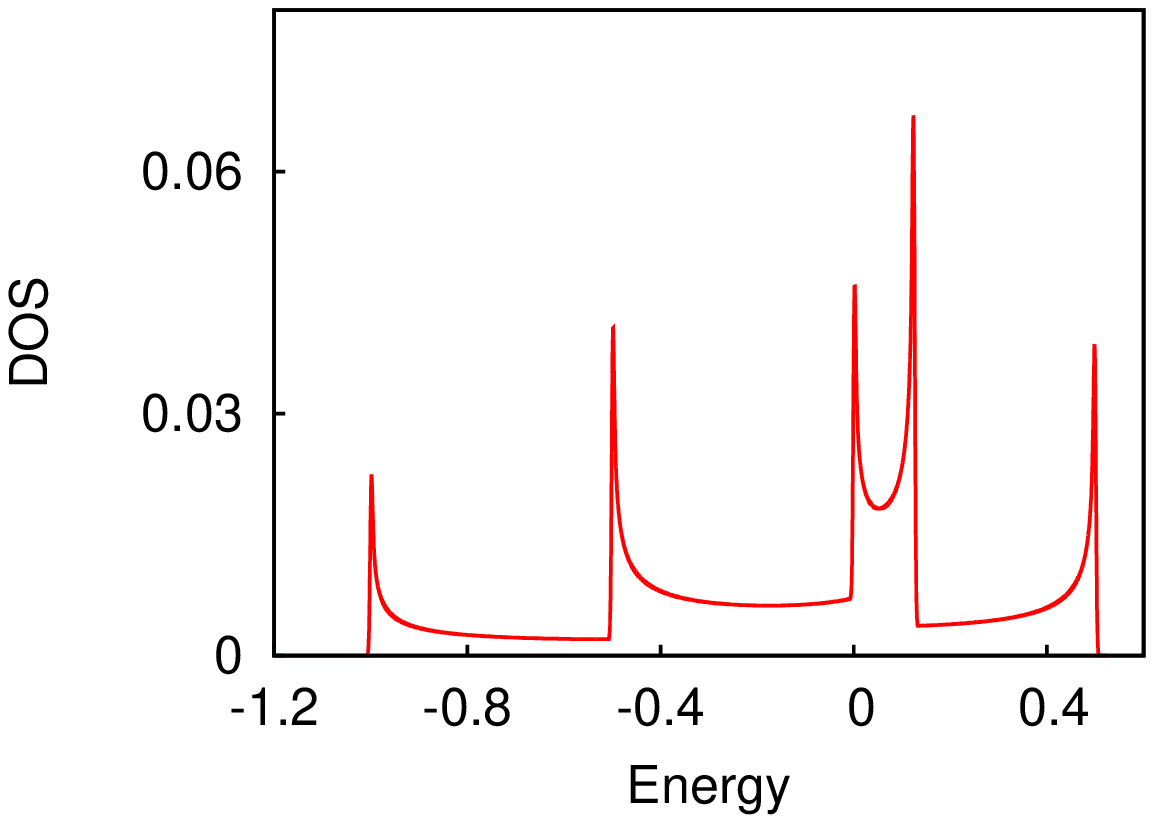}}
\vspace{2mm}
\leftline{ \includegraphics[scale=0.4]{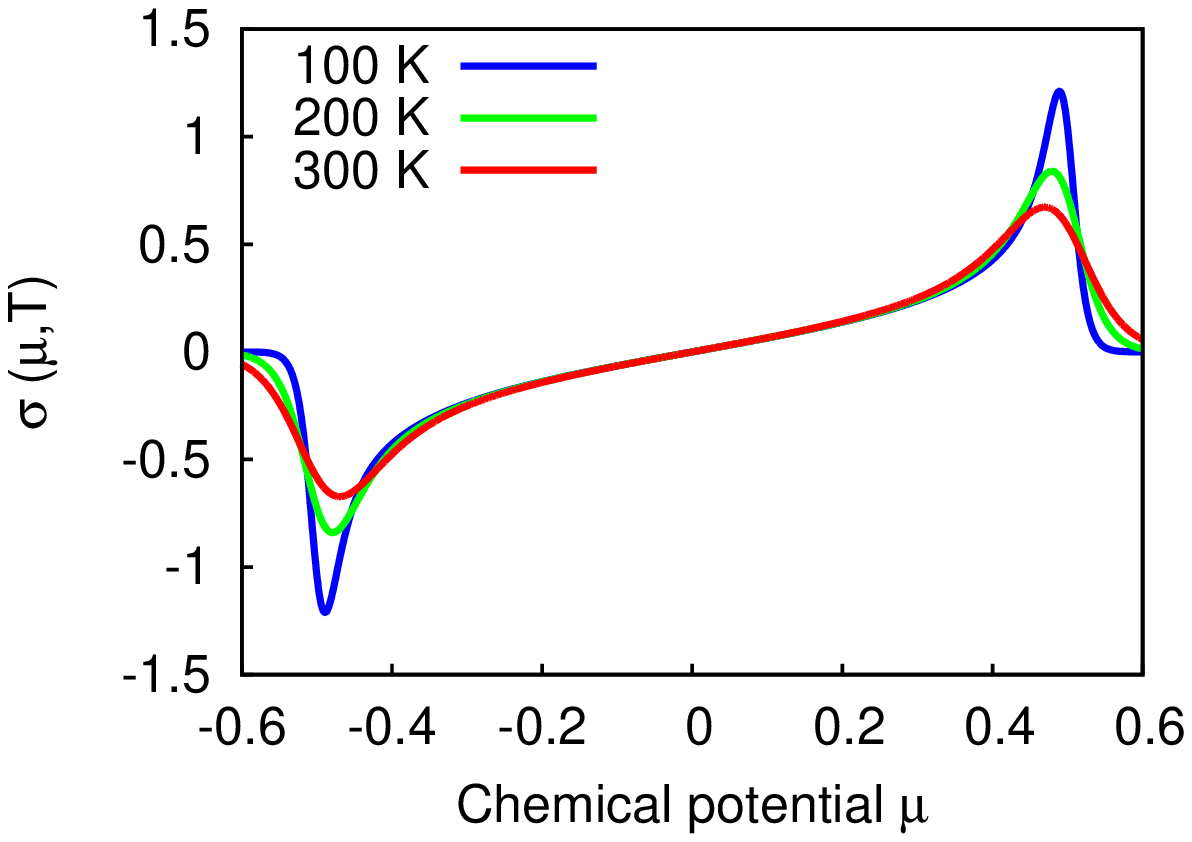}
\includegraphics[scale=0.4]{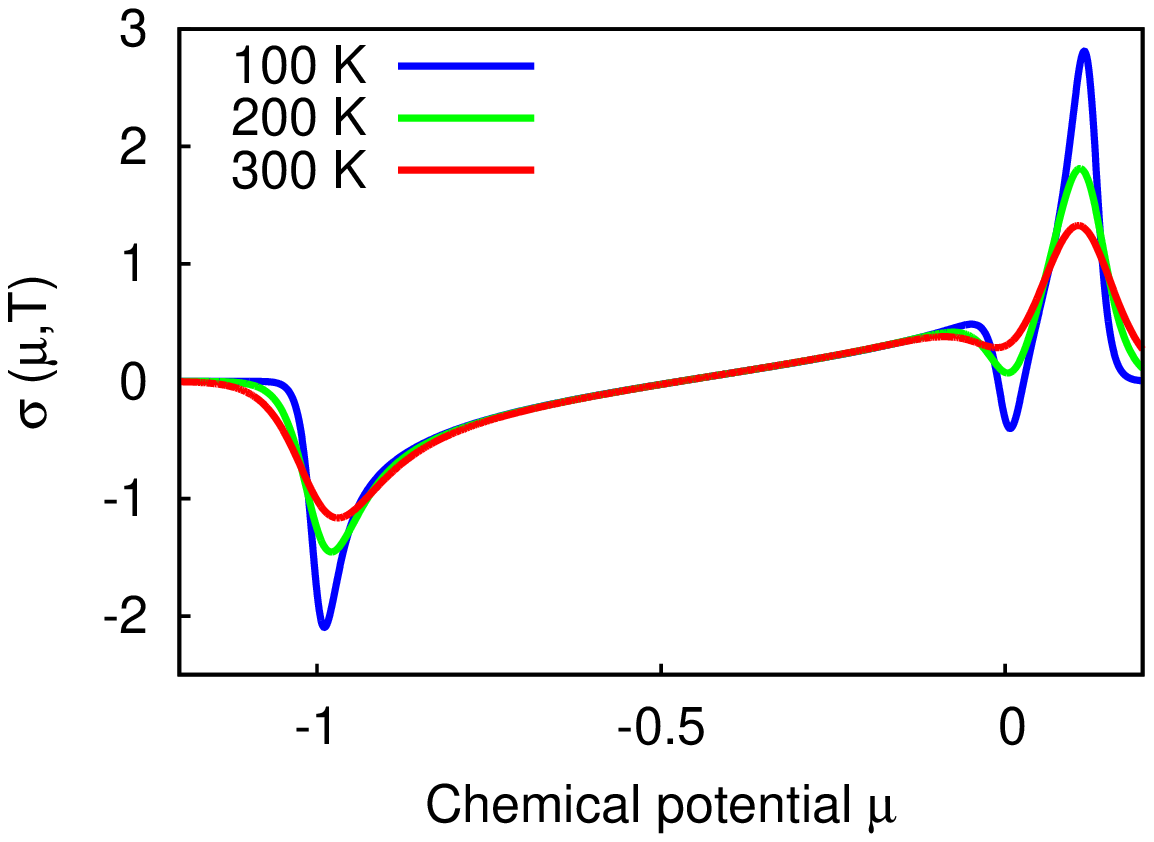}
\includegraphics[scale=0.4]{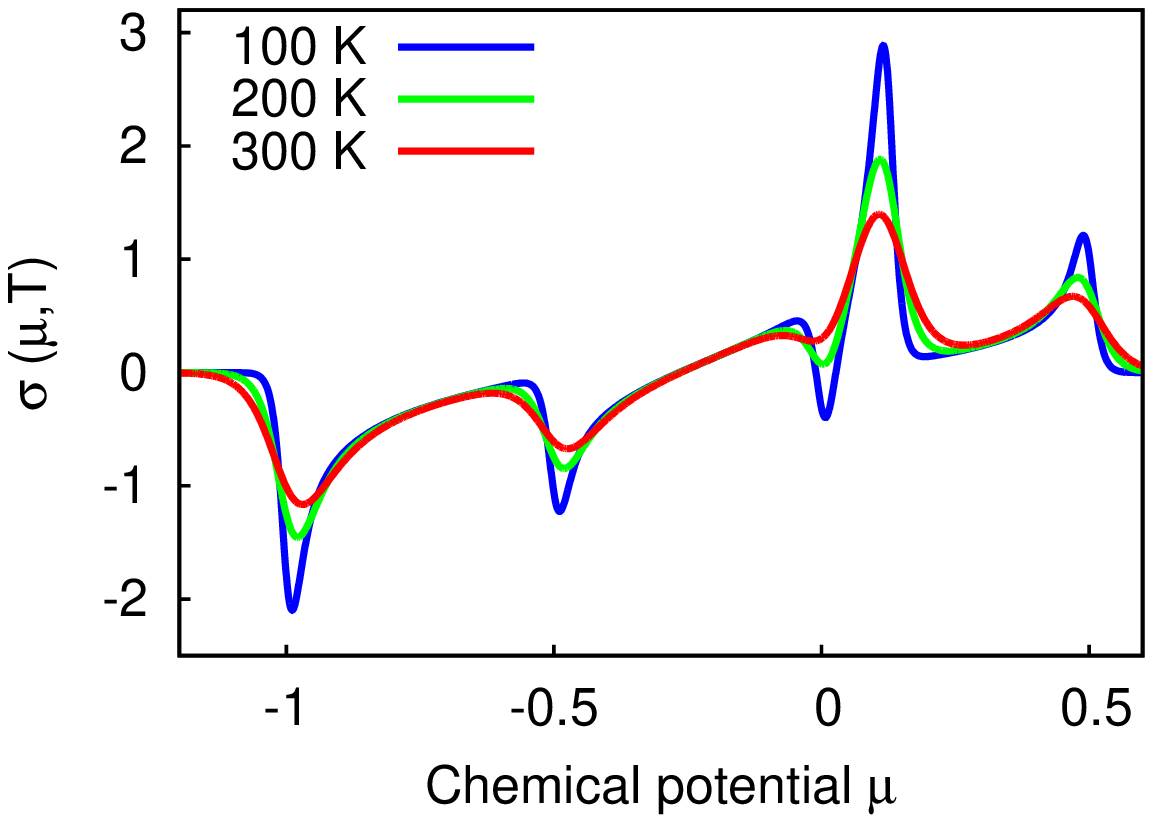}}
\vspace{2mm}
\leftline{ \includegraphics[scale=0.4]{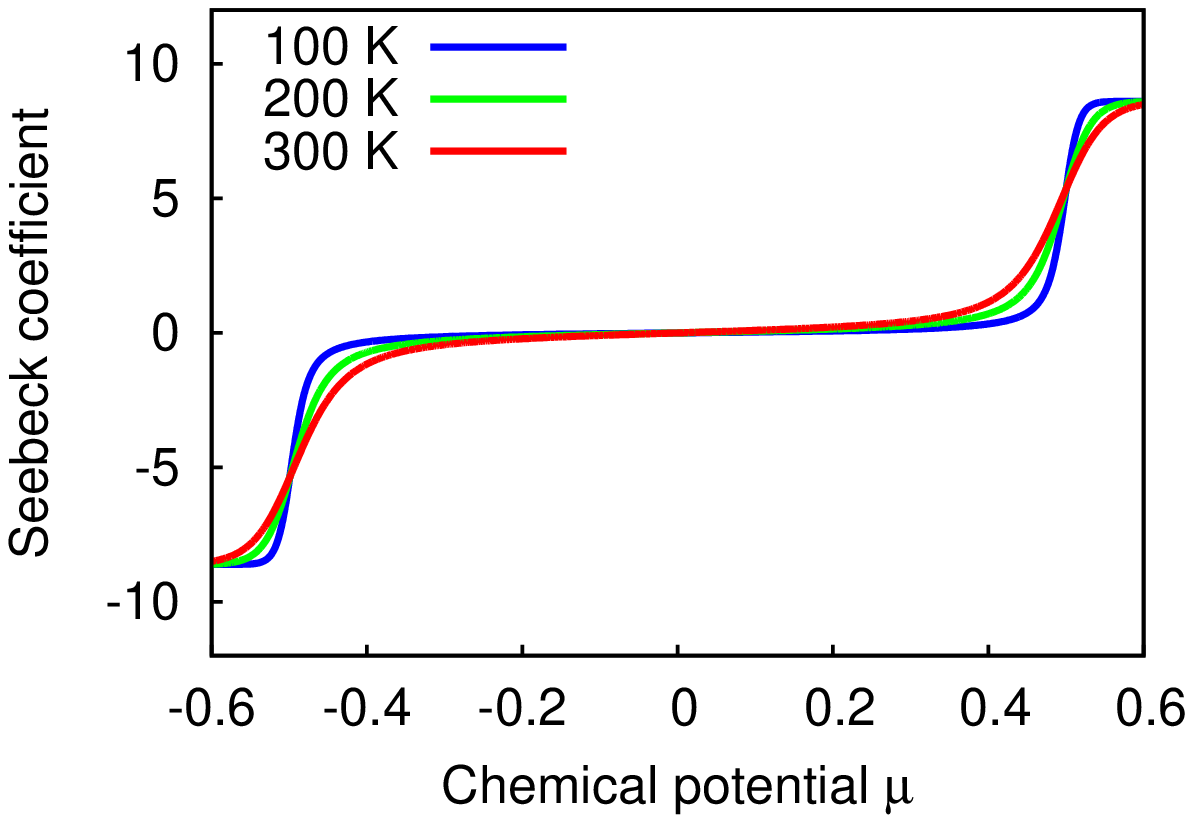}
\includegraphics[scale=0.4]{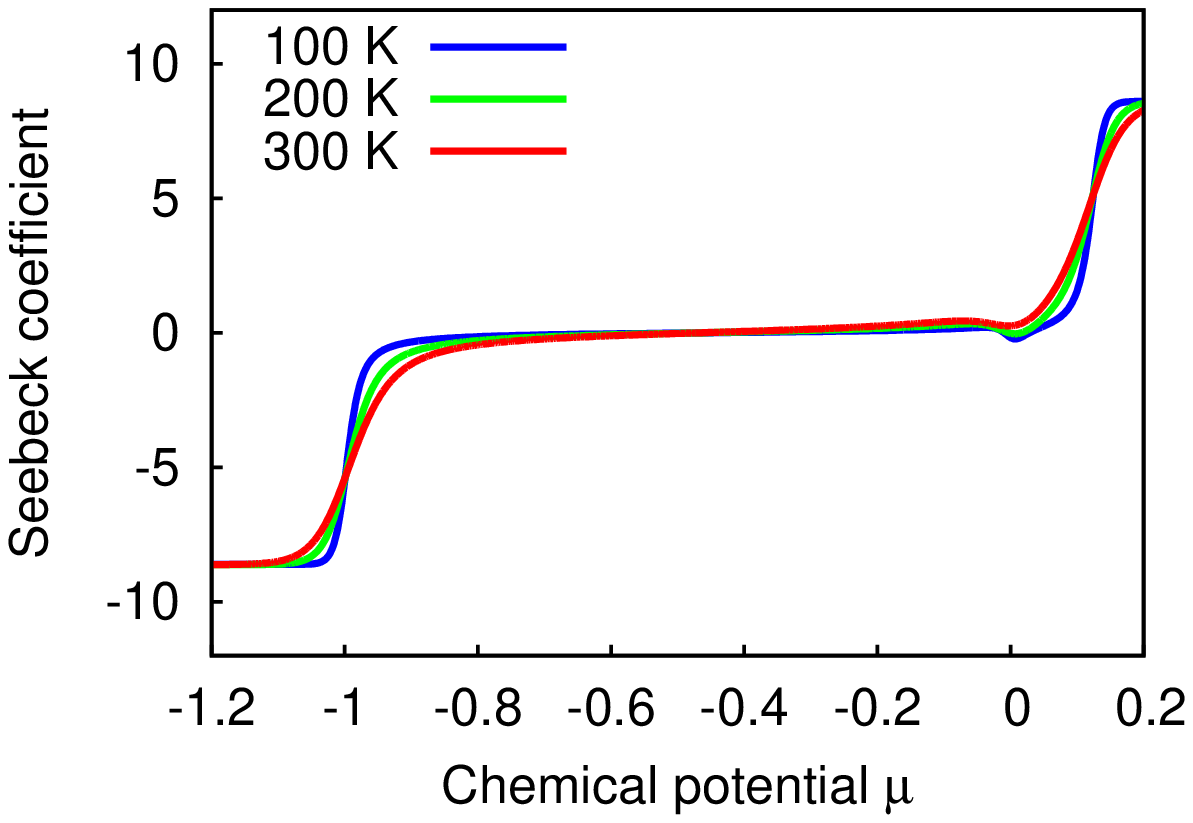}
\includegraphics[scale=0.4]{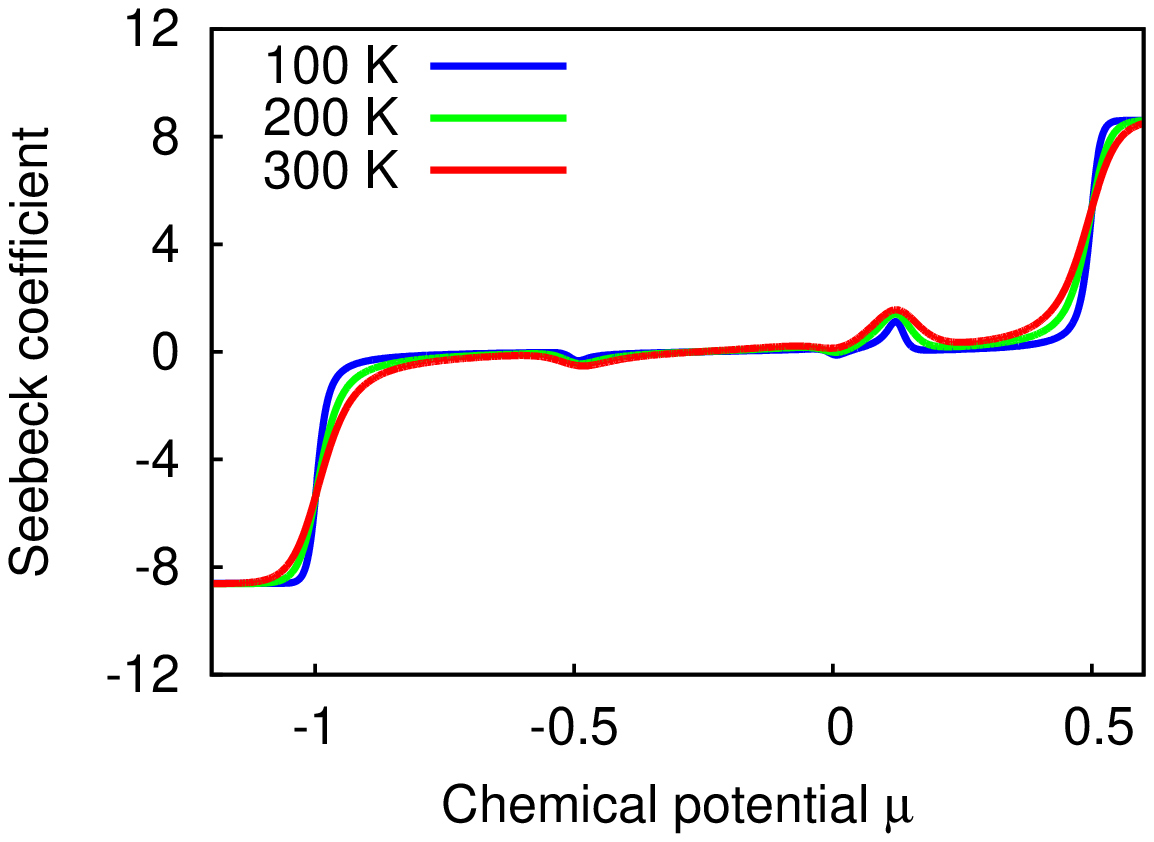}}
\caption{Model band structures in 1D Brillouin zone with the density of states (DOS),
electrical conductivity ($\sigma(\mu,T)$) obtained from the TDF and the Seebeck
coefficient  from the approximate Mott's formula [in arbitrary units].
Grey lines in the band structure plots guide an eye to the vanishing band-velocities regions.}
\label{b8}
\end{figure*}

The transport distribution function (TDF) is a 3D tensor $\Sigma_{ij}(E)$   
defined as a function of the energy $E$ as follows:
\begin{eqnarray}
\Sigma_{ij}(E) & = & \frac{1}{V} \; \sum_{n,{\bf k}} \; v_i(n,{\bf k}) v_j(n,{\bf k})
\; \tau_{n{\bf k}} \; \delta (E-E_{n,{\bf k}}),  \nonumber \\
v_i(n,{\bf k}) & = & \frac{1}{\hbar} \; \frac{\partial E_{n,{\bf k}}}{\partial k_i},
\nonumber
\end{eqnarray}
where $V$ is the system volume,  
$v_i(n,{\bf k})$ is the band velocity, $E_{n,{\bf k}}$ is the band dispersion,
and $\tau_{n{\bf k}}$ is the relaxation time dependent on the band index $n$ and 
the reciprocal space ${\bf k}$.
The TDF enters an expression for the 3D tensor of the Seebeck coefficients $S_{ij}$,  
which depends on the chemical potential $\mu$ and temperature T, and is defined as:
\begin{eqnarray}
S_{ij}(\mu,T) & = & \frac{1}{eT} \; K_{ij,1} (\mu,T) \; K_{ij,2}^{-1} (\mu,T), \nonumber 
\end{eqnarray}
where
\begin{eqnarray}
K_{ij,1} (\mu,T) & = &
\int_{-\infty}^{\infty} dE \; \left( -\frac{\partial f(E,\mu,T)}{\partial E} \right) \;
(E-\mu) \; \Sigma_{ij}(E), \nonumber \\
K_{ij,2} (\mu,T) & = &
\int_{-\infty}^{\infty} dE \; \left( -\frac{\partial f(E,\mu,T)}{\partial E} \right) \;
\Sigma_{ij}(E). \nonumber
\end{eqnarray}
In the above formulae, the function $f(E,\mu,T)$ is the Fermi-Dirac distribution 
and $e$ is the electron charge.

\begin{figure*}
\leftline{ \includegraphics[scale=0.7]{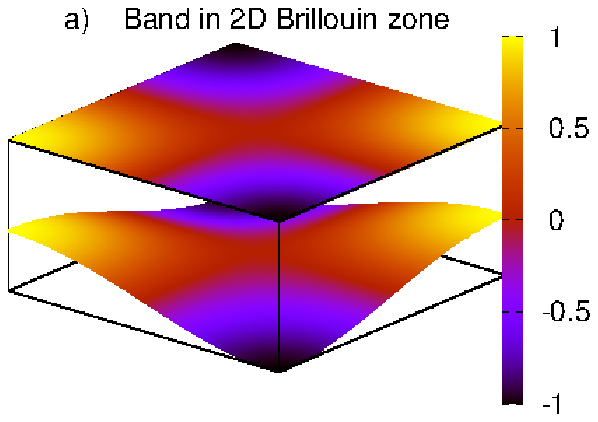}
\includegraphics[scale=0.4]{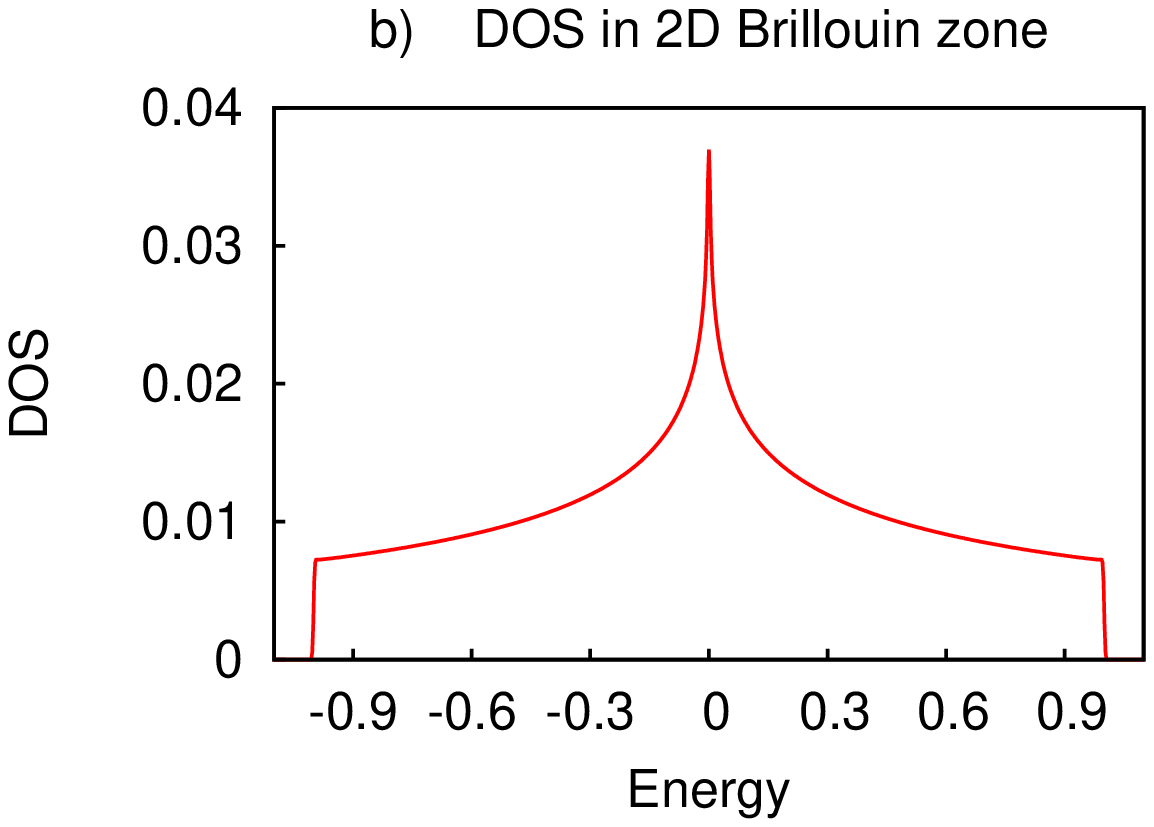}}
\vspace{2mm}
\leftline{ \includegraphics[scale=0.7]{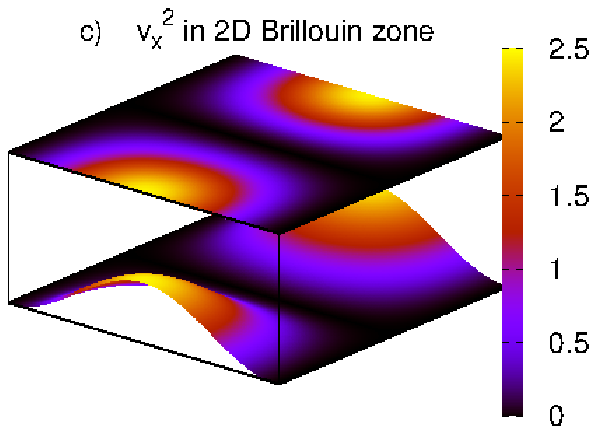}
\includegraphics[scale=0.4]{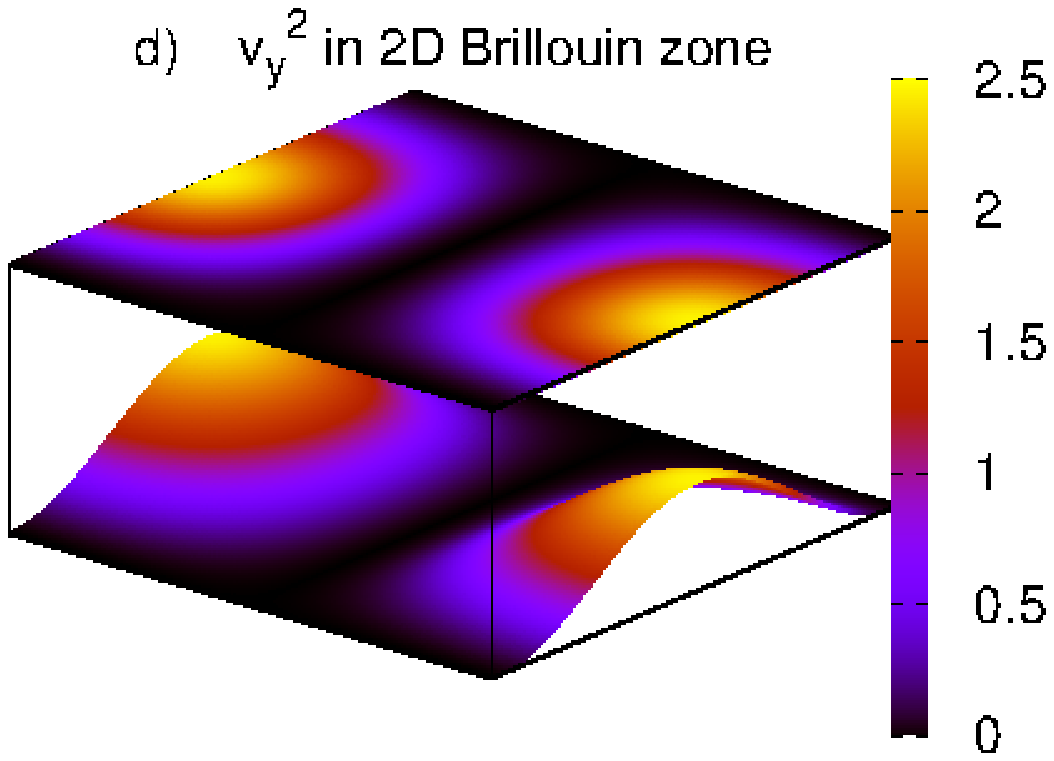}
\includegraphics[scale=0.7]{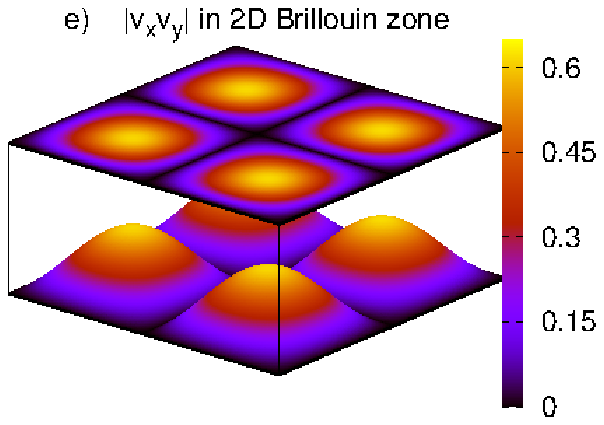}}
\vspace{2mm}
\leftline{ \includegraphics[scale=0.4]{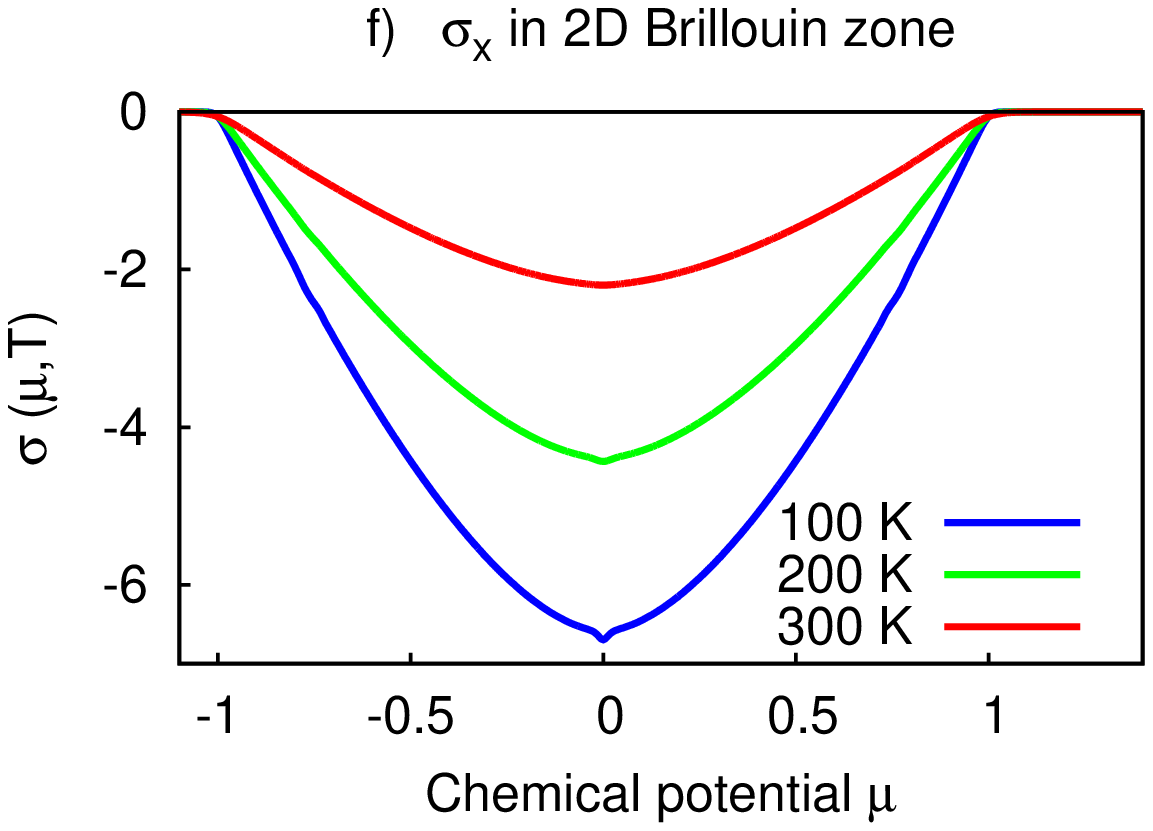}
\includegraphics[scale=0.4]{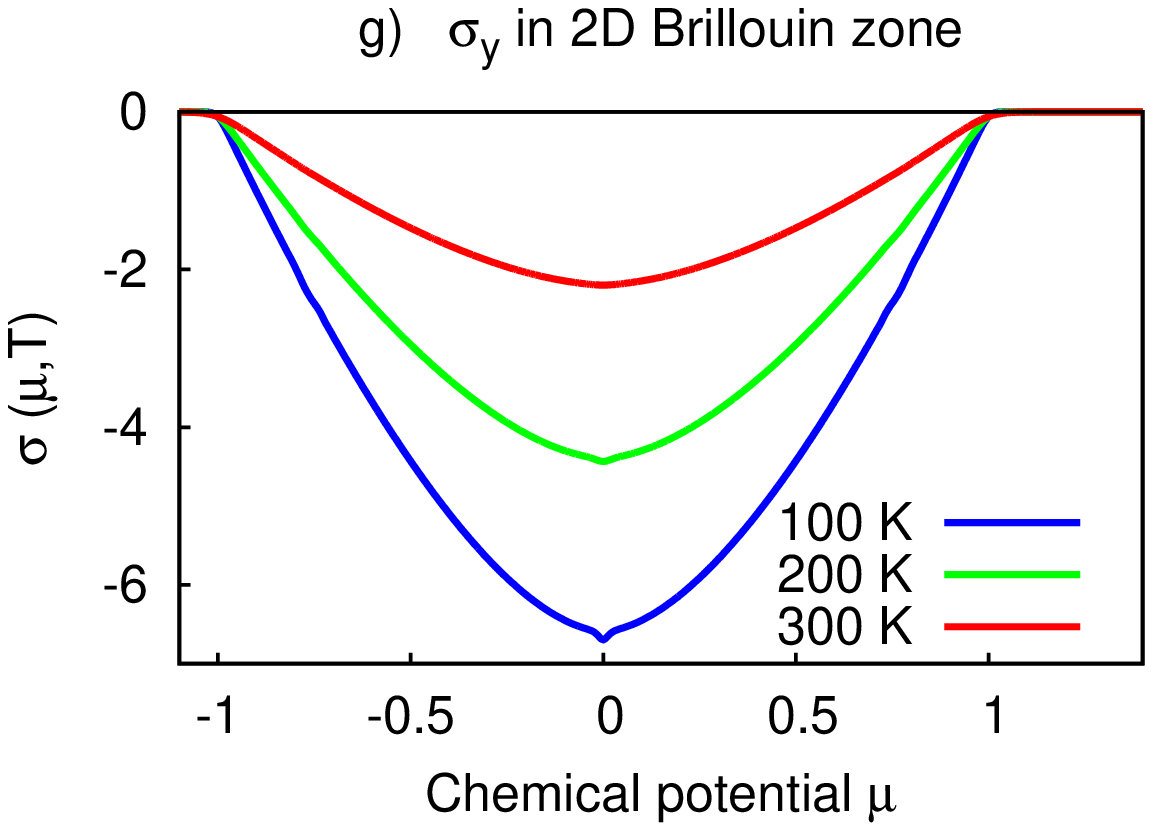}
\includegraphics[scale=0.4]{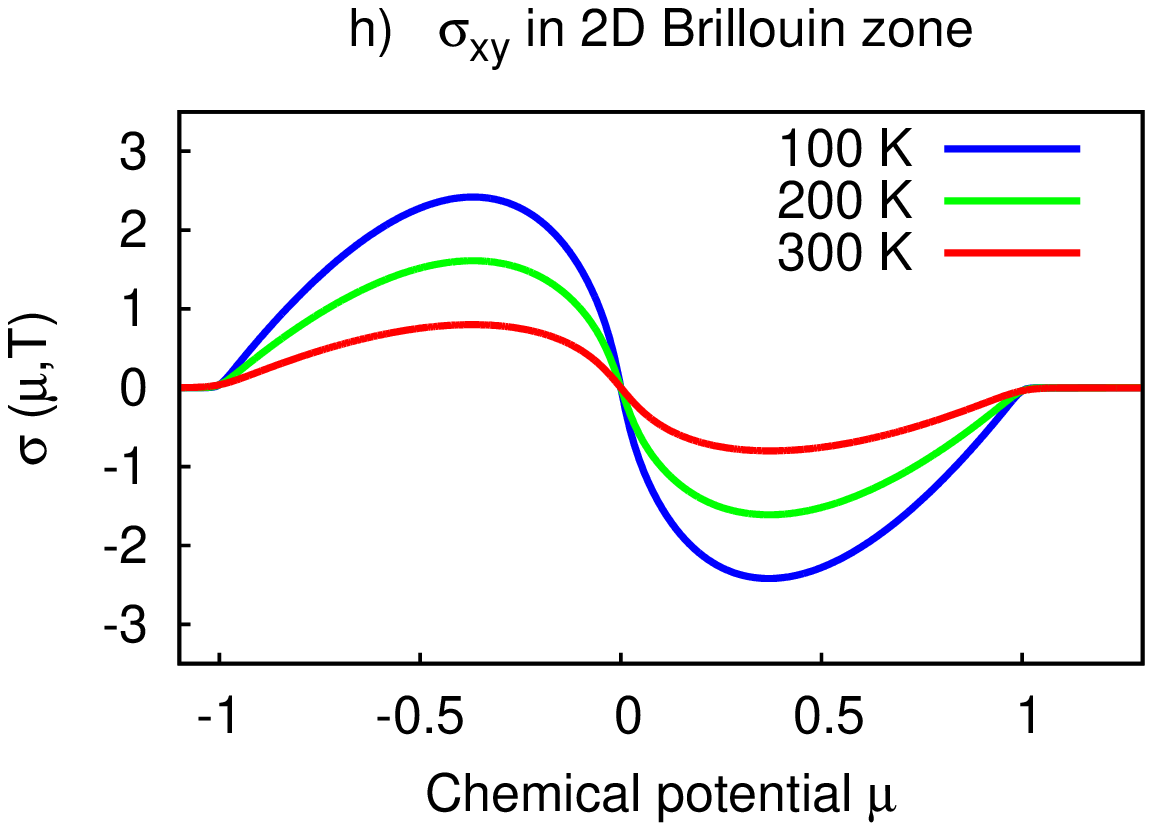}}
\vspace{2mm}
\leftline{ \includegraphics[scale=0.4]{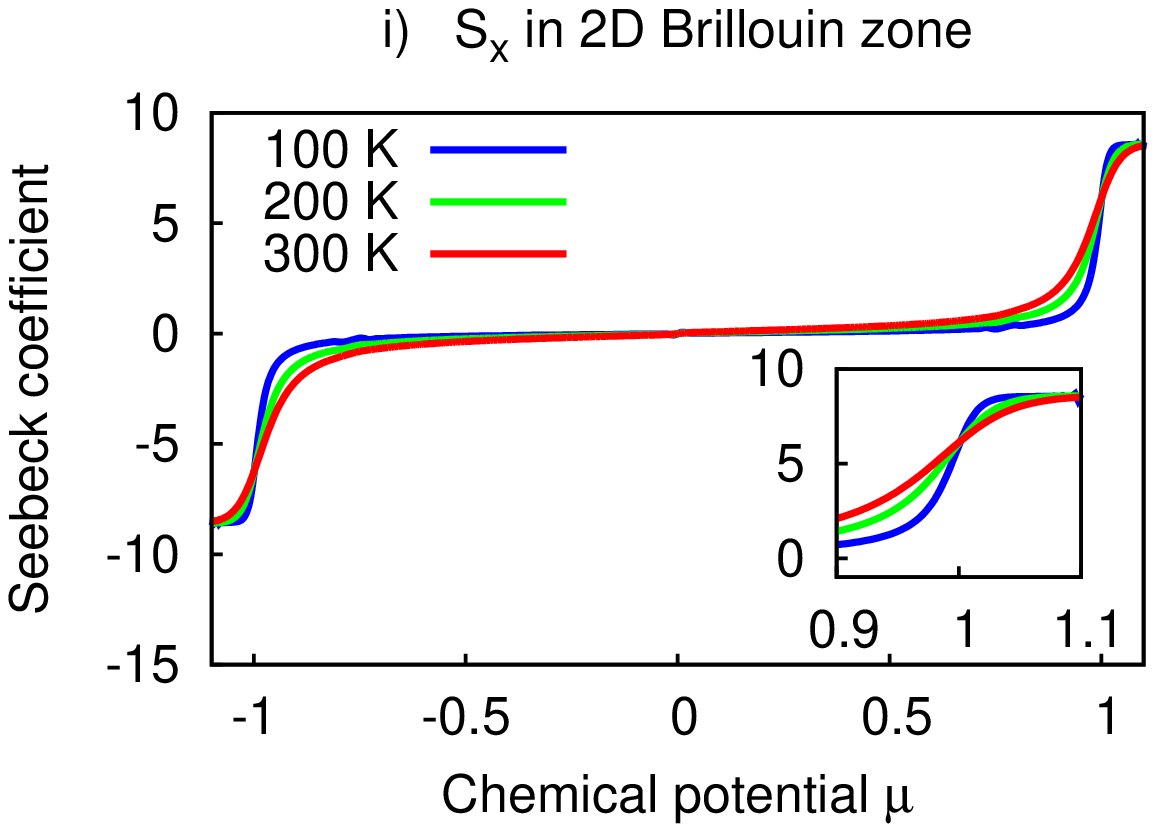}
\includegraphics[scale=0.4]{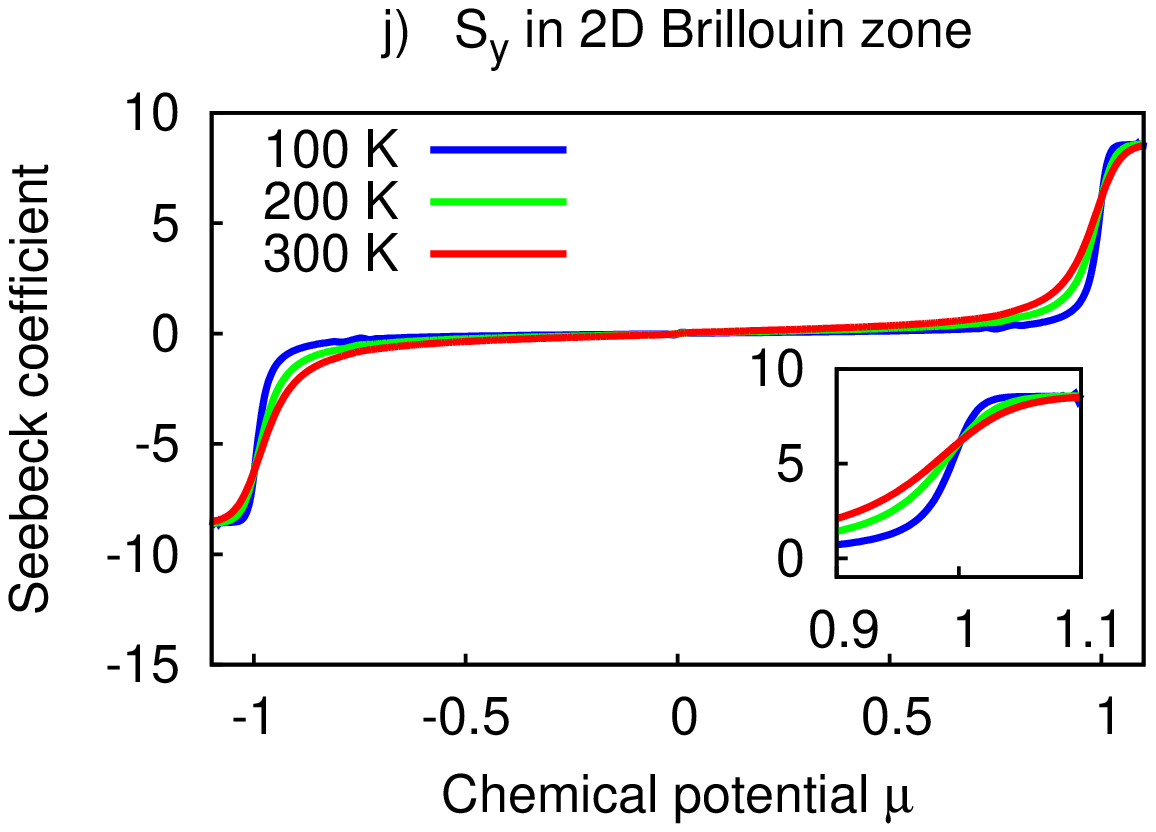}
\includegraphics[scale=0.4]{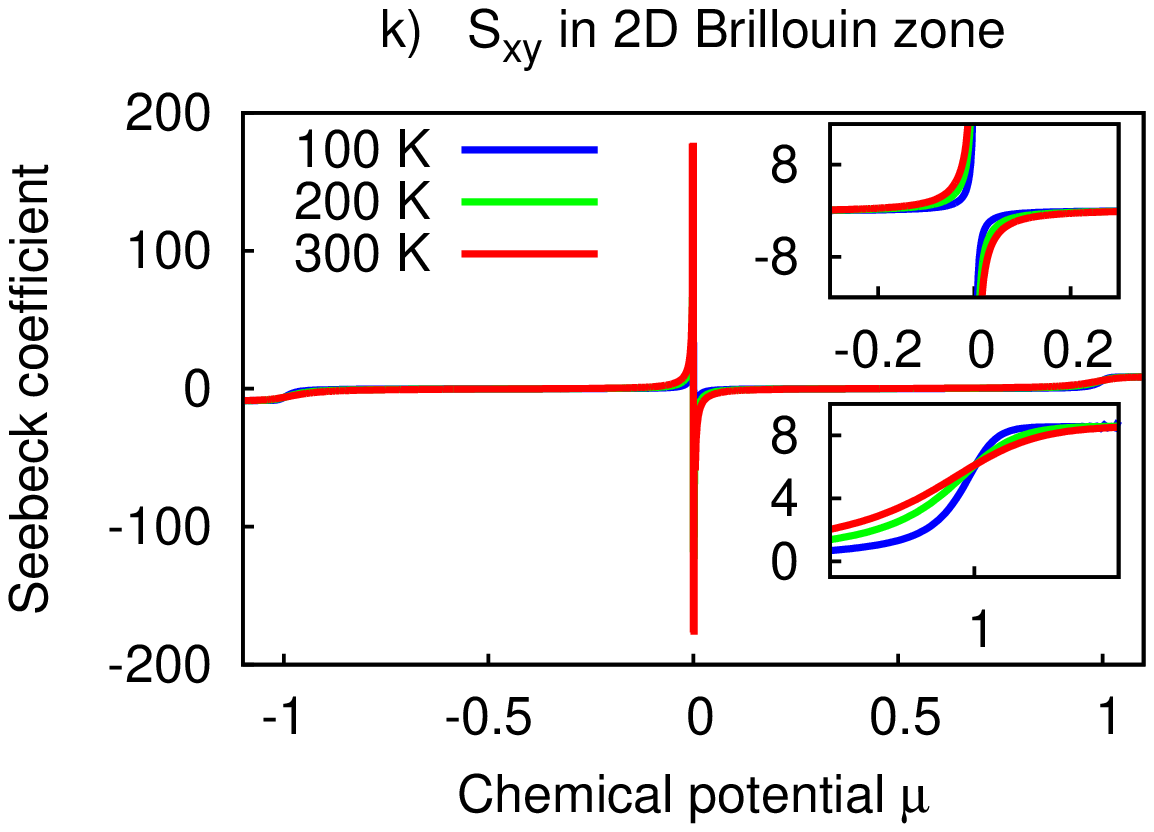}}
\caption{2D Brillouin zone, a) model band structure ($sin(\pi x/2)sin(\pi y/2)$) 
b) the density of states (DOS),
c)-e) the band velocities products: $v_x^2$, $v_y^2$ and $v_{x}v_{y}$,
f)-h) the corresponding electrical conductivities: $\sigma_x$, $\sigma_y$, $\sigma_{xy}$
obtained from the TDF and
i)-k) the Seebeck coefficients: $S_x$, $S_y$, $S_{xy}$
calculated from the approximate Mott's formula [in arbitrary units].}
\label{b9}
\end{figure*}

We begin, for simplicity,
with the model bands in 1D and skip the indeeces $ij$. 
In Fig.~7, we propose three band structures: 1) isolated symmetric band, 
2) isolated asymmetric band, 3) two bands of the previous models together,
which partially overlap in the energetic range. 
The constant relaxation time $\tau$ is asumed. 
We plot also the band velocities and the kernels entering the expression for the thermopower.
From the formula for $\Sigma_ij(E)$ and Fig.~7, it is obvious that the TDF is vanishing
when the band velocities are negligible.
Importantly, the kernel $K_2$ is smaller than the kernel $K_1$. Thus the Seebeck coefficient
as a function of the chemical potential is growing with the energetic range 
where the band extrema or termination occur. Summation over the band index
in the TDF expression makes the superposition of the effects from all bands   
- where the band velocities vanish for one band and the other not, the overall effect
is smoothed in the thermopower curve. Therefore, after summation over bands, only  
the effects from the highest and the lowest energetic parts of the band manifold rest. 

\begin{figure*}
\leftline{ \includegraphics[scale=0.6]{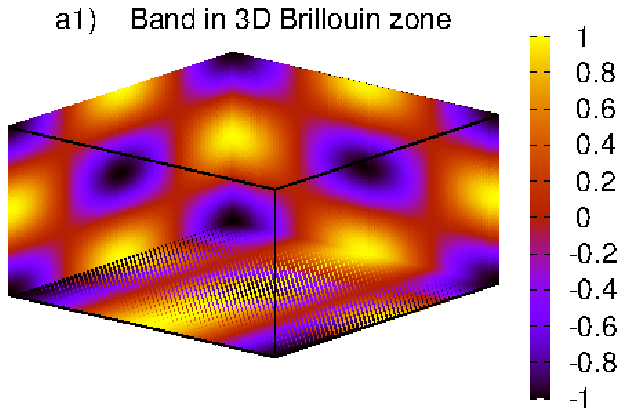}
\includegraphics[scale=0.35]{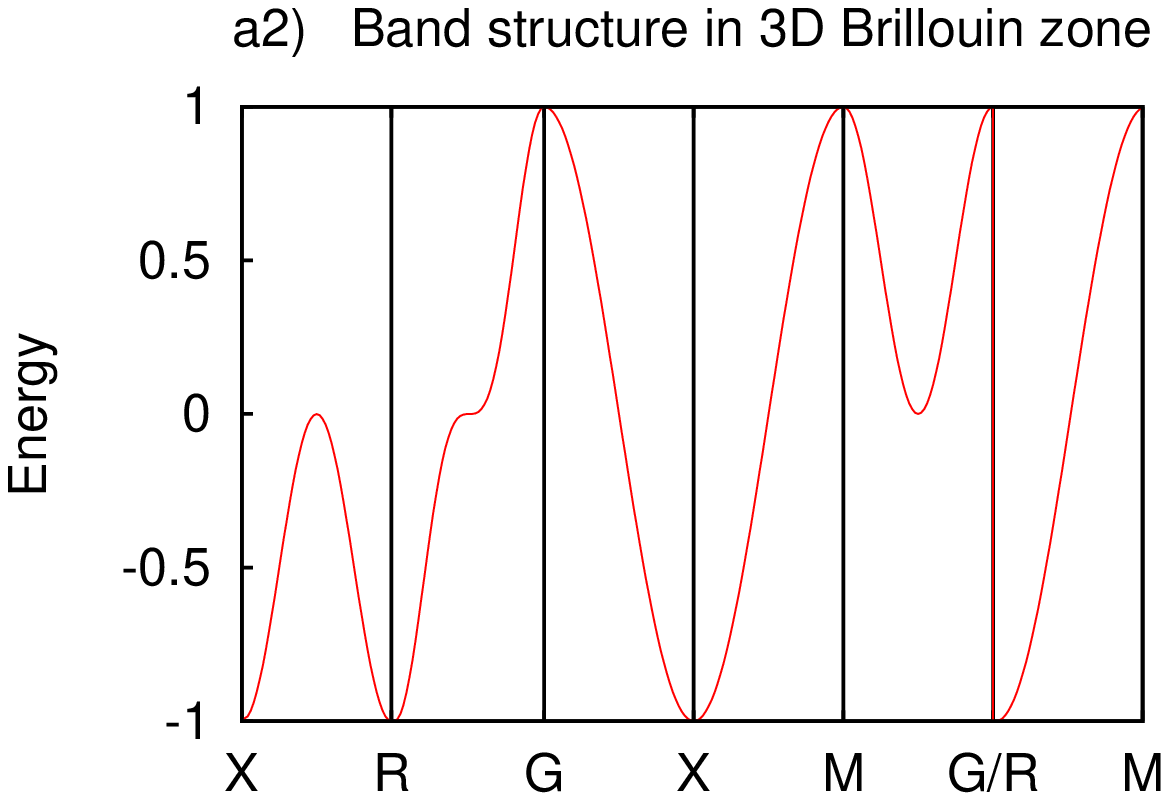}
\includegraphics[scale=0.35]{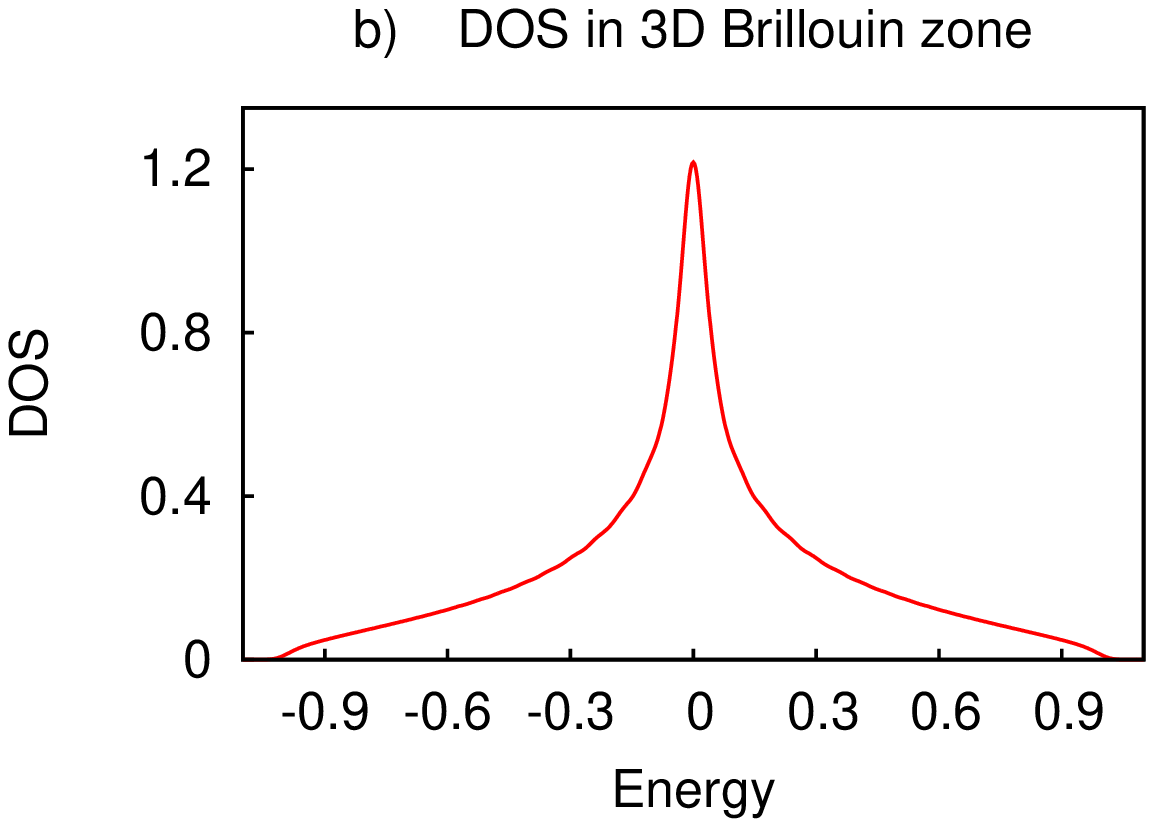}}
\vspace{2mm}
\leftline{ \includegraphics[scale=0.6]{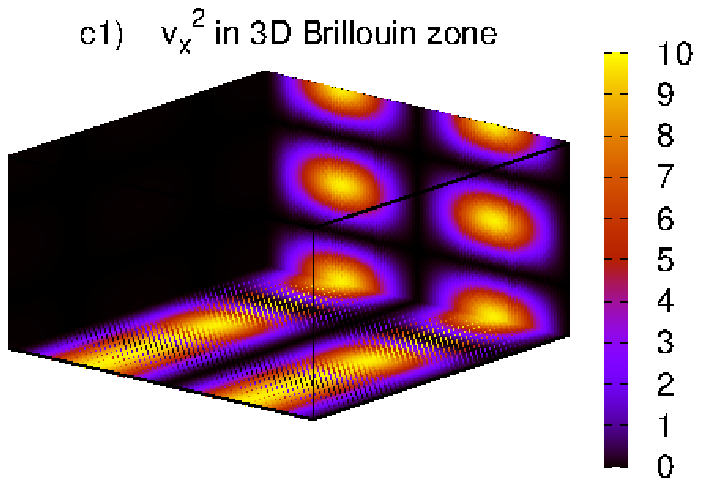}
\includegraphics[scale=0.35]{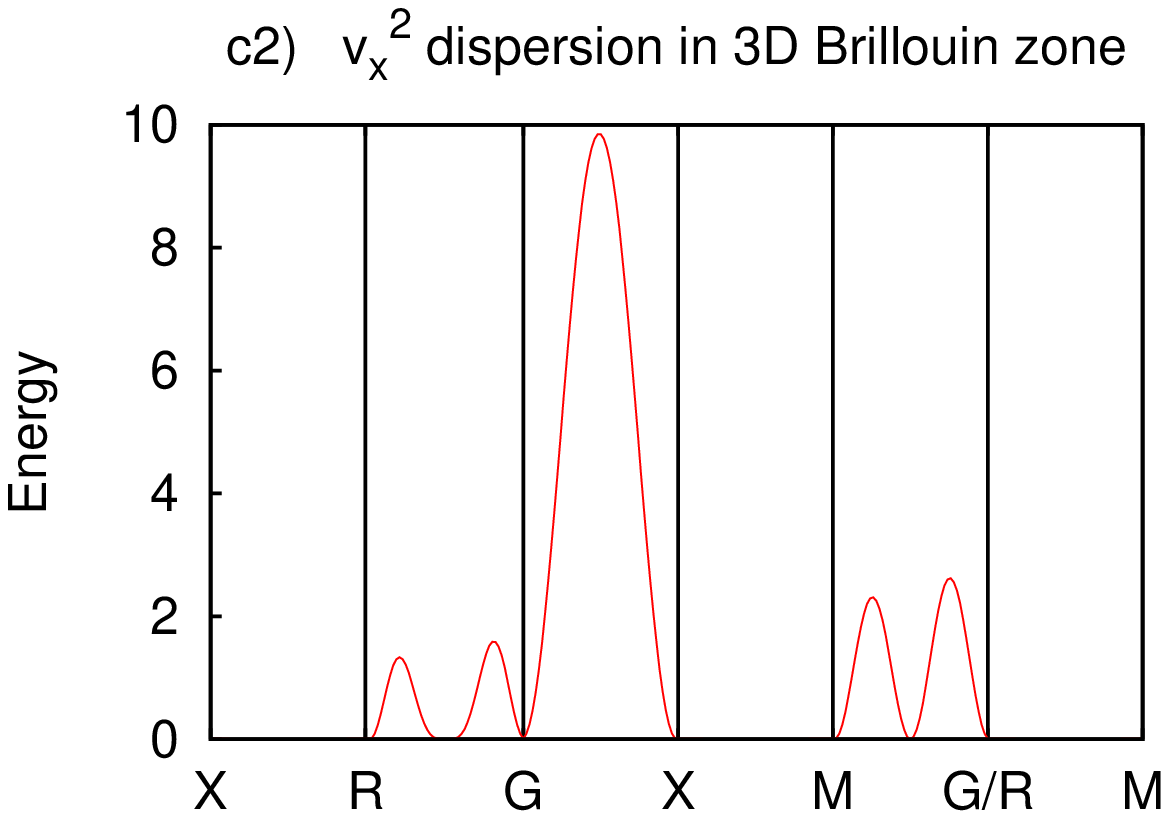}
\includegraphics[scale=0.35]{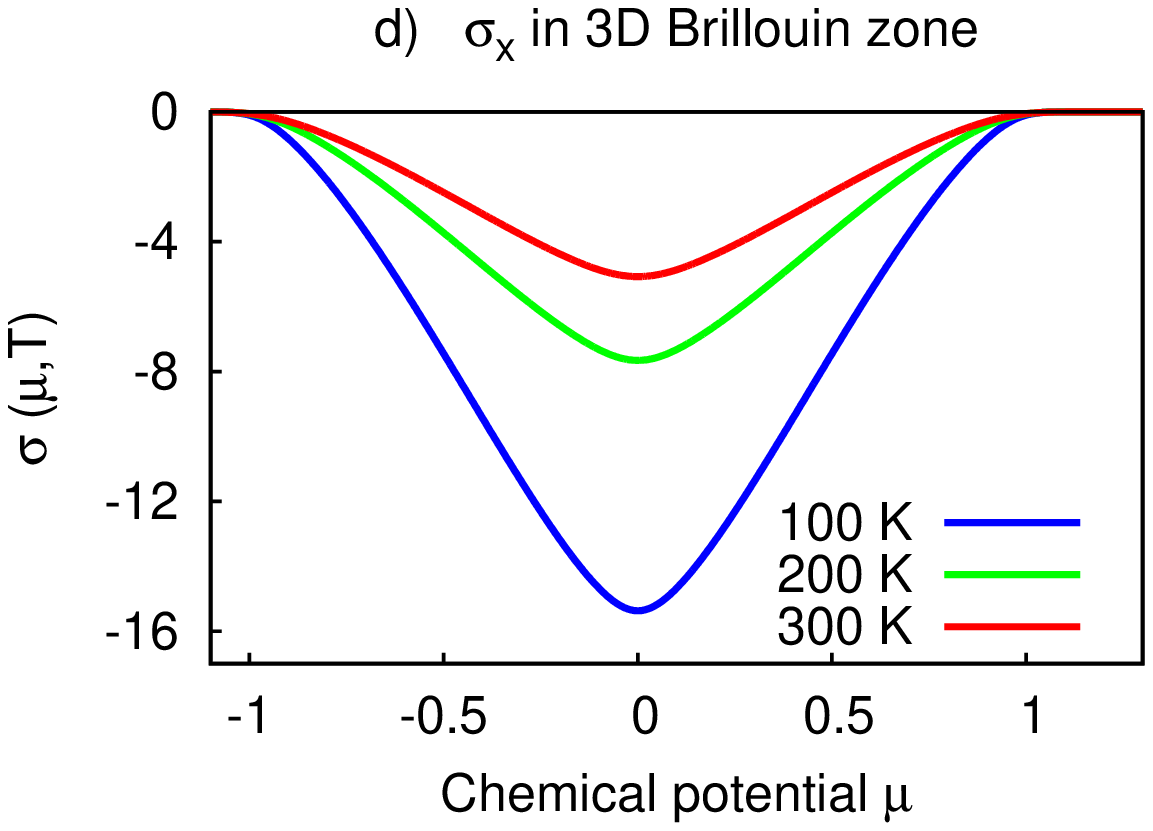}
\includegraphics[scale=0.35]{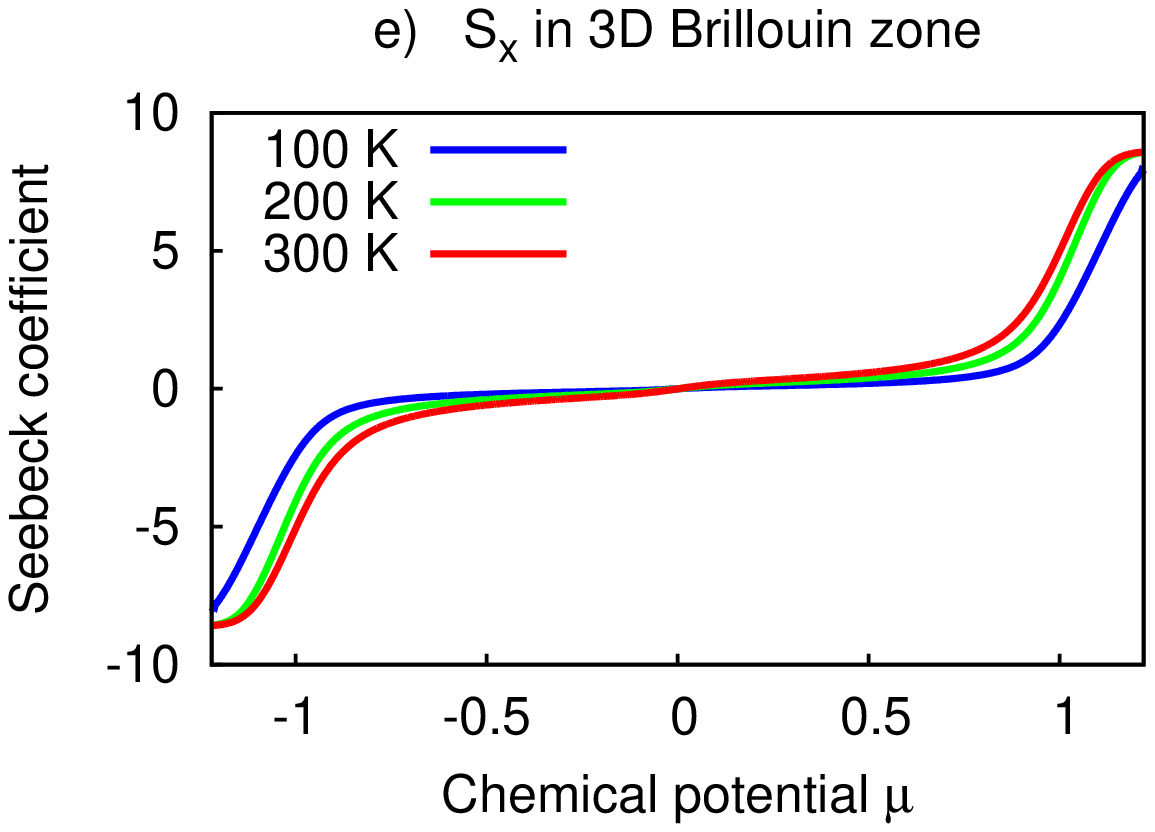}}
\caption{3D Brillouin zone, a1)-a2) model band structure ($cos(\pi x)cos(\pi y)cos(\pi z)$)
b) the density of states (DOS),
c1)-c2) the band velocities $v_x^2$ in the 3D plot and the BZ-dispersion plot,
d) the corresponding electrical conductivity $\sigma_x$ obtained from the TDF
and e) the Seebeck coefficient $S_x$
calculated from the approximate Mott's formula [in arbitrary units].}
\label{b10}
\end{figure*}
\begin{figure*}
\leftline{ \includegraphics[scale=0.35]{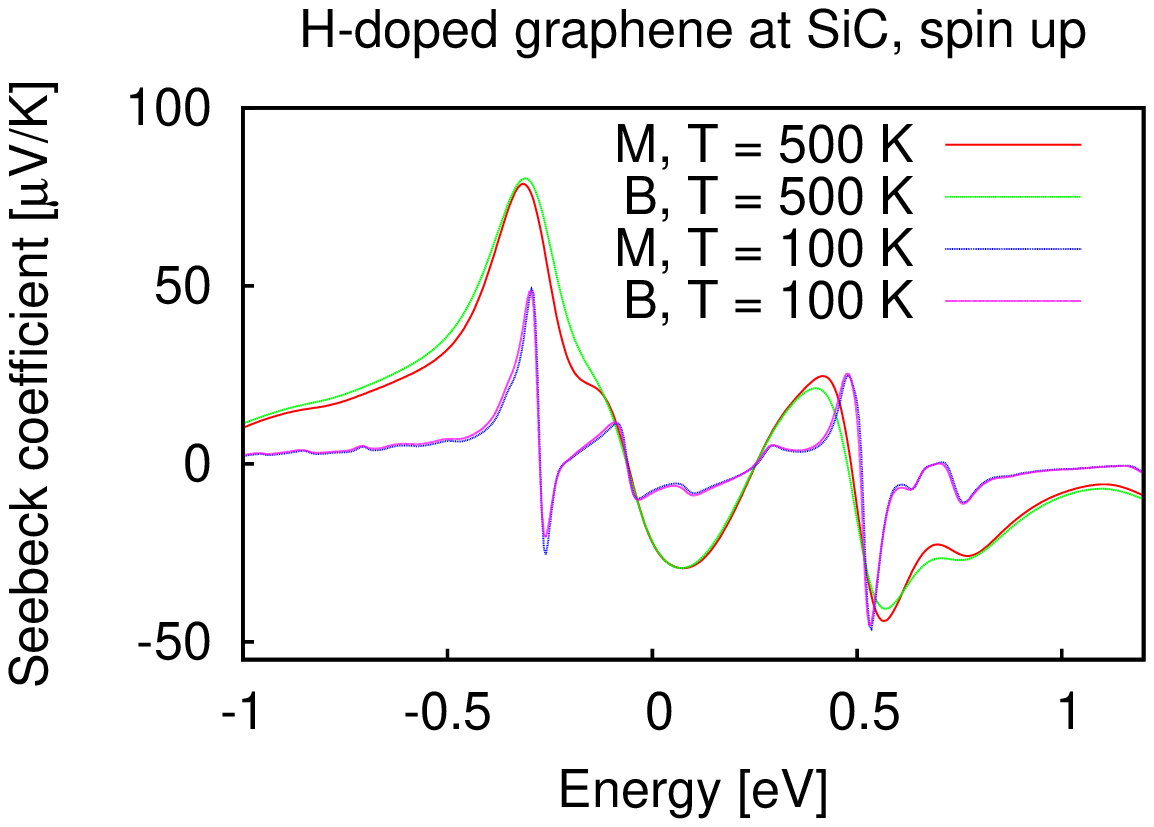}
\includegraphics[scale=0.35]{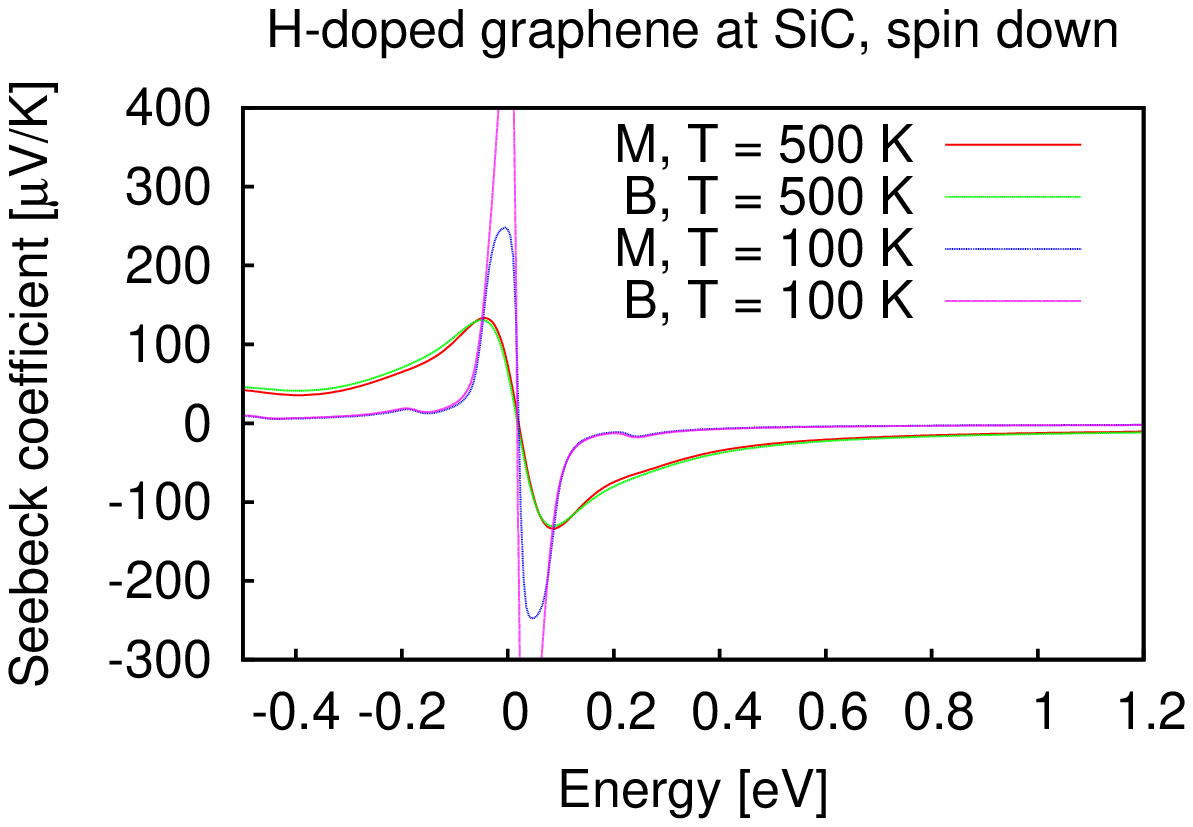}
\includegraphics[scale=0.35]{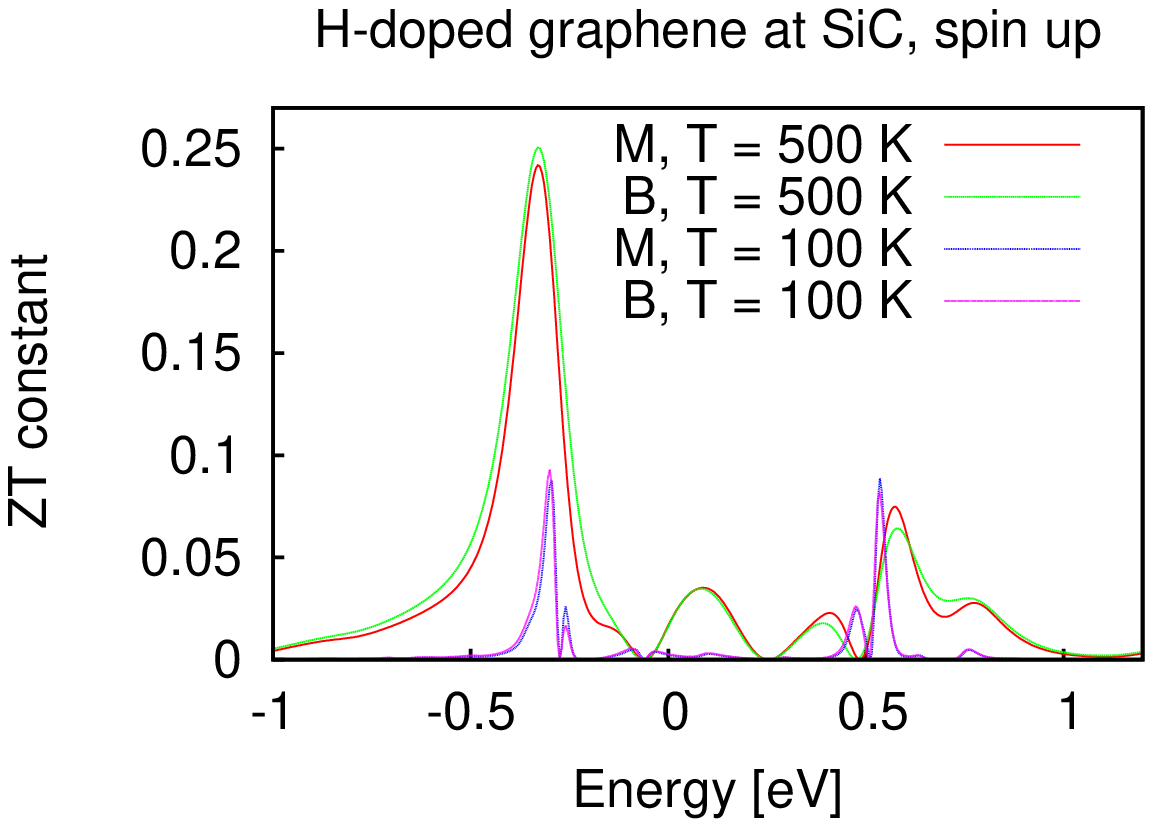}
\includegraphics[scale=0.35]{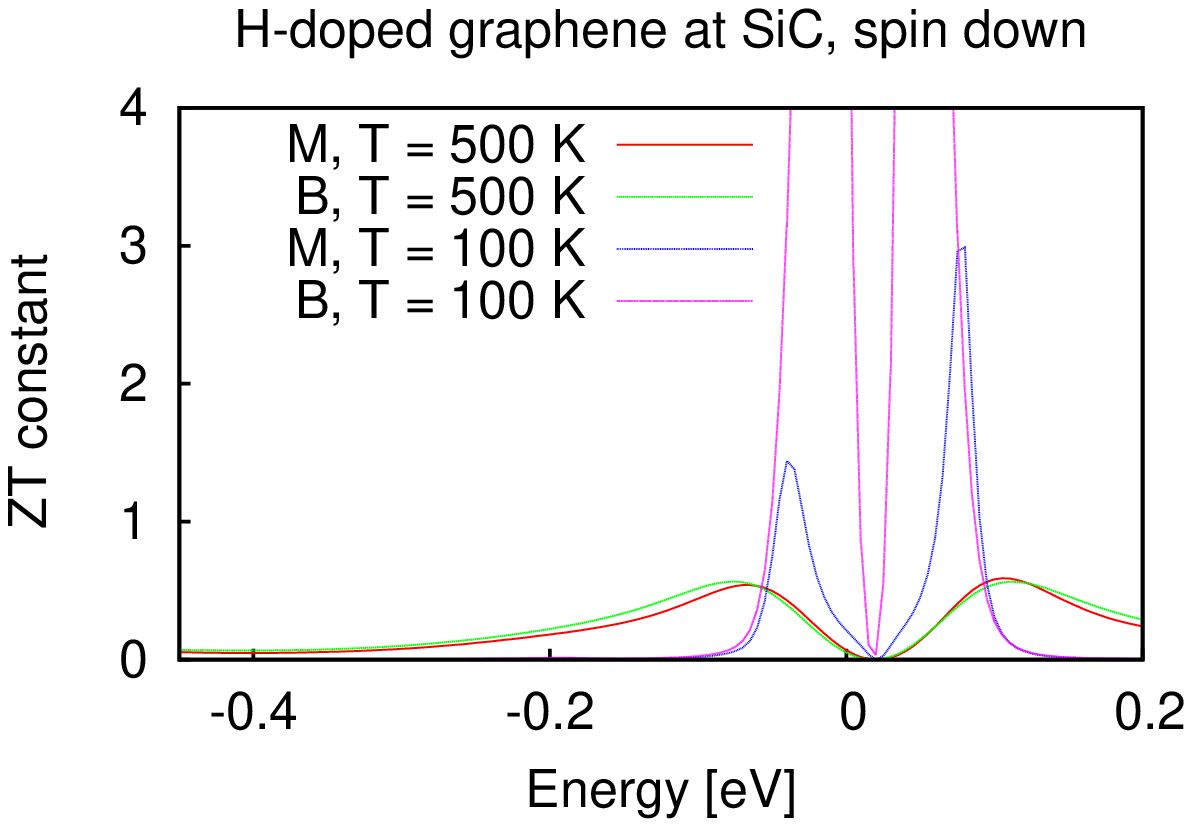}}
\caption{Comparison of the BoltzWann-code results (B) and Mott's formula (M)
for the H-doped graphene deposited at the SiC C-terminated substrate.}
\label{b11}
\end{figure*}

The effect is stronger for low temperatures but in a very tiny energetic range, which
depends also on the band width of the given system.
For higher temperatures, the thermopower growth is
pronounced at wider energetic range and could be easier measured.

If we did not assume that the relaxation time was constant,
the described features would be either better pronounced or smoothed
depending on the $\tau$ dispersion. On the other hand, the band velocities do not change
under the relaxation-time model.
Thus, one can draw the conclusion that the band termination \--
within the energetic range \-- will enhance the thermopower in any case.

In order to extend the analysis to 2D and 3D cases and compare the peaks of
the thermopower curve with the location of the Van Hove singularities, we use
also the approximate Mott relation which contains the full temperature dependent
electron conductivity $\sigma(\mu,T)$.
This formula was introduced by L\"ofwander and Fogelstr\"om
in Ref. \cite{mott} with an application to the disorder effects in graphene.
These approximate Mott formula reads:
\begin{eqnarray}
S & \approx & -(\pi^2/3)(k_B^2 T/e)\;\frac{d\;[ln\;\sigma(\mu,T)]}{d\;\mu}
\nonumber
\end{eqnarray}

In Fig.~8, we show again the 1D analysis of the thermopower components but
for the above Mott's formula.
In the band structure pictures, in the first row, we drow
the lines which guide an eye to the local extrema corresponding to the vanishing
band velocities. In the second row, we display the corresponding density of states
(DOS). Further, in the third row, we show the electrical conductivity obtained from
the TDF. In the last row, we plot the Seebeck coefficient fom the Mott's formula.
The growth of thermopower coinsides only with these Van Hove singularities which
correspond to the parts of the TDF where vanishing band velocities are not summed
over the Brillouin zone with the non-vanishing band velocities.

In Fig.~9 and Fig.~10, we present similar analysis like in Fig.~8, with the
difference that the bands are defined in 2D and 3D, with the corresponding
model band formulae $sin(\pi x/2)sin(\pi y/2)$ and
$cos(\pi x)cos(\pi y)cos(\pi z)$, respectively. The 3D band is very symmetric, therefore
all components of the tensors $\hat{sigma}$ and $\hat{S}$ are equal, since the same
occurs for the band velocities $v_x=v_y=v_z$.
All the conclusions drawn from the 1D model are still valid here. With an addition
that in 2D and 3D, the anisotropy is plaussible, and we see this effect in Fig.~9 k).
This is due to the fact that the energy isosurfaces (for E=0) overlap in the Brillouin
zone with the zeros of $v_x v_y$ function. The thermopower growth effect does not occur
for x and y diagonal components because the energy isosurfaces are larger than the
band velocities isosurfaces.
Thus non-vanishing components of the band velocities contribute to the TDF
from the perpendicular direction to the polarization of chosen element of the thermopower tensor
when the summation over BZ is performed.

In the end, we present a comparison of the two approaches: our Boltzmann-equations based scheme
and approximate Mott formula \cite{mott}. These methods are applied to the case of H-doped graphene
deposited at the SiC C-terminated surface. The results for the temperatures of 100 and 500 K are
displayed in Fig.~11. In the Mott's formula, we used the electrical conductivity calculated
from the DFT bands. The two approaches give the same or very similar curves. The difference
is in the height of the peaks at thermoelectrically enhanced parameters. Both methods lead 
to the same conclusions. 

The phenomenon is general for many materials, especially for doped semiconductors
where it should occur at the top of the valence band and the bottom of the conduction band.
It applies also at the superconducting gap in the graphene-superconductor junction,
as discussed by Wysoki\'nski and Spa\l{}ek in Ref. \cite{Spalek},
and in the superconductor-ferromagnet junction in the presence of the magnetic field,
as predicted by Ozaeta et al. in Ref. \cite{Ozaeta}.
The same effect has been found recently in carbon nanowires by Tan et al.
in Ref. \cite{C-wire}, where large Seebeck effect showed up at the boundary
of the zero-transmission windows.
Interestingly, large growth of the thermopower should be observed in materials which have many
flat bands, almost molecular-like states, such as molecular crystals or 2D organic semiconductors
possess. In the perovskites, the flat band regions at $\Gamma$-X and X-M lines will certainly
increase the thermopower, although these features are screened in the summation over
the band index and {\bf k}-space in the TDF expression. Similar effects have been observed
in Ref. \cite{perovskites} for the electron doped SrTiO$_3$ and KTaO$_3$ and are expected to 
occur in topological insulators.

\section{Conclusions} 

We investigated the thermoelectric properties of halfly H- and F-doped graphene.  
These systems are spin-polarized with the magnetic moments located at the undoped C-atoms.
The effect of deposition on the C-face $4H$-SiC(0001) surface with two buffer layers
have been also examined. The accurate calculations were performed for the band structures
obtained from the density-functional theory and interpolated using the maximally-localized
Wannier functions, finally to be embedded in the Boltzmann equations.  
We found very interesting effect, which manifests with a huge growth of the thermopower
and the ZT efficiency at the energies where the band manifolds terminate. 
The enhancement of the thermoelectric parameters at the band edges is a general property, 
not restricted to the graphene systems. This phenomenon deserves further
investigations, in order to pave the way for two-digital values of ZT, and deny 
an inconvenient truth about the thermoelectrics that there are limitations for 
the search of efficient materials \cite{ZT1}.
The interesting and technologically important fact is that the thermoelectric parameters 
at the band-termination energies can be very high at low temperatures, 
which is not plaussible in the ordinary high-ZT materials. Similar effects have been observed 
for gapped graphene \cite{Sharapov,Hao}, graphene-superconductor \cite{Spalek} and
graphene-ferromagnet \cite{Ozaeta} junctions, and also graphene nanowires \cite{C-wire}. \\

{\bf Acknowledgements} \\

We thank Prof. J\'ozef Barna\'s for helpful discussion.
This work has been supported by the European Funds for Regional Development
within the SICMAT Project (Contract No. UDA-POIG.01.03.01-14-155/09), 
Operational Program Innovative Economy (Contract No. POIG.01.01.02-00-097/09 
"TERMET - New structural materials with enhanced thermal conductivity") and
the National Science Center in Poland (the Project No. DEC-2012/04/A/ST3/00372).
Calculations have been performed in the Interdisciplinary Centre of
Mathematical and Computer Modeling (ICM) of the University of Warsaw
within the grants G51-2 and G47-7,
and by part supported by PL-Grid Infrastructure. \\

\end{document}